\documentclass[12pt]{article}
\usepackage[scheme=plain]{ctex}
\usepackage{amscd}
\usepackage{bm}
\usepackage{bbm}
\usepackage{mathrsfs}
\usepackage{amssymb}
\usepackage{graphics}
\usepackage{graphicx}
\usepackage{amsfonts}
\usepackage{amsmath}
\usepackage{array}
\usepackage{multirow}
\usepackage{booktabs}
\usepackage{subfigure}
\usepackage{float} 

\UseRawInputEncoding
\usepackage[numbers,sort&compress]{natbib}
\begin{document}
\date{}
\title{Interactions between rarefaction waves and dispersive shock waves in the Gerdjikov--Ivanov equation with non-convex hydrodynamics}
\author{Hua-Ying Ren, Rui Guo$\thanks{Corresponding author:
gr81@sina.com}$\
\\
\\{\em
School of Mathematics, Taiyuan University  of} \\
{\em Technology, Taiyuan 030024, China}
} \maketitle
\begin{abstract}
This paper systematically investigates admissible configurations of the interaction between a rarefaction wave (RW) and a dispersive shock wave (DSW) by constructing a two-step piecewise constant initial condition in the Gerdjikov--Ivanov (GI) equation with non-convex hydrodynamics. Owing to the two-valued mapping
associated with the Riemann invariants, the GI equation also supports contact DSWs (CDSWs), which are
included in the interaction analysis. Using Whitham modulation theory, we elucidate the self-similar descriptions of the incoming waves, derive the boundary matching conditions, and obtain analytical expressions of the full evolution before, during, and after the interaction through generalized hodograph transformation. The results show that the interaction of a RW with either a DSW or a CDSW can exhibit two distinct scenarios: in one case, the two waves exchange spectral parameters after the interaction and fully separate, propagating independently with modified phases; in the other case, the wave structures merge upon interaction and subsequently evolve into either a soliton train or a small-amplitude harmonic wave for a classical DSW, and into an algebraic soliton train for a CDSW. The theoretical results are in good agreement with numerical simulations. This work can provide theoretical support for wave interaction phenomena in non-convex dispersive systems.

\vspace{5mm}\noindent\emph{Keywords}: Dispersive shock wave; Rarefaction wave; Interaction; Gerdjikov--Ivanov equation; Whitham modulation theory

\end{abstract}

\maketitle

\vspace{7mm}\noindent\textbf{1 Introduction}
\hspace*{\parindent}
\renewcommand{\theequation}{1.\arabic{equation}}\setcounter{equation}{0}\\

The Gerdjikov--Ivanov (GI) equation
\begin{equation}
 i u_t+u_{xx}+i u^2u_x^*+\frac{1}{2}|u|^4u=0,
 \label{eq:GI}
\end{equation}
where $u=u(x,t)$ is complex valued and $^*$ denotes complex conjugation, is an integrable equation which can describe the propagation of optical pulses in fibers by considering
dispersion, quintic nonlinearity and self-steepening effects ~\cite{ck1,ck2,ck3,ck4,ck5}. It is the third type of derivative nonlinear Schrödinger (DNLS) equation which is gauge equivalent to the Kaup--Newell and Chen--Lee--Liu equations~\cite{ck6,ck7,ck8,ck9}. The integrable structure has supported extensive studies of localised solutions, including solitons and rogue waves obtained by inverse scattering and Darboux transformations ~\cite{ck10,ck11}. The present work concerns the solutions for nonlocal, large-scale modulated wave structures.

Equation (1.1) can be represented in a hydrodynamic form 
\begin{equation}
\begin{aligned}
\begin{cases} 
	\rho _t+2\rho _xv+2\rho v_x+\rho \rho _x&=0,\\
	v_t+2vv_x-\left( \rho v \right) _x-\rho \rho _x&=\frac{1}{\sqrt{\rho}}\left( \frac{\rho _{xx}}{2\sqrt{\rho}}-\frac{\rho _{x}^{2}}{4\rho ^{\frac{3}{2}}} \right) _x,
\end{cases} 
\end{aligned}
\end{equation}
by taking Madelung transformation
$$
u(x,t)=\sqrt{\rho(x,t)}e^{i\varphi \left( x,t \right)},\ v(x,t)=\varphi _x\left( x,t \right),
$$
where $\rho = \rho(x,t)$ and $v = v(x,t)$ are analogues of density and velocity in hydrodynamics, respectively. We consider a class of two-step initial value problems in which the initial values consist of three constant states separated by two transition points $x_1=0$ and $x_2=\sigma d,$ where $\sigma \in \left\{ 1,-1 \right\}$ and $d>0$. We restrict attention to the configurations in which one transition that has expansive distribution generates a continuous rarefaction fan, whereas the other that has compressive distribution leads to wave breaking or gradient catastrophe in dispersionless limit. In the presence of dispersion, the latter is regularised into a dispersive shock wave (DSW) through the interplay between nonlinearity and dispersion. Thus, a rarefaction wave (RW) is a slowly varying, non-oscillatory expansion, while a DSW exhibits an oscillatory domain with a soliton train at the leading edge and a small-amplitude harmonic wave at the trailing edge~\cite{ck12,ck13,ck14,ck15,ck16,ck17,ck18,ck19,ck20}. Let $
x_{min}=\min\{x_1,x_2\}, x_{max}=\max\{x_1,x_2\},$
the two-step initial condition is written as
\begin{equation}
\begin{aligned}
\boldsymbol{U}\left( x,0 \right) =\begin{cases}
	\boldsymbol{U}_L,&		x<x_{min},\\
	\boldsymbol{U}_M,&		x_{min}<x<x_{max},\\
	\boldsymbol{U}_R,&		x>x_{max},\\
\end{cases}\ \quad \boldsymbol{U}_j=\left( \rho _j,v_j \right) ^T 
\end{aligned}
\end{equation} At early times, the two resulting wave regions remain spatially separated by an intermediate constant state. If the leading boundary of one wave reaches the adjacent boundary of the other wave, the intermediate plateau disappears and a nonlinear interaction region is formed. From that time onward, the solution is no longer described by two independent structures: the oscillatory wave propagates on a continuously varying mean field of the RW, and the boundaries of the interaction region must be determined by matching its modulation solution with those in the non-interacting regions.

The interaction between RWs and DSWs, also interpreted as the refraction of DSWs, has been investigated for the evolution of an initial elevation or depression shelf in Korteweg--de Vries (KdV) equation, with analytical solutions for the full interaction process obtained through hodograph transformations~\cite{ck21}. A generalisation of head-on interaction of a RW and a DSW to bidirectional hydrodynamic system associated with the defocusing nonlinear Schrödinger (NLS) equation was solved in Ref.~\cite{ck22}. Beyond this kind of interaction, head-on and overtaking collisions between two DSWs have been systematically studied~\cite{ck23,ck24,ck25}; the interaction region gives rise to a modulated two-phase quasi-periodic solution. Further studies have examined the interaction of localised solitons with the mean fields of DSWs and RWs, elucidating conditions for soliton transmission or trapping and deriving the corresponding amplitude and phase shifts~\cite{ck26,ck27}. These results establish the interaction theory in the dispersive conservation laws with hyperbolic convex fluxes.

In this work we study the interaction of a RW and a DSW in the GI equation which differs from the classical convex model considered previously. Its dispersionless limit, characterized by the wave intensity $\rho$ and velocity $v$, yields a non-convex two-component hydrodynamic system which fails to satisfy the conditions of strict hyperbolicity and genuine nonlinearity. Consequently, the GI equation supports a much richer family of nonlinear wave patterns, including algebraic solitons, contact DSWs (CDSWs) and combined waves in addition to RWs and DSWs~\cite{ck28,ck29,ck30,ck31}. The CDSWs were originally categorized as a novel form of sinusoidal undular bore in Ref.~\cite{ck28} and were subsequently investigated for equations such as modified KdV, modified NLS (mNLS), and the Landau-Lifshitz equations. Here, we investigate not only the interaction of a RW with a classical DSW, but also the interaction involving a CDSW. Our analysis therefore aims to determine how non-convexity generates the admissible wave configurations, and to explore the differences in the resulting wave patterns compared with those in convex cases.

Whitham modulation theory is based on the averaging of conservation laws over a periodic wave whose
parameters vary slowly in space and time in accordance with the rule~\cite{ck32} 
$$
\langle \mathcal{F} \rangle \approx \frac{1}{L}\int_{0}^{L} \mathcal{F}(x',t)\,\mathrm{d}x'.
$$
Within the framework, Gurevich and Pitaevskii formulated a general approach to constructing a theoretical description of the formation and evolution of DSW, also called Gurevich--Pitaevskii problem~\cite{ck33}. Their construction represents the oscillatory region as a slowly modulated periodic wave matched to the adjacent dispersionless states through its soliton and small-amplitude edges. The theory has been used to construct analytical solutions for classical DSWs and CDSWs~\cite{ck34,ck35,ck36,ck37,ck38,ck39,ck40,ck41,ck42}. In particular, the single step initial value problem for the GI equation has been classified and shown to generate RWs, DSWs, CDSWs, and combined wave structures~\cite{ck43}. Wave breaking for the GI equation under square-root initial profiles has also been studied~\cite{ck44}. These results provide the basis for the two-step problem. However, they cannot determine the solution after the two initially separated wave regions overlap. And the issue for the GI equation is that the oscillatory part of the interaction can be either a DSW or a CDSW.

Here, we will make use of Whitham modulation theory for the interaction problem. Before the waves meet, the solution can be described by two self-similar structures separated by a constant plateau. After the plateau vanishes, the modulated periodic solution must satisfy boundary conditions inherited from both non-interacted waves. One boundary degenerates to the slowly varying RW, while the other retains the DSW or CDSW. Thus, the solution of interaction is no longer self-similar. To describe the interaction process, we will employ the generalized hodograph method proposed by Tsarev to solve the associated Whitham modulation equations~\cite{ck45}. 

The remainder of this paper is organised as follows: Section 2 introduces the periodic solutions of the GI equation and the fundamentals of Whitham modulation theory. Section 3 and 4 present the construction of initial conditions and the classification of interaction scenarios, followed by a detailed derivation of the analytical results covering all stages of the interaction process. Section 5 summarises the main conclusions of this work and outlines prospects for further study.

\vspace{7mm}\noindent\textbf{2 Zero-phase and one-phase modulated solutions for the Gerdjikov-Ivanov equation}
\hspace*{\parindent}
\renewcommand{\theequation}{2.\arabic{equation}}\setcounter{equation}{0}\\

The zero-phase and one-phase modulated solutions have been comprehensively discussed in Ref.~\cite{ck43, ck44}. Only the main results are presented here.
The zero-phase solution has the form
 \begin{equation}
\rho =-\frac{1}{2}\left( l_++l_- \right) \pm \sqrt{l_+l_-},\quad v=\frac{1}{2}\left( l_++l_- \right) ,
\end{equation}
where $l_+$, $l_-$ are Riemann invarients, which are obtained via diagonalization in the dispersionless limit of the hydrodynamic formulation (1.2). And the zero-phase Whitham equations can be written as
\begin{equation}
\begin{aligned}
&\frac{\partial l_{\pm}}{\partial t}+V_{\pm}\frac{\partial l_{\pm}}{\partial x}=0,\\
V_+=&\frac{3}{2}l_++\frac{1}{2}l_-,\quad V_-=\frac{1}{2}l_++\frac{3}{2}l_-,
\end{aligned}
\end{equation}

The one-phase solution of Eq.~(1.1) can be represented in terms of the Jacobi elliptic function and has the form
\begin{equation}
\rho\left(x,t\right)=\frac{\rho _2\left( \rho _4-\rho _1 \right) -\rho _4\left( \rho _2-\rho _1 \right) \text{cn}^2\left( \frac{1}{4}\sqrt{\left( \rho _3-\rho _1 \right) \left( \rho _4-\rho _2 \right)}\left( x-Vt \right) ,m \right)}{\rho _4-\rho _2+\left( \rho _2-\rho _1 \right) \text{sn}^2\left( \frac{1}{4}\sqrt{\left( \rho _3-\rho _1 \right) \left( \rho _4-\rho _2 \right)}\left( x-Vt \right)  ,m \right)}
\end{equation}
for $\rho _1\le \rho \le \rho _2$ and 
\begin{equation}
\rho\left(x,t\right)=\frac{\rho _3\left( \rho _1-\rho _4 \right) +\rho _1\left( \rho _4-\rho _3 \right) \text{cn}^2\left(\frac{1}{4}\sqrt{\left( \rho _3-\rho _1 \right) \left( \rho _4-\rho _2 \right)}\left( x-Vt \right)  ,m \right)}{\rho _1-\rho _3+\left( \rho _3-\rho _4 \right) \text{sn}^2\left(\frac{1}{4}\sqrt{\left( \rho _3-\rho _1 \right) \left( \rho _4-\rho _2 \right)}\left( x-Vt \right)  ,m \right)}
\end{equation}
for $\rho _3\le \rho \le \rho _4$, where
\begin{equation}
V=-\frac{1}{4}\sum_{j=1}^4{\rho _j},~m=\frac{\left( \rho _4-\rho _3 \right) \left( \rho _2-\rho _1 \right)}{\left( \rho _4-\rho _2 \right) \left( \rho _3-\rho _1 \right)},
\end{equation}
and $\rho _i,\ i=1,\ 2,\ 3,\ 4$ can be represented by parameters $l_i~(l_i<0)$ as
\begin{equation}
\begin{aligned}
\rho_1=(\sqrt{-l_2}-\sqrt{-l_1}+\sqrt{-l_3}+\sqrt{-l_4})^2/2,~
\rho_2=(\sqrt{-l_1}-\sqrt{-l_2}+\sqrt{-l_3}+\sqrt{-l_4})^2/2,\\
\rho_3=(\sqrt{-l_1}+\sqrt{-l_2}-\sqrt{-l_3}+\sqrt{-l_4})^2/2,~
\rho_4=(\sqrt{-l_1}+\sqrt{-l_2}+\sqrt{-l_3}-\sqrt{-l_4})^2/2,
\end{aligned}
\end{equation}
or
\begin{equation}
\begin{aligned}
\rho_1=(\sqrt{-l_2}-\sqrt{-l_1}+\sqrt{-l_3}-\sqrt{-l_4})^2/2,~
\rho_2=(\sqrt{-l_1}-\sqrt{-l_2}+\sqrt{-l_3}-\sqrt{-l_4})^2/2,\\
\rho_3=(\sqrt{-l_1}+\sqrt{-l_2}-\sqrt{-l_3}-\sqrt{-l_4})^2/2,~
\rho_4=(\sqrt{-l_1}+\sqrt{-l_2}+\sqrt{-l_3}+\sqrt{-l_4})^2/2.
\end{aligned}
\end{equation}
The periodic solutions have a period of
\begin{equation}
\mathfrak L=\frac{2\text{K}(m)}{\sqrt{\left( l_{3}-l_{1} \right) \left( l_{4}-l_{2} \right)}}.
\end{equation}
The Whitham equation for the Riemannn invariants $l_i~(i = 1, 2, 3, 4) $ are expressed by
\begin{equation}
\frac{\partial l_i}{\partial t}+V_i\frac{\partial l_i}{\partial x}=0,\ i=1,\ 2,\ 3,\ 4,
\end{equation}
where $V_i$ are expressed as
\begin{equation}
V_i=\left( 1 - \frac { \mathfrak { L } } { \partial _i \mathfrak { L } } \partial _i \right) V,
\end{equation}
and their explicit expression give
\begin{equation}
\begin{aligned}
V_1=\frac{1} {2}\sum_{i=1}^4{l_{i}+\frac{\left( l_{1}-l_{2} \right) \left( l_{1}-l_{4} \right) \text{K}( m )}{\left( l_{1}-l_{4} \right) \text{K}( m ) -\left( l_{2}-l_{4} \right) \text{E}( m )}},\\
V_2=\frac{1} {2}\sum_{i=1}^4{l_{i}+\frac{\left( l_{1}-l_{2} \right) \left( l_{2}-l_{3} \right) \text{K}( m )}{\left( l_{3}-l_{2} \right) \text{K}( m ) -\left( l_{3}-l_{1} \right) \text{E}( m )}},\\
V_3=\frac{1} {2}\sum_{i=1}^4{l_{i}+\frac{\left( l_{3}-l_{4} \right) \left( l_{3}-l_{2} \right) \text{K}( m )}{\left( l_{3}-l_{2} \right) \text{K}( m ) -\left( l_{4}-l_{2} \right) \text{E}( m )}},\\
V_4=\frac{1} {2}\sum_{i=1}^4{l_{i}+\frac{\left( l_{3}-l_{4} \right) \left( l_{1}-l_{4} \right) \text{K}( m )}{\left( l_{4}-l_{1} \right) \text{K}( m ) -\left( l_{3}-l_{1} \right) \text{E}( m )}},
\end{aligned}
\end{equation}
where $\text{K}(m)$ and $\text{E}(m)$ denote the complete elliptic integrals of the first and second kinds. 

In the limit of $m \rightarrow  1 $ $(l_2 \rightarrow  l_3)$, the Whitham velocities $V_i$ are degenerated to
$$
V_1=\frac{3}{2}l_1+\frac{1}{2}l_4,~
V_2=V_3=\frac{1}{2}l_1+l_3+\frac{1}{2}l_4,~
V_4=\frac{3}{2}l_4+\frac{1}{2}l_1.
$$
In the limit of $m \rightarrow  0 $, the Whitham velocities $V _i $ are degenerated to
$$
V_1=\frac{3}{2}l_1+\frac{1}{2}l_2,~V_2=\frac{3}{2}l_2+\frac{1}{2}l_1,~
V_3=V_4=2l_4+\frac{\left( l_1-l_2 \right) ^2}{2\left( l_1+l_2-2l_4 \right)},~l_3\rightarrow  l_4,
$$
or
$$
V_1=V_2=2l_1-\frac{\left( l_4-l_3 \right) ^2}{2\left( l_3+l_4-2l_1 \right)},~V_3=\frac{3}{2}l_3+\frac{1}{2}l_4,~V_4=\frac{3}{2}l_4+\frac{1}{2}l_3,~l_2\rightarrow  l_1.
$$

Using Eqs. (2.2), (2.3), (2.7), (2.8) and (2.10), we establish the Gurevich--Pitaevskii matching conditions as follows to realise a smooth transition between the zero-phase and one-phase solutions
\begin{equation}
\begin{aligned}
	l_1&=l_-,~l_2=l_+~~\text{at\,\,}m=0,\\
	l_1&=l_-,~l_4=l_+~~\text{at\,\,}m=1,
\end{aligned}
\end{equation}
or
\begin{equation}
\begin{aligned}
	l_3&=l_-,~l_4=l_+~~\text{at\,\,}m=0,\\
	l_1&=l_-,~l_4=l_+~~\text{at\,\,}m=1.
\end{aligned}
\end{equation}

Within the framework of GI equation, the Riemann invariants take negative values, and both the RWs and DSWs evolved from step initial data propagate along the negative x-axis. Nevertheless, they possess distinct characteristic velocities, which naturally gives rise to an overtaking process: the faster wave structure gradually catches up with the slower one, driving the two initially independent waves into a nonlinear interaction regime.

Based on the foregoing results, we will systematically proceed to investigate the interaction between a RW and a DSW or a CDSW, which exhibits two regimes: overtaking followed by separation and overtaking followed by merging.

\vspace{5mm}\noindent\textbf{3 Overtaking followed by separation}
\hspace*{\parindent}
\renewcommand{\theequation}{3.\arabic{equation}}\setcounter{equation}{0}\\

For the case of overtaking followed by separation, we consider three distinct initial wave configurations with the DSW or CDSW positioned at the origin. In the first configuration, the RW lies to the right of the DSW, and overtakes the DSW during their propagation and after interaction, the two waves separate and travel apart. In the second configuration, the RW lies to the left of the DSW and the DSW possesses a faster speed and overtakes the RW. Similarly, the two waves interact transiently before fully separating and evolving independently thereafter. In particular, we investigate the interaction between the RW and CDSW, which replaces the DSW in this configuration.

\vspace{5mm}\noindent\textbf{3.1 RW overtakes DSW}
\hspace*{\parindent}\\

Configuration: Suppose the DSW and RW are generated at 
$(0,~0)$ and $(d,~0)$ on the $(x,~t)$ plane at $t=0$, respectively. Then both waves propagate leftward, with their respective expanding regions given by $x_1^-(t)<x(t)<x_1^+(t)$ and $x _2^-(t)<x(t)<x_2^+(t)$. We construct the following initial data:
\begin{equation}
\begin{aligned}
l_+(x,0)=
\begin{cases} 
l_+^L, & x<0,\\
 l_+^R, & x>0, 
\end{cases}
\quad \text{and} \quad
l_-(x,0)=
\begin{cases} 
l_-^L, & x<d, \\
l_-^R, & x>d,
\end{cases}
\end{aligned}
\end{equation}
where $l_-^L<l_-^R<l_+^R<l_+^L$ and$~d>0$. Then the corresponding density $\rho$ and velocity $v$ take the following piecewise functional forms (see Fig. 1)
\begin{equation}
\begin{aligned}
\rho(x,0)=
\begin{cases} 
-\frac{1}{2}\left( l_+^L+l_-^L \right) \pm \sqrt{l_+^Ll_-^L}, &x<0,\\
-\frac{1}{2}\left( l_+^R+l_-^L \right) \pm \sqrt{l_+^Rl_-^L}, & 0<x<d,\\
-\frac{1}{2}\left( l_+^R+l_-^R \right) \pm \sqrt{l_+^Rl_-^R}, & x>d, 
\end{cases}
\end{aligned}
\end{equation}
\begin{equation}
\begin{aligned}
v(x,0)=
\begin{cases} 
\frac{1}{2}\left( l_+^L+l_-^L \right), & x<0, \\
\frac{1}{2}\left( l_+^R+l_-^L \right),& 0<x<d, \\
\frac{1}{2}\left( l_+^R+l_-^R \right), & x>d. 
\end{cases}
\end{aligned}
\end{equation}

$(i)$ Before interaction $(0<t<t_1)$:

The DSW region in $x_1^-(t)<x(t)<x_1^+(t)$ is described by a self-similar modulated solution with
\begin{equation}
\begin{aligned}
&l_1= l_-^L ,~l_2 = l_+^R ,~l_4 = l_+^L,~x=V_3(l_-^L ,l_+^R , l_3 , l_+^L )t,\\
V _ { 3 } (l_1 ,l_2 , l_3 , l_4 ) &= \frac{1}{2} (l_-^L +l_+^R +~l_3 + l_+^L) + \frac{(l_+^R - l_3) (l_3 - 
      l_+^L) \text{K}(m)} {(-l_+^R + 
       l_+^L) \text{E}(m) + (l_+^R - 
       l_3) \text{K}(m)},
\end{aligned}
\end{equation}
where $m = \frac { (l_+^R-l_-^L) ( l_+^L - l_3 ) } { ( l_+^L - l_+^R ) ( l_3 - l_-^L ) }$. We then obtain the boundaries of the DSW
\begin{equation}
x _ { 1 } ^{-} = \frac {(l_-^L + l_+^L + 2 l_+^R)} {2} t , ~x _ { 1 } ^{+} =  \frac { l_-^{L2}+4l_-^Ll_+^L-8l_+^{L2}-2l_-^Ll_+^L+4l_+^Ll_+^R+l_+^{R2}} {2(l_-^L - 2l_+^L + l_+^R)} t .
\end{equation}

The RW region $x_2^-(t)<x(t)<x_2^+(t)$ centred at $x=d$, which is asymptotically governed by the zero-phase Whitham equations
\begin{equation}
\begin{aligned}
 &l_+= l_+^R , \\
 \frac{x-d}{t}=&V _ { - } (l_+ ,l_- ) = \frac{3l_-+l_+}{2}.
\end{aligned}
\end{equation}
The boundaries $x_{2}^{\pm}$ have the following forms
\begin{equation}
x _ { 2 } ^{-} = d + \frac { 3 l_-^L + l_+^R } { 2 }t ,\   x _ { 2 } ^{+} = d + \frac { 3 l_-^R + l_+^R } { 2 } t .
\end{equation}
\begin{figure}[htbp]
\centering
\setcounter{subfigure}{0}
\subfigure[]{\includegraphics[width=0.33\linewidth]{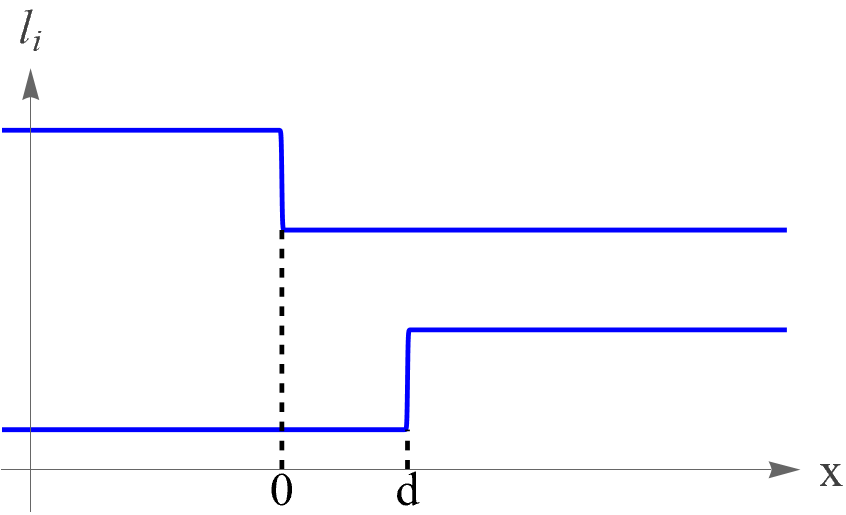}}\hfill
\subfigure[]{\includegraphics[width=0.33\linewidth]{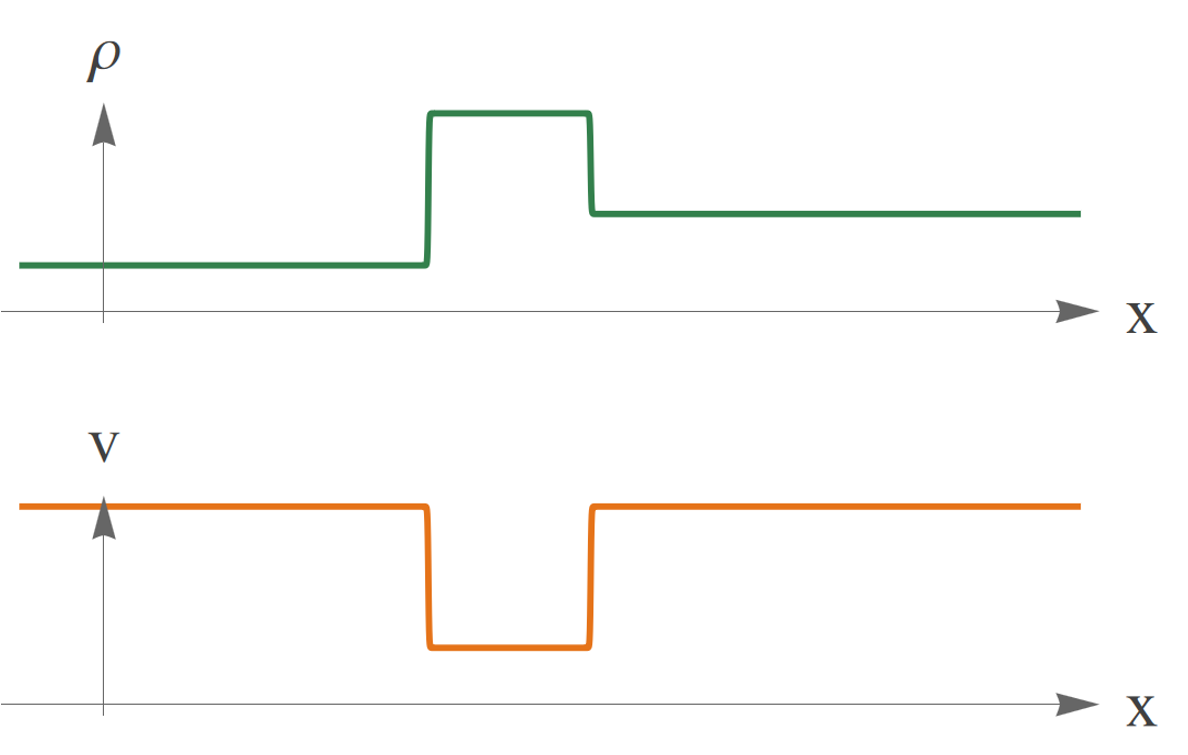}}\hfill
\subfigure[]{\includegraphics[width=0.33\linewidth]{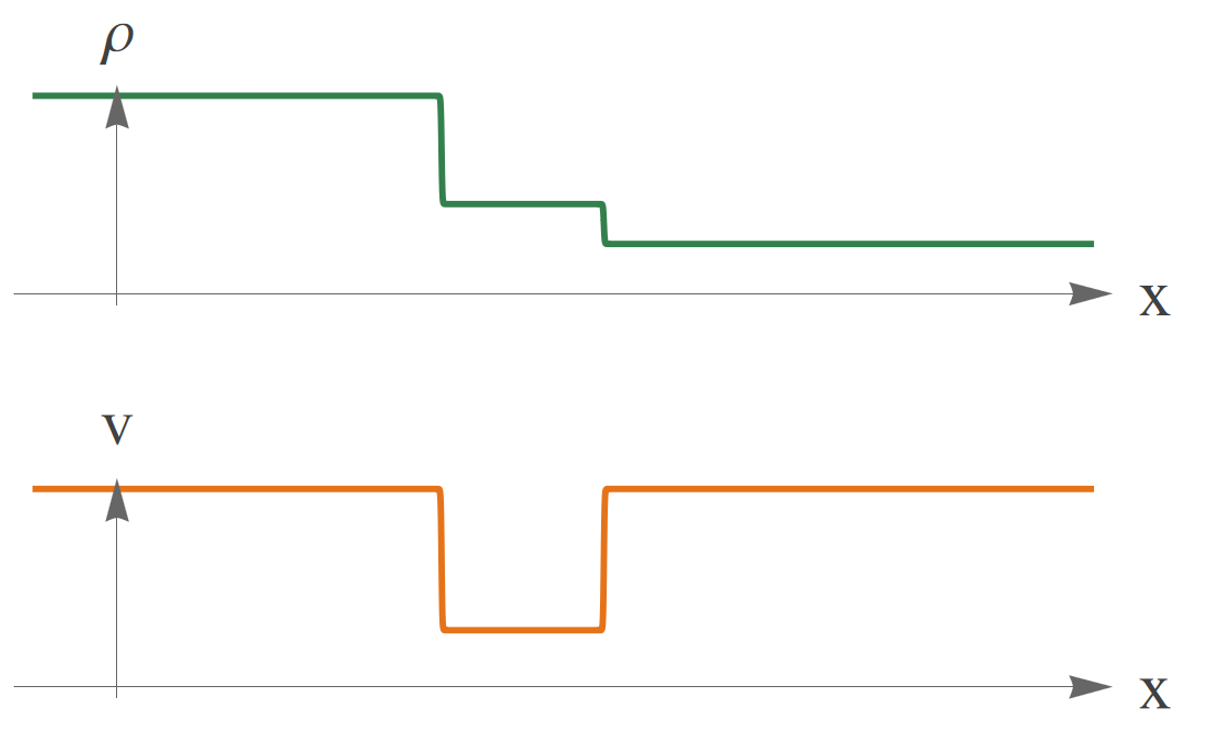}}
\flushleft{\footnotesize
\textbf{Fig.~$\bm{1}$.} The initial configuration (3.1) of the Riemann invariants, along with two sets of initial density and velocity corresponding to the constant states of the Riemann problem.}
\end{figure}
\begin{figure}[htbp]
\centering
\setcounter{subfigure}{0}
\subfigure[]{\includegraphics[width=0.33\linewidth]{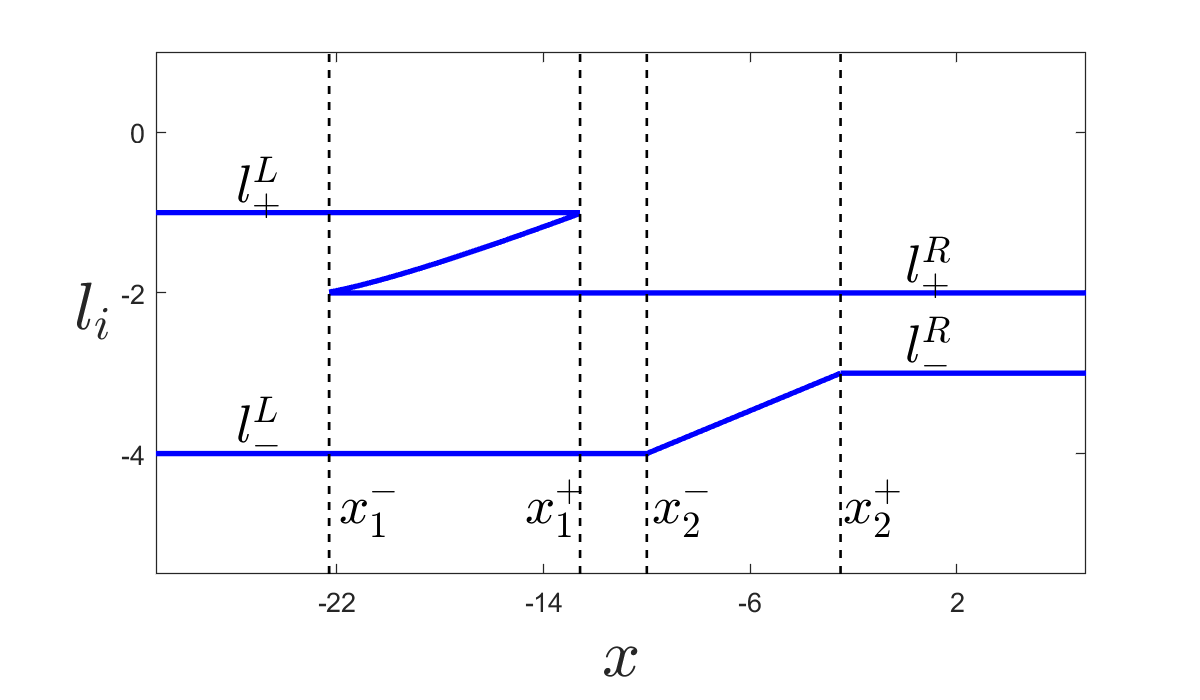}}\hfill
\subfigure[]{\includegraphics[width=0.33\linewidth]{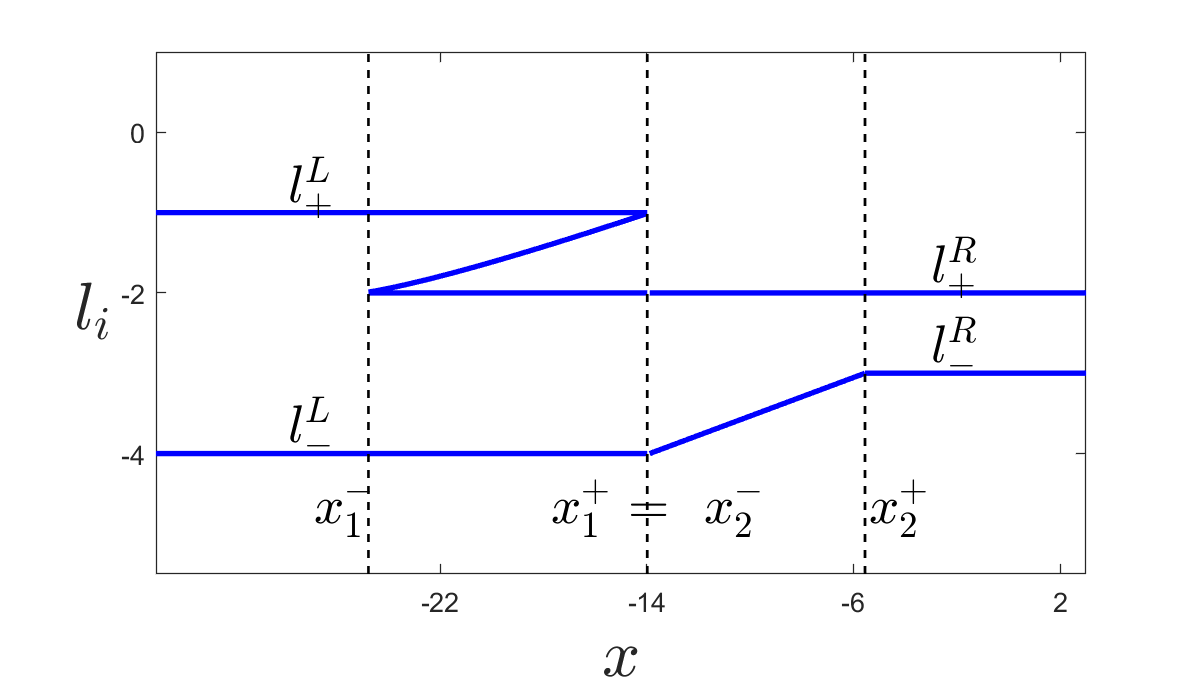}}\hfill
\subfigure[]{\includegraphics[width=0.33\linewidth]{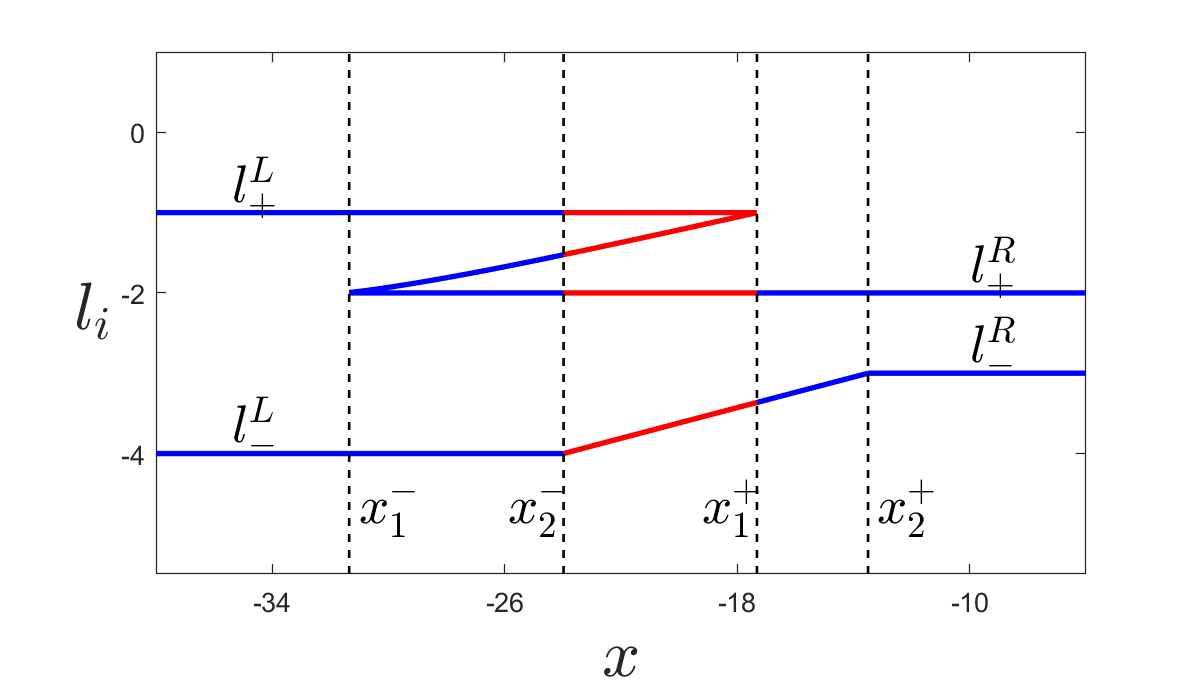}}\hfill
\subfigure[]{\includegraphics[width=0.33\linewidth]{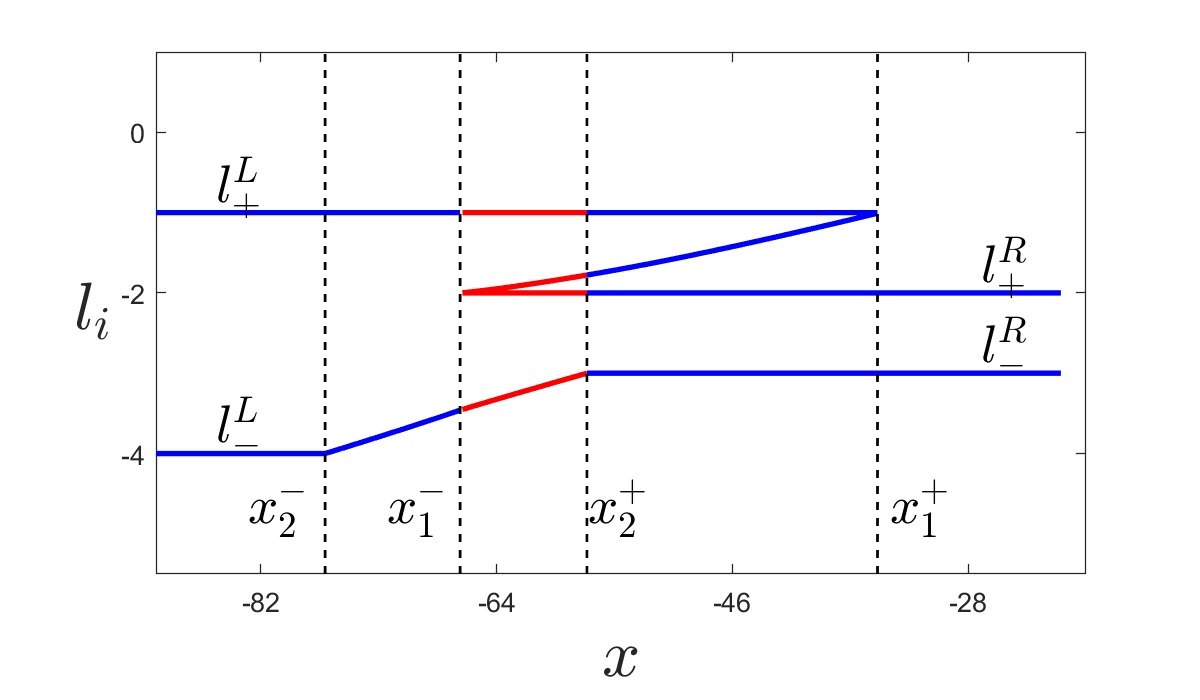}}\hfill
\subfigure[]{\includegraphics[width=0.33\linewidth]{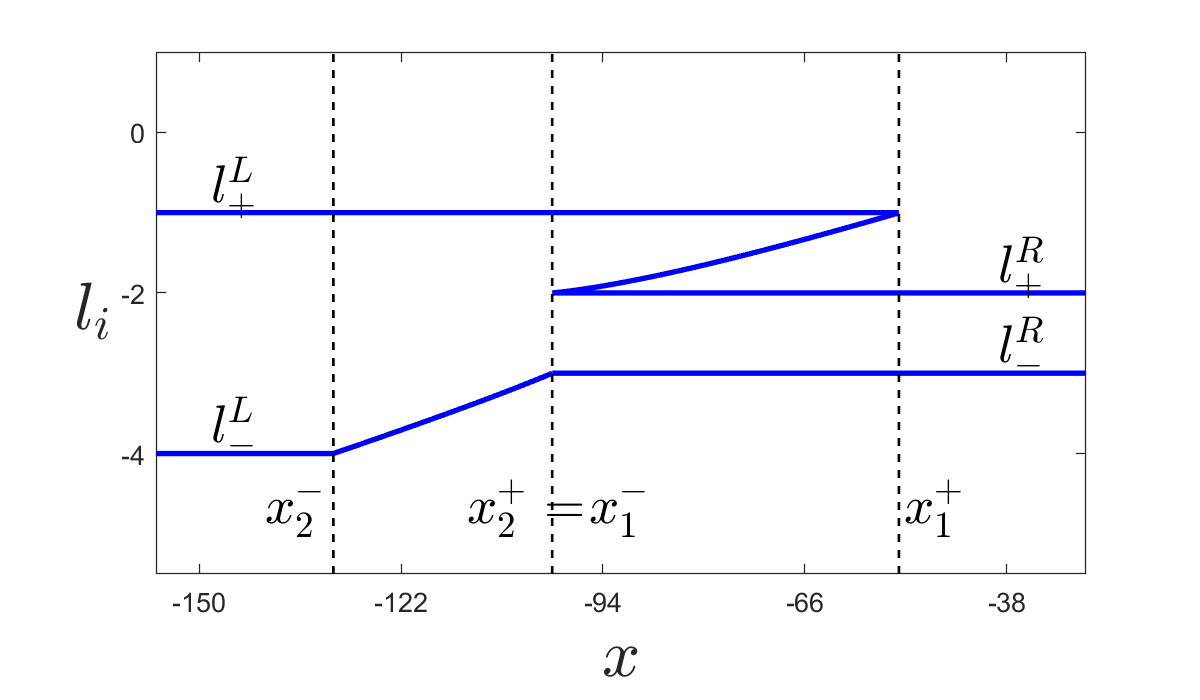}}\hfill
\subfigure[]{\includegraphics[width=0.33\linewidth]{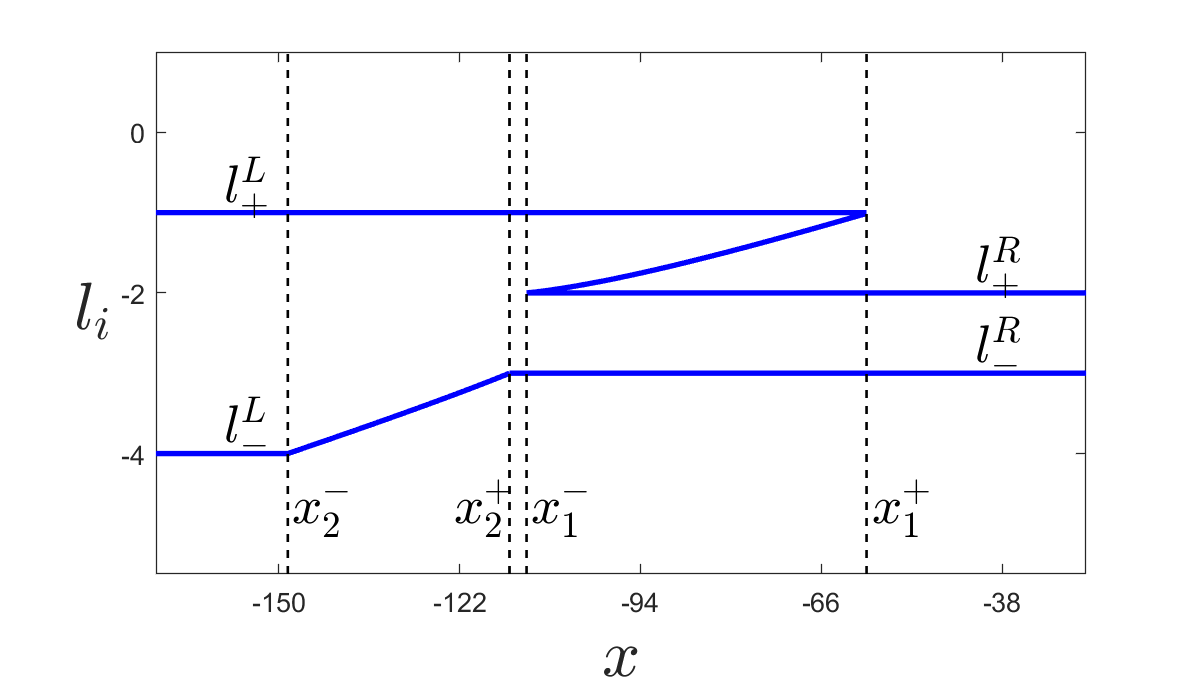}}
\subfigure[]{\includegraphics[width=0.71\linewidth]{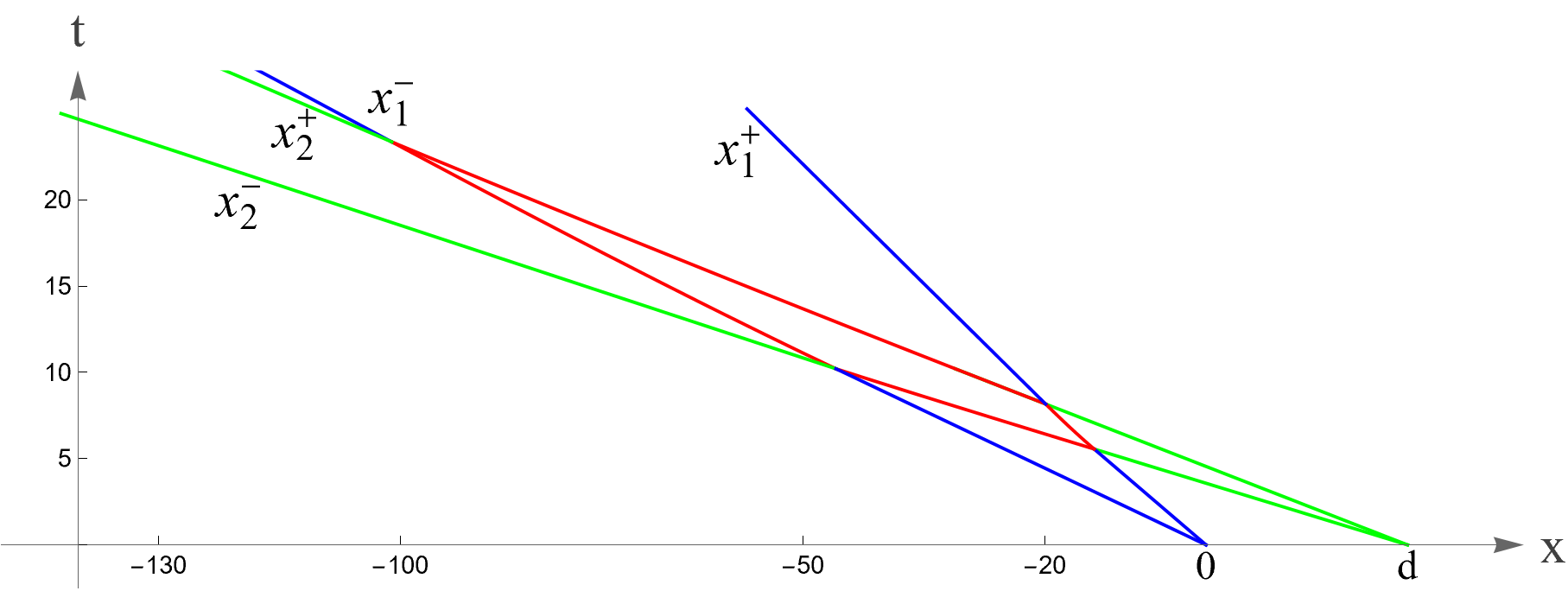}}\hfill
\flushleft{\footnotesize
\textbf{Fig.~$\bm{2}$.} (a)-(f) illustrate the evolution of Riemann invariants, with solid red lines denoting the interaction region. (g) presents the boundaries of distinct wave regions throughout the evolution: blue lines indicate the boundaries of the DSW, green lines indicate the boundaries of the RW, and red lines indicate the boundaries of the interaction region.}
\end{figure}
The evolution of the Riemann invariants before the interaction is shown in Fig. $2(a)$. It can be observed that the leading edge of RW catches up with the trailing edge of the DSW when $ x_{1}^{+}(t_{1}) = x_{2}^{-}(t_{1})$ (see Fig. $2(b)$) and simple calculation we obtain
\begin{equation}
t _ {1} =- \frac { d(l_-^L-2l_+^L+l_+^R)} { (l_-^L-l_+^L)(l_-^L-4l_+^L+3l_+^R) } , ~x _ {1} = -d \frac {l_-^{L2}+4l_-^Ll_+^L-8l_+^{L2}-2l_-^Ll_+^R+4l_+^Ll_-^R+l_+^{R2}} { 2(l_-^L-l_+^L)(l_-^L-4l_+^L+3l_+^R)} .
\end{equation}

$(ii)$ Being interaction $(t_1<t<t_2)$:

In the non-interacting region, the DSW and RW follow their descriptions before interaction. In the interaction region, the modulated solution is still adopted, yet the boundary conditions are altered and the solution is no longer self-similar. The left boundary of the interaction region, denoted by $x_2^-$, should be matched with the DSW domain, while its right boundary $x_1^+$ is required to match the RW domain. So that
\begin{equation}
\begin{cases} 
l_ { 1 } = l_-^L ,~ l_ { 2 } = l_+^R,~ l_ { 4 } = l_+^L,~
{x}= V _ { 3 } (l_1 ,l_2 , l_3 , l_4 )t,~at~x = x_{2}^{-}(t),\\
l_ { 2 } = l_+^R , ~l_ { 3 } = l_ { 4 } = l_+^L,~
x = V _- (l_1 ,l_2 )t+d,~at~x = x_{1}^{+}(t).
\end{cases} 
\end{equation}
It is clear that the Whitham system (2.10) turns to
\begin{equation}
\frac{\partial l_i}{\partial t}+V_i\frac{\partial l_i}{\partial x}=0,\ i=1,\ 3,
\end{equation}
which can be obtained a linear system
\begin{equation}
\frac{\partial x}{\partial l_1}-V_3\frac{\partial t}{\partial l_1}=0,~\frac{\partial x}{\partial l_3}-V_1\frac{\partial t}{\partial l_3}=0
\end{equation}
by the hodograph transform. Making in (3.11) the change of variables
\begin{equation}
x - V _ { 1 } t = W _ { 1 } ,~x - V _ { 3 } t = W _ { 3 },
\end{equation}
we obtain
\begin{equation}
\frac{\partial _iW_j}{W_i-W_j}=\frac{\partial _iV_j}{V_i-V_j},\quad i\ne j,\ i,j=1,3.
\end{equation}
Using the boundary condition at $x = x_{1}^{+}$ and $x = x_{2}^{-}$, Eqs. (3.12) take the following form
\begin{equation}
x - V _- (l_1 ,~l_+^L ) t = W _ { 1 } ( l_ { 1 } ,~l_+^L ),~x - V_3(l_-^L,~l_+^R,~l_3,l_+^L) t = W _ { 3 } ( l_ -^{ L } ,~l_3 ).
\end{equation}
According to the match conditions (3.9), we get
\begin{equation}
W _ { 1 } (l_ { 1 } ,~l_+^L) = d,~W _ { 3 } ( l_-^L,~l_ { 3 } ) = 0.
\end{equation}
It can be seen that 
$W_i$ and $V_i$ are fully symmetric, which implies that $W_i$ admits an expression analogous to that of $V_i$
\begin{equation}
W_i=\left( 1 - \frac { \mathfrak { L } } { \partial _i \mathfrak { L } } \partial _i \right) f  ( l _ { 1 } , l _ { 3 } ) ,
\end{equation}
where $f ( l _ { 1 } , l _ { 3 } )$ is a arbitrary function. Substituting Eqs. (3.16) and (2.11) into Eq. (3.13), we find that 
the function $f(l_1, l_3)$ satisfies the Euler-Darboux-Poisson (EDP) equation~\cite{ck46}
\begin{equation}
2 ( l_ { 3 } - l_ { 1 } ) \partial _ { 13 } ^{2}g = \partial _ { 3 } g  - \partial _ { 1 }  g ,~ \partial _ { j } \equiv \partial / \partial l _ { j } .
\end{equation}
Its general solution can be expressed as
\begin{equation}
g=\int_{l_{-}^{L}}^{l_1}{\frac{\phi _1\left( \zeta \right) \,d\zeta}{\sqrt{\left( \zeta -l_1 \right) \left( l_3-\zeta \right)}}}+\int_{l_{+}^{L}}^{l_3}{\frac{\phi _2\left( \zeta \right) \,d\zeta}{\sqrt{\left( \zeta -l_1 \right) \left( l_3-\zeta \right)}}},
\end{equation}
where $\phi _1\left( \zeta \right)$ and $\phi _2\left( \zeta \right)$ are arbitrary functions~\cite{ck46}.
Then translating $W_i$ into the boundary conditions (3.15) gives
\begin{equation}
\begin{aligned}
f ( l _ { 1 } ,~l_+^L ) - \frac { \mathfrak { L } ( l _ { 1 } ,~l_+^R ,~l_+^L ,~l_+^L ) } { \partial _ { 1 } \mathfrak { L } ( l_ { 1 } ,~l_+^R ,~l_+^L ,~l_+^L ) } \partial _ { 1 } f ( l_ { 1 } ,~l_+^L ) = d ,\\
f ( l_ -^{ L } ,~l_3 ) - \frac { \mathfrak { L } (  l_-^L ,~l_+^R ,~l_3 ,~l_+^L ) } { \partial _ { 3 } \mathfrak { L } ( l_-^L ,~l_+^R ,~l_3 ,~l_+^L ) } \partial _ { 3 } f ( l_ -^{ L } ,~l_3 ) = 0,
\end{aligned}
\end{equation}
which are integrated to give the boundary value of the function $f(l_1, l_3)$
\begin{equation}
\begin{aligned}
&f ( l_ { 1 } ,l_{+}^L ) = \frac { C _ { 1 } } { \sqrt { l_+^L - l_ { 1 } } } + d ,\\
&f( l_-^L , l_ { 3 } ) = C _ { 2 } \mathfrak { L } ( l_-^L ,~l_+^R ,~l_3 ,~l_+^L ) ,
\end{aligned}
\end{equation}
where $C_{1}, C_2$ are arbitrary constants. It is convenient to set $C_2=0$, and we obtain $\phi _2\left( \zeta \right)\equiv 0$. Upon inserting Eq. (3.18) into the first condition in (3.20), we obtain
\begin{equation}
\int_{l_{-}^{L}}^{l_1}{\frac{\phi _1\left( \zeta \right) \,d\zeta}{\sqrt{\left( \zeta -l_1 \right) \left( l_3-\zeta \right)}}}=\frac { C _ { 1 } } { \sqrt { l_+^L - l_ { 1 } } }+d.
\end{equation}
Via Abel transform formula, one derives
\begin{equation}
\phi _1\left( \zeta \right) =\frac{1}{\pi \sqrt{\zeta -l_{-}^{L}}}\left( C_1\sqrt{\frac{l_{+}^{L}-l_{-}^{L}}{l_{+}^{L}-\zeta}}+d\sqrt{l_{+}^{L}-\zeta} \right) .
\end{equation}
From the second condition in (3.20), we get $\phi _1\left( l_-^L \right)\equiv 0$, thus straightforward simplification gives
\begin{equation}
\begin{aligned}
 & f ( l_ { 1 } , l_ { 3 } ) = - \frac { d } { \pi } \int _ { - 1 } ^{l_ { 1} } \frac { \sqrt { l-l_-^L } } { \sqrt { ( l_+^L- l) ( l_ { 3 } - l ) ( l_ { 1 } -l ) } } dl \\&  = \frac {- 2 d ( l_-^L -l_+^L ) } { \pi \sqrt { ( l_+^L - l_ { 1 } ) ( l _ { 3 }-l_-^L) } } ( \Pi _ { 1 } ( s , z ) -\text{K}( z ) ) ,
 \end{aligned}
\end{equation}
where $\Pi_{1}(s,z)$ is the complete elliptic integral of the third kind and

$$
z = \frac { ( l_+^L- l_ { 3 } ) ( l_ { 1 } -l_-^L ) } { ( l_+^L- l_ { 1 } ) ( l_ { 3 } -l_-^L ) } ,~ s = - \frac { l_ { 1 } -l_-^L } { l_+^L - l_ { 1 } } .
$$

Thus far, we have derived the modulation solution for the interaction region between the RW and the DSW
\begin{equation}
\begin{aligned} 
&l_ 2= l_+^R ,~l_ { 4 } = l_+^L,  \\ 
x - V _ { 1 , 3 } ( l_ { 1 } ,~l_+^R & ,~l_ { 3 } ,~l_+^L ) t= \left( 1 - \frac { \mathfrak { L } } { \partial _ { 1 , 3 } \mathfrak { L } } \partial _ { 1 , 3 } \right) f ( l _ { 1 } , l _ { 3 } ).
\end{aligned}
\end{equation}
The profiles are presented in Fig.~2(c) and 2(d). When $x_2^+(t_2)=x_1^-(t_2)$, the RW fully overtakes the DSW (see Fig. 2(e)). From Eqs. (3.12), we obtain
\begin{equation}
t_2 = \frac { 2 d(l_-^L-l_+^R)\text{E}(r)  } { \pi (l_-^R - l_+^R ) \sqrt { (l_-^R - l_+^L)(l_-^L-l_+^R) } }, 
\end{equation}
where $r=\frac { (l_-^R-l_-^L)(l_+^L-l_+^R)  } { (l_-^R - l_+^L)(l_-^L-l_+^R) }$.
The corresponding coordinate $x_2 = x_{2}^{+}(t_2) = x_{1}^{-}(t_2)$ is given by

\begin{equation}
x_2 = \frac{l_-^R + l_+^L + 2l_+^R} {2} t_2 +\frac{2d(l_-^L-l_+^L)(\text{K}(r)-\Pi_1(-\frac{ l_-^R -l_-^L } { l_+^L - l_-^R},r))}{\pi \sqrt{(l_-^R -l_-^L)(l_-^L -l_+^R)}}.
\end{equation}

$(iii)$ After interaction $(t>t_2)$:

After separation, the DSW is no longer centered at $x=0$. Instead, it admits a modulated solution of the following form
\begin{equation}
l_ { 1 } = l_-^R,~l_ { 2 } =l_+^R,~l_ { 4 } =l_+^L,~
x=V_{3}t+W_3(l_-^R,l_3),
\end{equation}
and the function $W_3(l_-^R, l_3)$ is represented by
\begin{equation}
\begin{aligned}
&W_3(l_-^R,l_3)=\frac{2d}{\pi \sqrt{(l_-^R -l_+^L)(l_-^L-l_3)}}((l_-^L-l_+^L)\Pi_1(\frac{-l_-^L+l_-^R}{l_-^R-l_+^L}
,{q})+\\ &\frac{(l_3-l_+^R)(-l_3+l_-^L)(l_-^R-l_+^L)\text{E}(q)+(-l_3+l_-^R)
(l_-^L-l_+^L)(-l_+^L+l_+^R)\mu(y)\text{E}(q)}{(-l_3+l_-^R)((-l_+^L+l_+^R)
\mu(y)+(l_3-l_+^R))}),
 \end{aligned}
\end{equation}
where
\begin{equation}
q = \frac{(-l_-^L+l_-^R)(l_3+l_+^L)}{(l_3-l_-^L)(l_-^R-l_+^L)}, ~y = \frac { (l_3-l_+^L) (l_-^L-l_+^R) } {(l_3-l_-^R)(l_+^L-l_+^R)},~\mu(y)=\frac{\text{E}(y)}{\text{K}(y)}.
\end{equation}
Following separation, the boundaries of the DSW take the form given below
\begin{equation}
\begin{aligned}
&~~~~~~~~~~~x_1 ^-=\frac{l_-^R + l_+^L + 2l_+^R} {2} t +W_3(l_-^R,l_+^R),  \\
x _ { 1 } ^{+} =  &\frac { l_-^{L2}+4l_-^Ll_+^L-8l_+^{L2}-2l_-^Ll_+^L+4l_+^Ll_+^R+l_+^{R2}} {2(l_-^L - 2l_+^L + l_+^R)} t+W_3(l_-^R,l_-^L). 
\end{aligned}
\end{equation}
\begin{figure}[htbp]
\centering
\setcounter{subfigure}{0}
\subfigure[]{\includegraphics[width=0.33\linewidth]{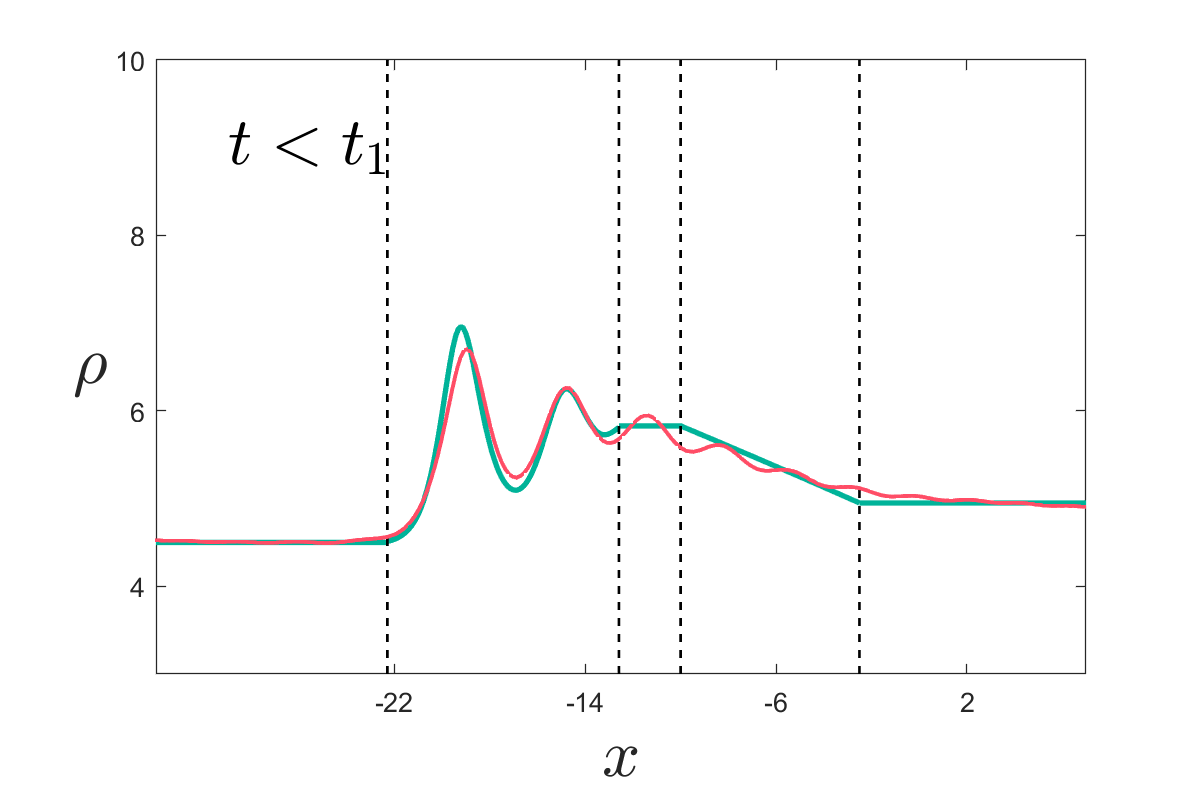}}\hfill
\subfigure[]{\includegraphics[width=0.33\linewidth]{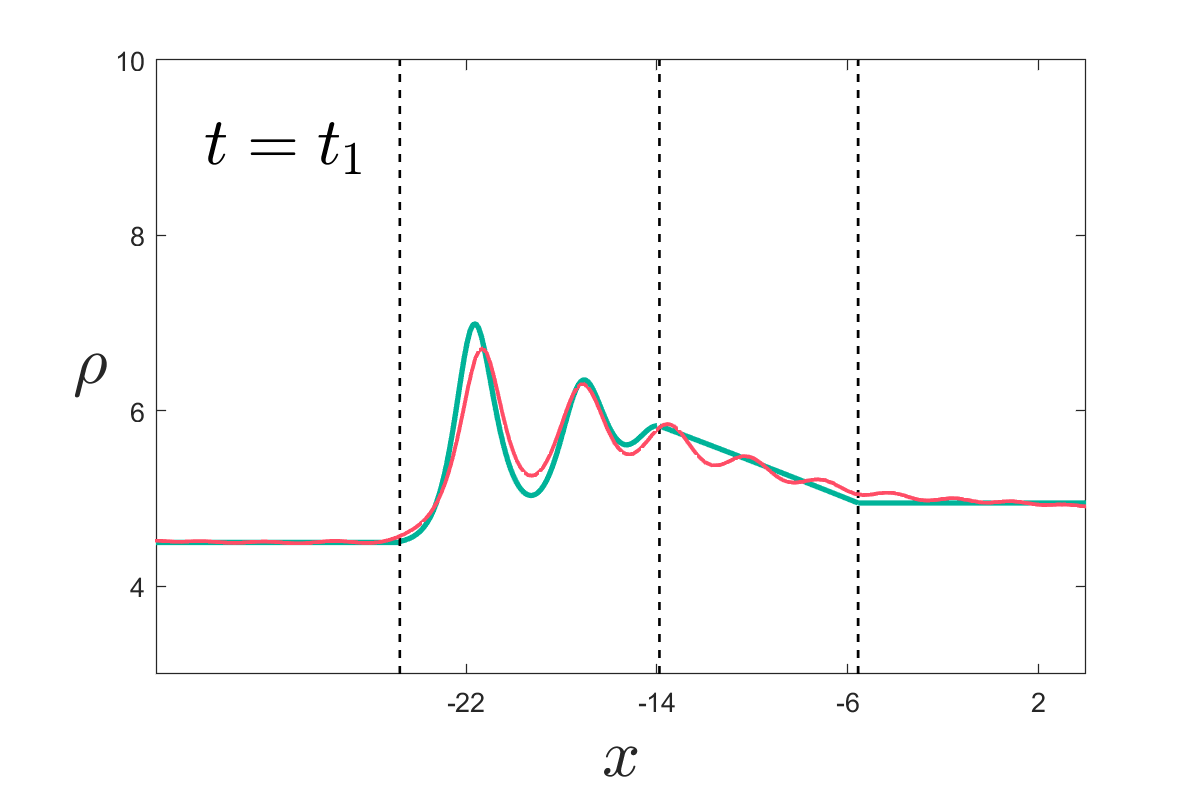}}\hfill
\subfigure[]{\includegraphics[width=0.33\linewidth]{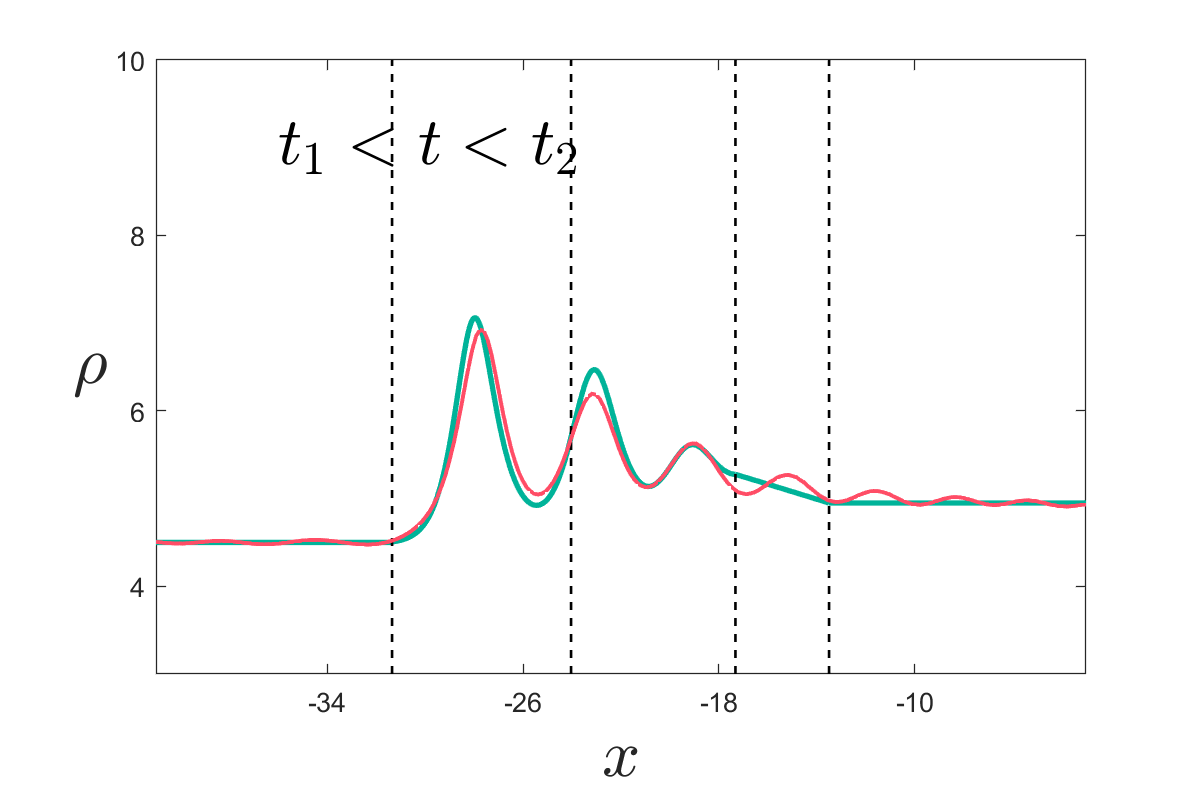}}\hfill
\subfigure[]{\includegraphics[width=0.33\linewidth]{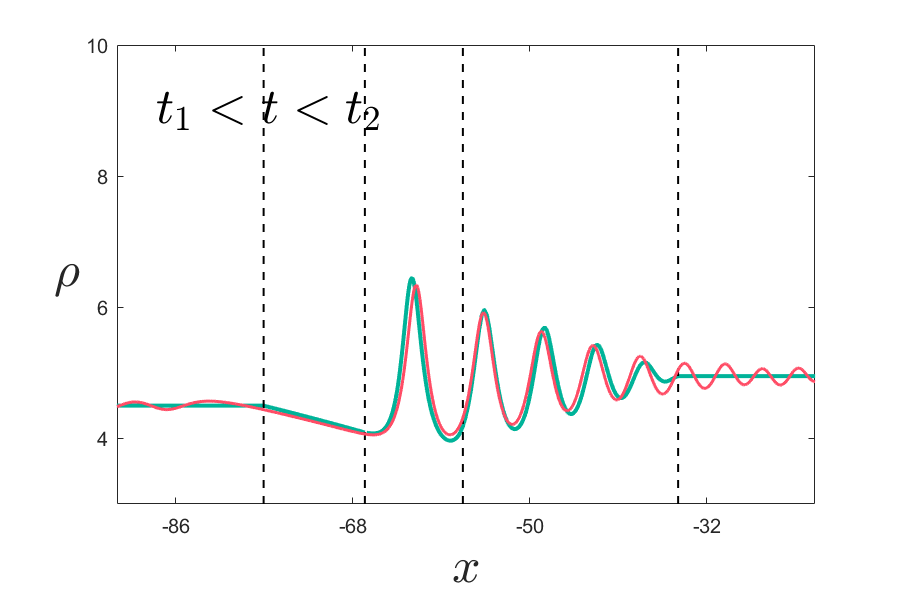}}\hfill
\subfigure[]{\includegraphics[width=0.33\linewidth]{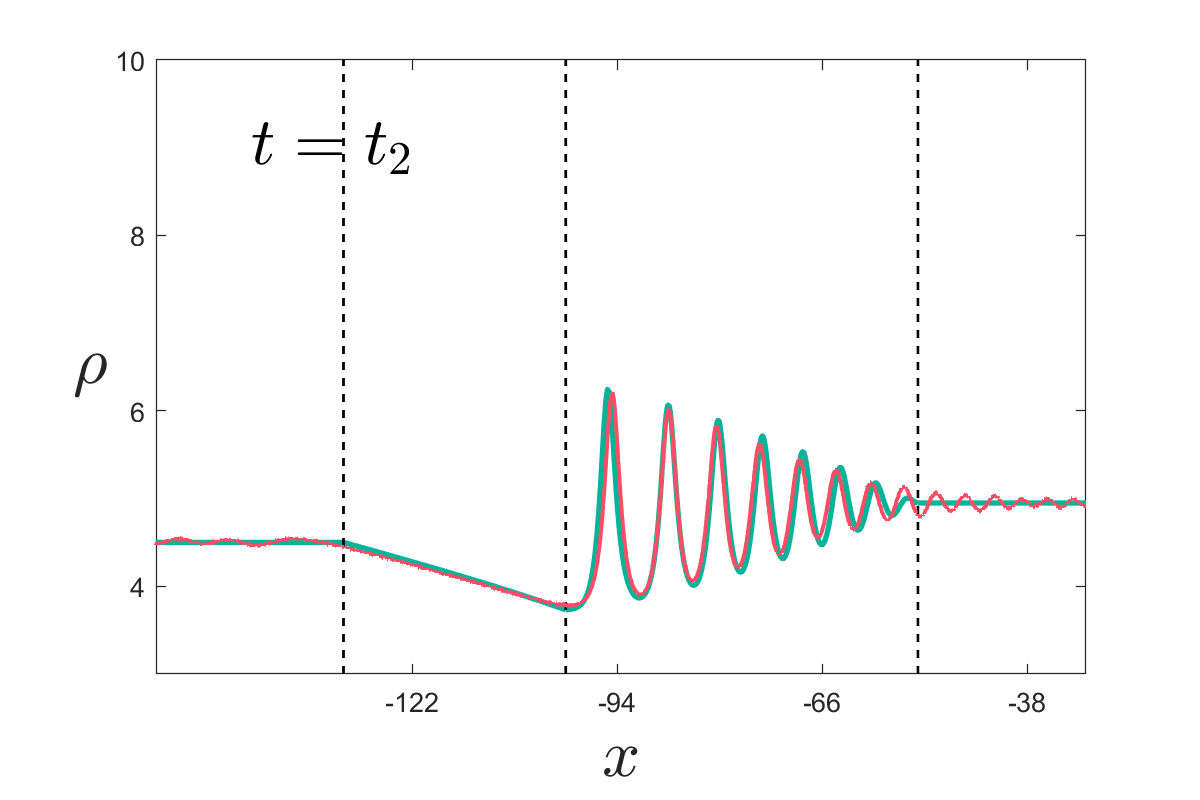}}\hfill
\subfigure[]{\includegraphics[width=0.33\linewidth]{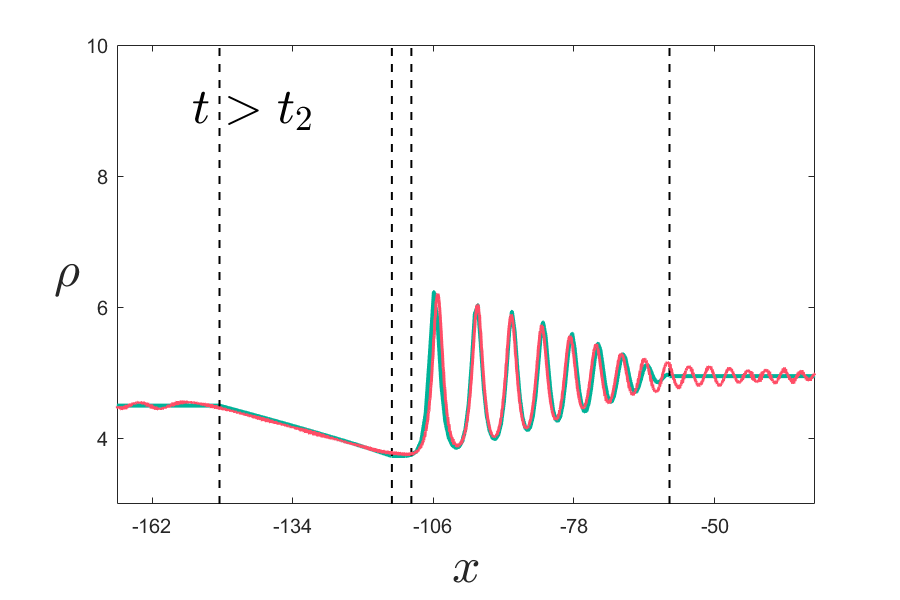}}
\flushleft{\footnotesize
\textbf{Fig.~$\bm{3}$.} The DSW analytical solutions constructed by mapping the Riemann invariants according to the relation given in Eq. (2.7) (green solid line) and the numerical simulation (red solid line) solutions on $x$.}
\end{figure}
\begin{figure}[htbp]
\centering
\setcounter{subfigure}{0}
\subfigure[]{\includegraphics[width=0.33\linewidth]{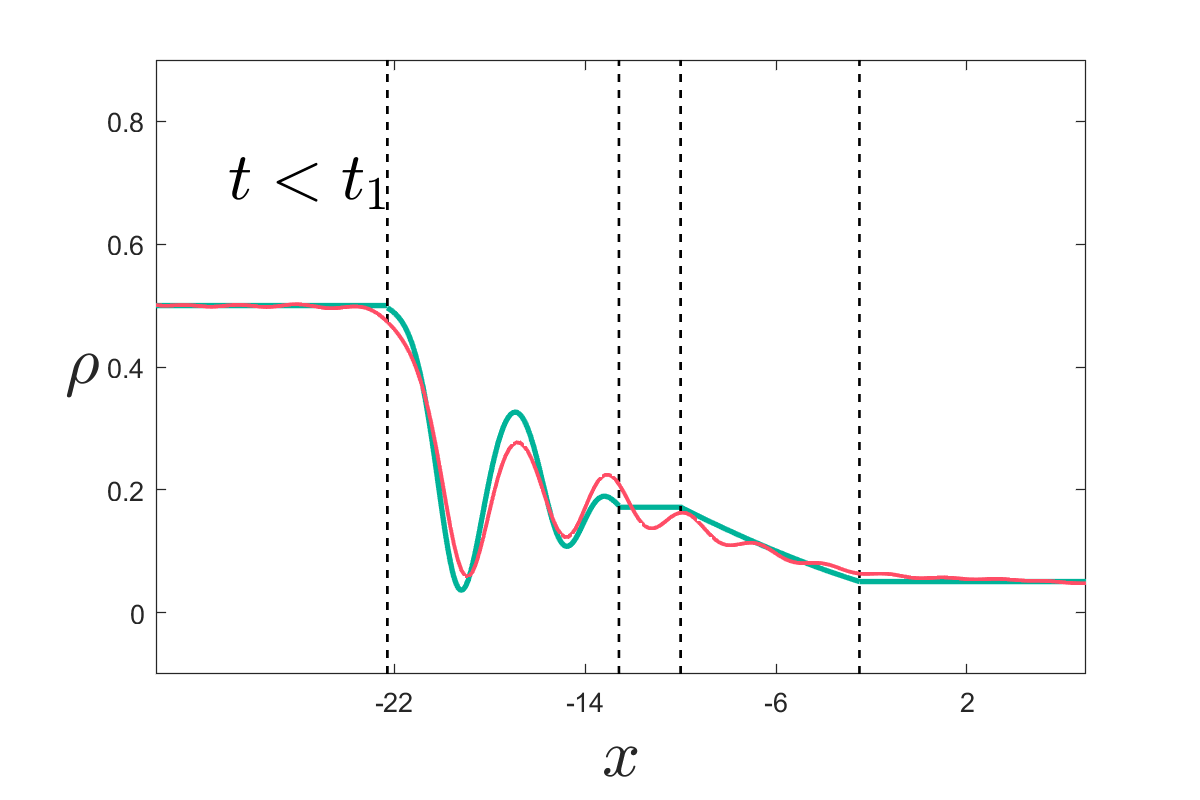}}\hfill
\subfigure[]{\includegraphics[width=0.33\linewidth]{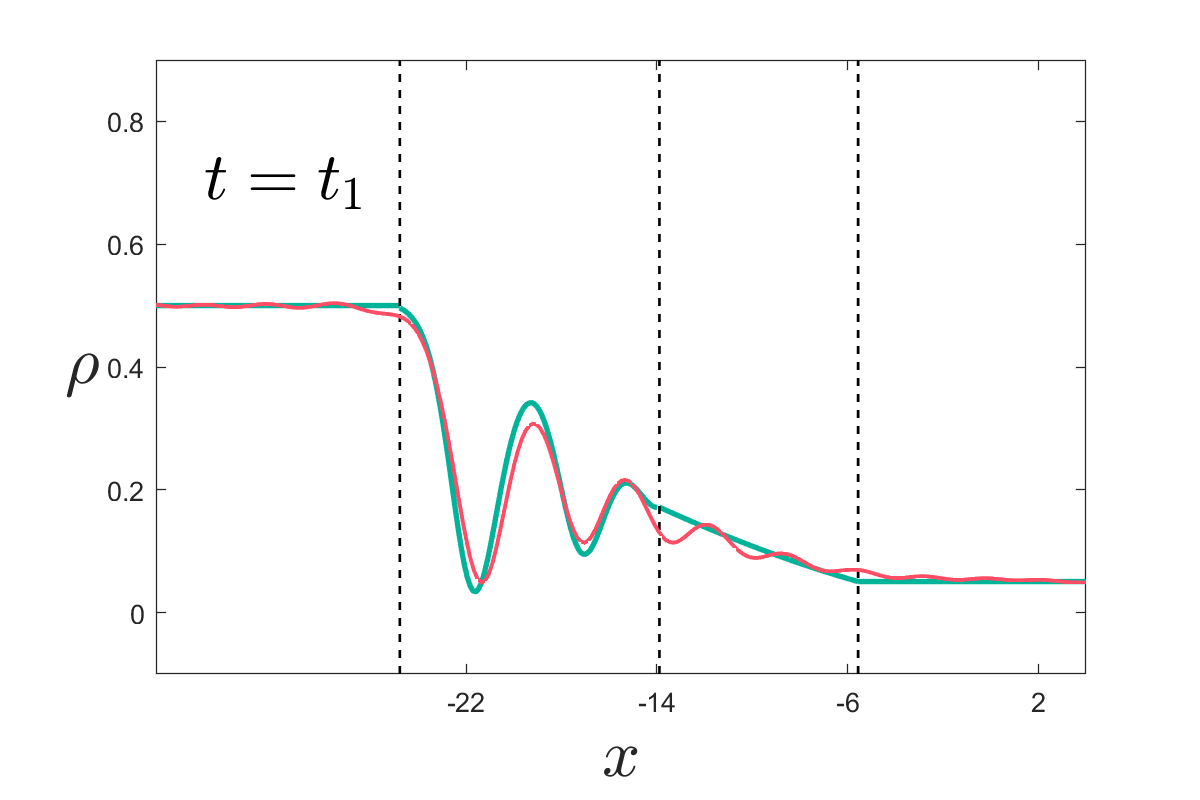}}\hfill
\subfigure[]{\includegraphics[width=0.33\linewidth]{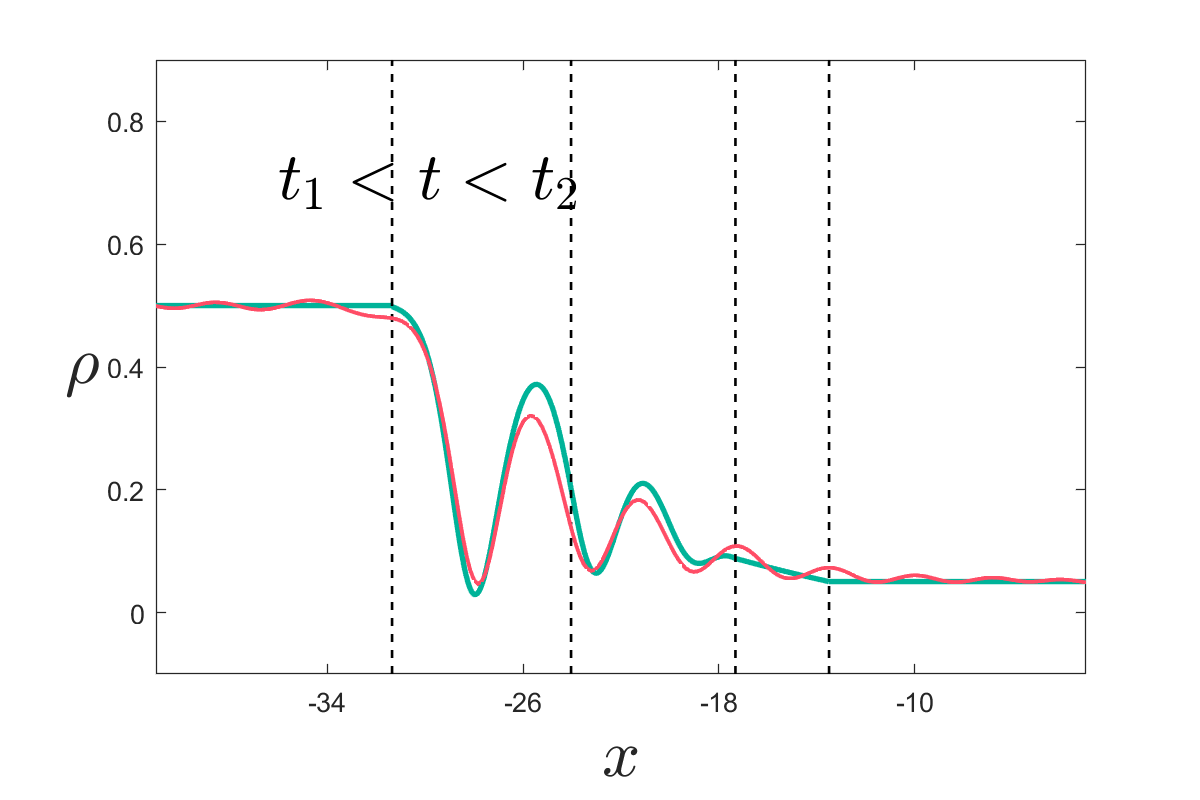}}\hfill
\subfigure[]{\includegraphics[width=0.33\linewidth]{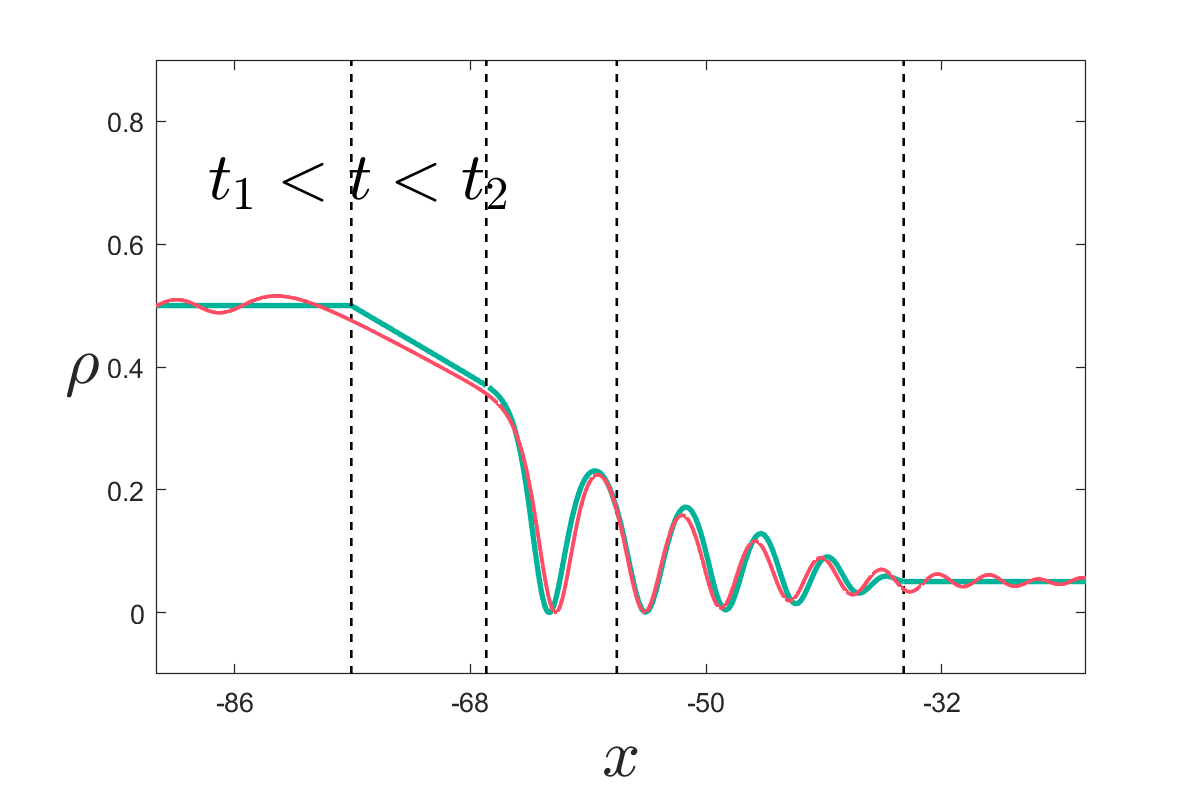}}\hfill
\subfigure[]{\includegraphics[width=0.33\linewidth]{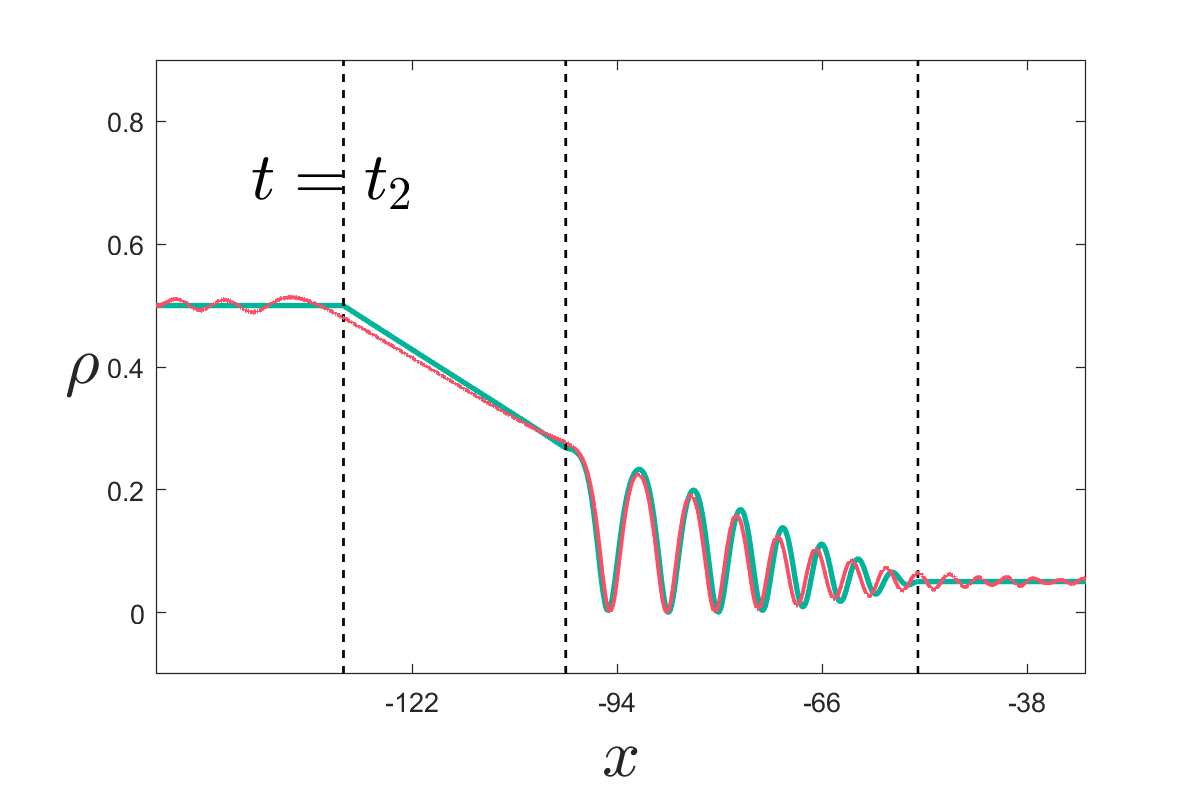}}\hfill
\subfigure[]{\includegraphics[width=0.33\linewidth]{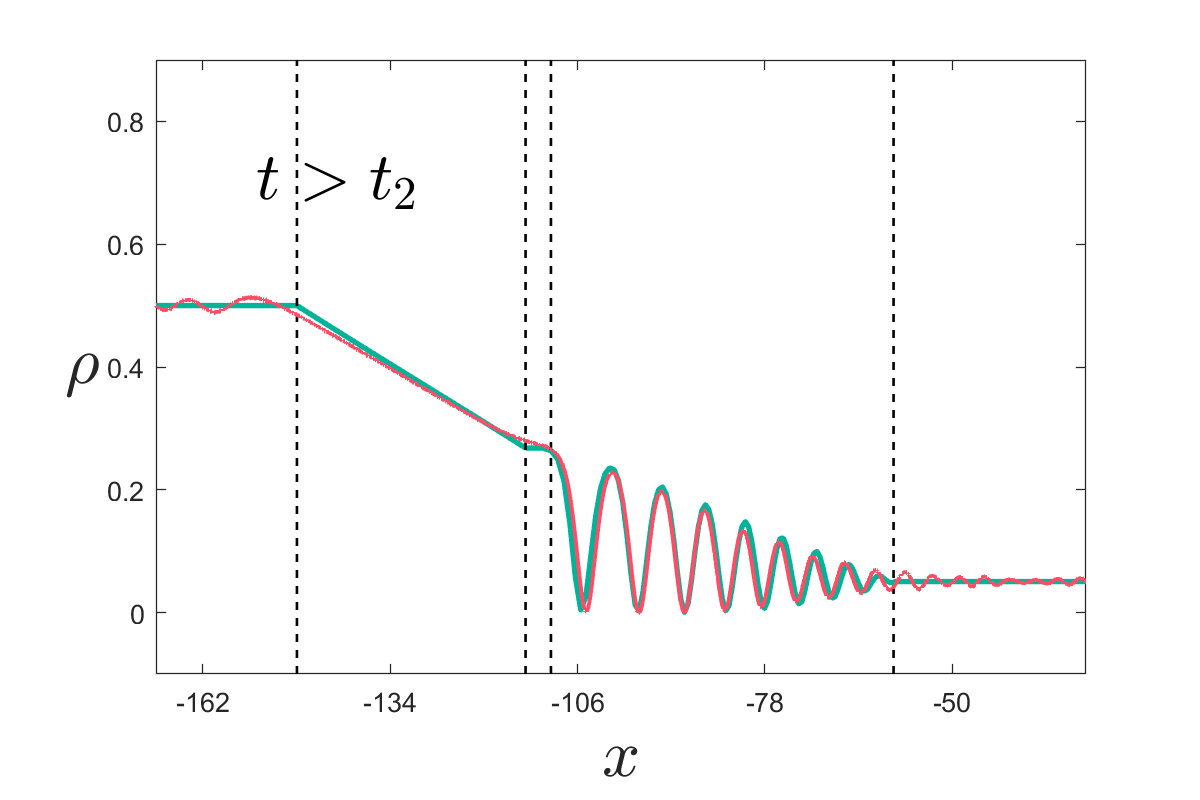}}
\flushleft{\footnotesize
\textbf{Fig.~$\bm{4}$.} The DSW analytical solutions constructed by mapping the Riemann invariants according to the relation given in Eq. (2.8) (green solid line) and the numerical simulation (red solid line) solutions on $x$.}
\end{figure}
The separated RW also possesses a new representation
\begin{equation}
l_+=l_+^L,~x = \frac { 3 l_ { - } + l_+^L} { 2 } t + W_1 ( l_ { - },l_+^L ),
\end{equation}
where $W_1( l_ { - },l_+^L )$ takes the form
\begin{equation}
W _ { 1 } ( l_ {-},l_+^L ) = \frac { 2d\left( (-l_-^L+l_+^L) ( \Pi _ { 1 } (p, r ) -\text K(r)) - 2(l_-^L-l_+^R) \text E (r) \right) } {\pi \sqrt {(l_- - l_+^L)(l_-^L-l_+^R)}}  , 
\end{equation}
where $r= \frac { ( l_--l_-^L) ( l_+^L-l_+^R) } {(l_--l_+^L)( l_-^L-l_+^R) } ,~p = - \frac { l_--l_-^L  } { l_+^L-l_- } .$
The boundaries give
$$
x _ { 2 } ^{-} = \frac { 3 l_-^L+ l_+^L} { 2 } t + W _ { 1 } ( l_ {-}^L,l_+^L ) ,~ x _ { 2 } ^{+} = \frac { 3 l_-^R+ l_+^L} { 2 } t + W _ { 1 } ( l_ {-}^R,l_+^L ) .
$$
The wave profile after separation is presented in Fig. 2(f). Fig. 2(g) illustrates the profiles of the boundaries of different wave regions during the evolution process.
To validate the Whitham modulation theory, we compare the analytical theoretical results with numerical simulations obtained via the split-step Fourier method. The two sets of density evolution profiles at distinct evolutionary stages are shown in Fig. 3 and Fig. 4.

\vspace{5mm}\noindent\textbf{3.2 DSW overtakes RW}
\hspace*{\parindent}\\

Configuration: Suppose the DSW and RW are generated at 
$(0,0)$ and $(-d,0)$ on the $(x,t)$ plane at $t=0$, respectively. They propagate leftwards, and the DSW situated to the right of the RW has a higher velocity. Their respective expanding regions given by $x_1^-(t)<x(t)<x_1^+(t)$ and $x _2^-(t)<x(t)<x_2^+(t)$. We have the following initial data:
\begin{equation}
\begin{aligned}
l_+(x,0)=
\begin{cases} 
l_+^L, & x<-d,\\
 l_+^R, & x>-d, 
\end{cases}
\quad \text{and} \quad
l_-(x,0)=
\begin{cases} 
l_-^L, & x<0, \\
l_-^R, & x>0,
\end{cases}
\end{aligned}
\end{equation}
where $ l_-^R<l_-^L<l_+^L<l_+^R,~d>0$. The density $\rho$ and velocity $v$ profiles corresponding to this construction are illustrated in the Fig. 5. Following a procedure analogous to that in the previous section, we can derive the three evolutionary stages. Given the identical derivation framework, we omit intermediate steps and provide the principal results.

\begin{figure}[htbp]
\centering
\setcounter{subfigure}{0}
\subfigure[]{\includegraphics[width=0.33\linewidth]{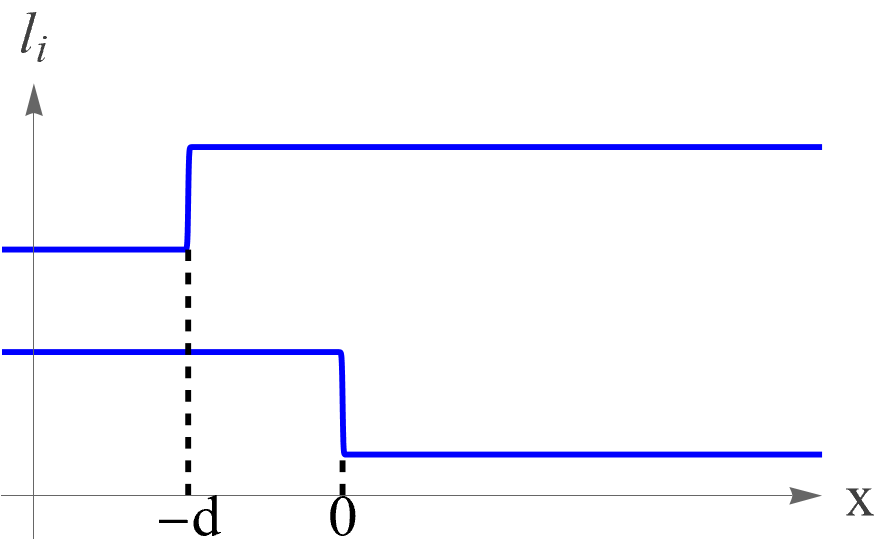}}\hfill
\subfigure[]{\includegraphics[width=0.33\linewidth]{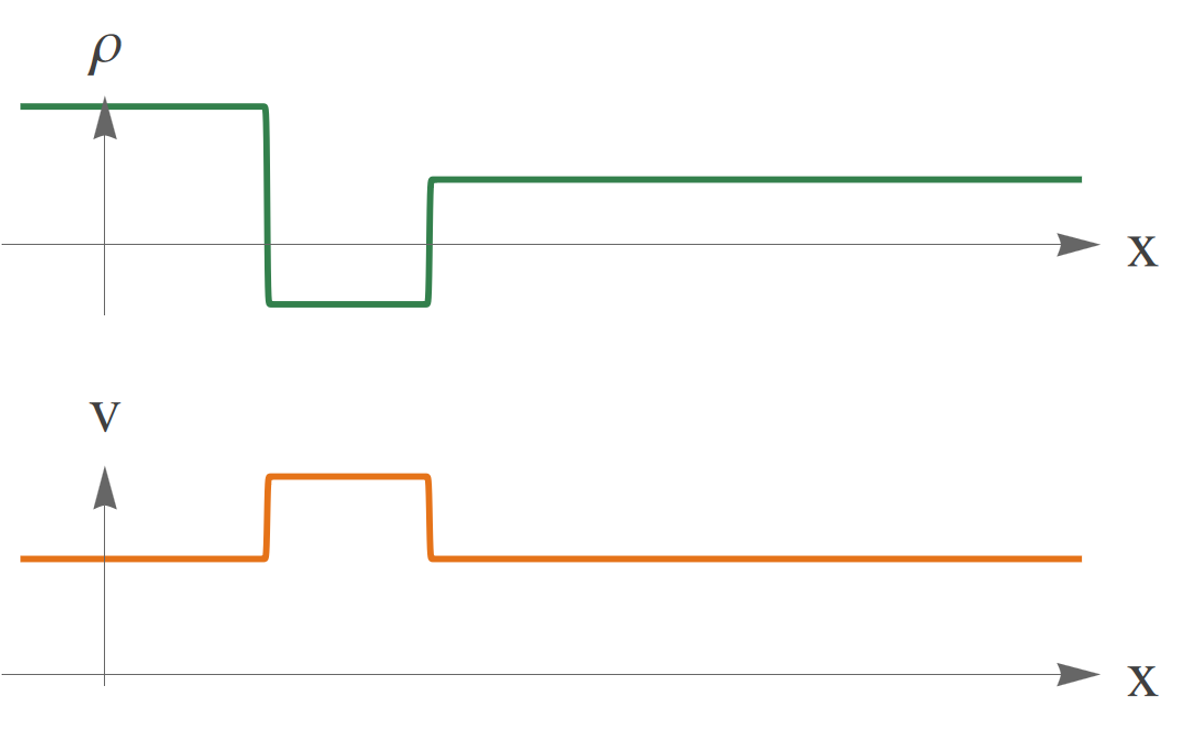}}\hfill
\subfigure[]{\includegraphics[width=0.33\linewidth]{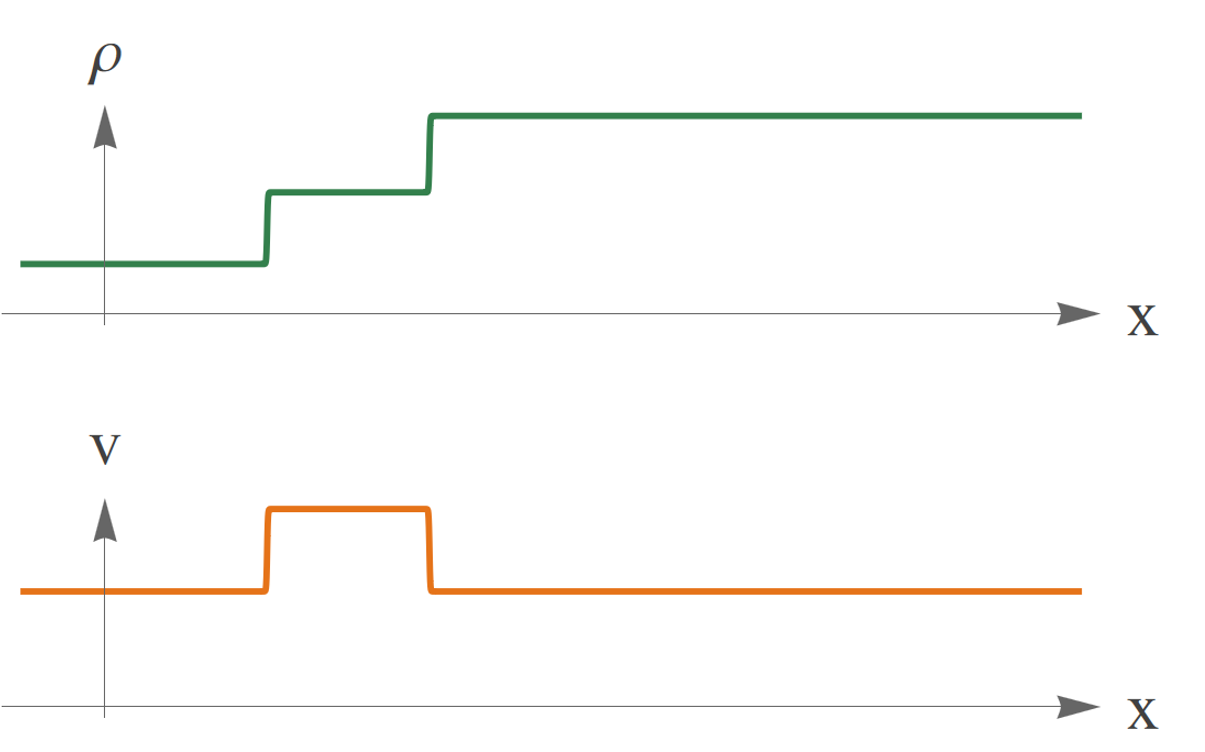}}
\flushleft{\footnotesize
\textbf{Fig.~$\bm{5}$.} The initial configuration (3.33) of the Riemann invariants, along with two sets of initial density and velocity corresponding to the constant states of the Riemann problem.}
\end{figure}
\begin{figure}[htbp]
\centering
\setcounter{subfigure}{0}
\subfigure[]{\includegraphics[width=0.71\linewidth]{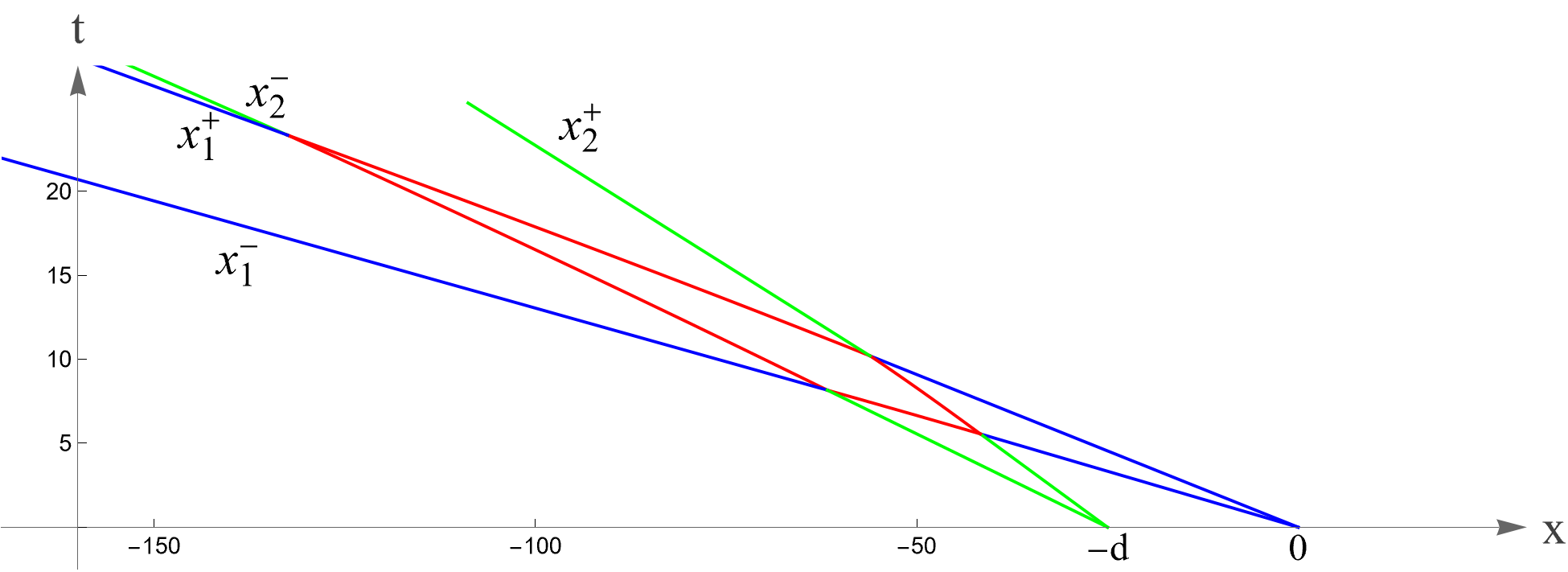}}\hfill
\subfigure[]{\includegraphics[width=0.33\linewidth]{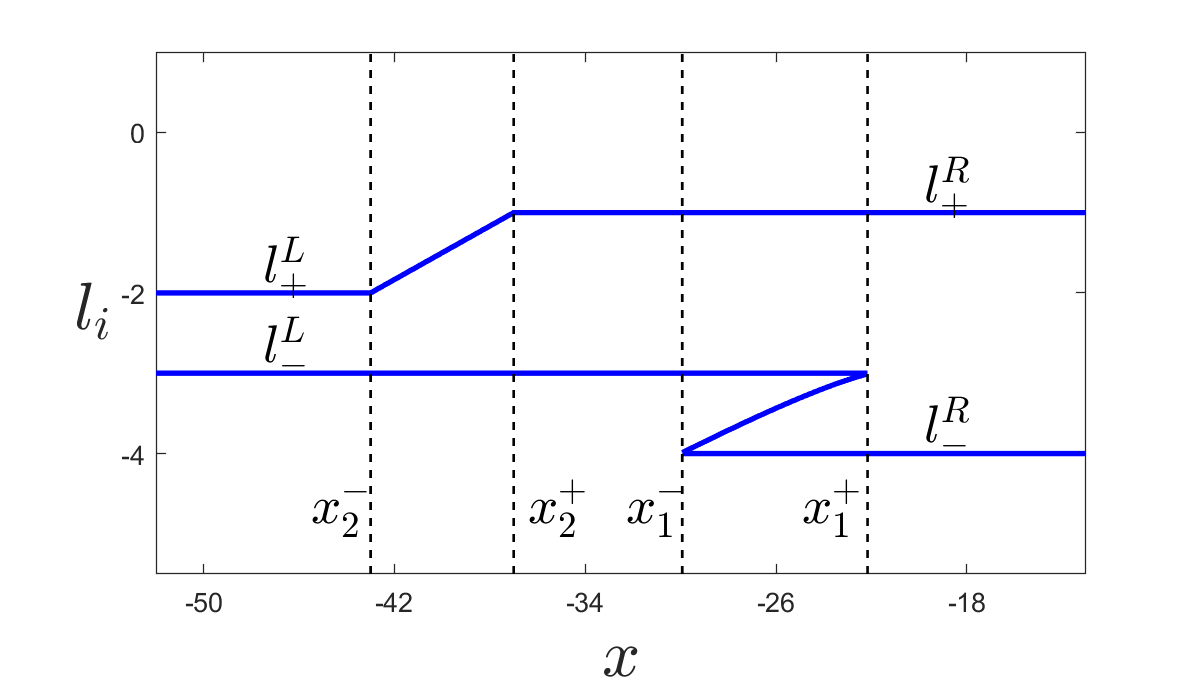}}\hfill
\subfigure[]{\includegraphics[width=0.33\linewidth]{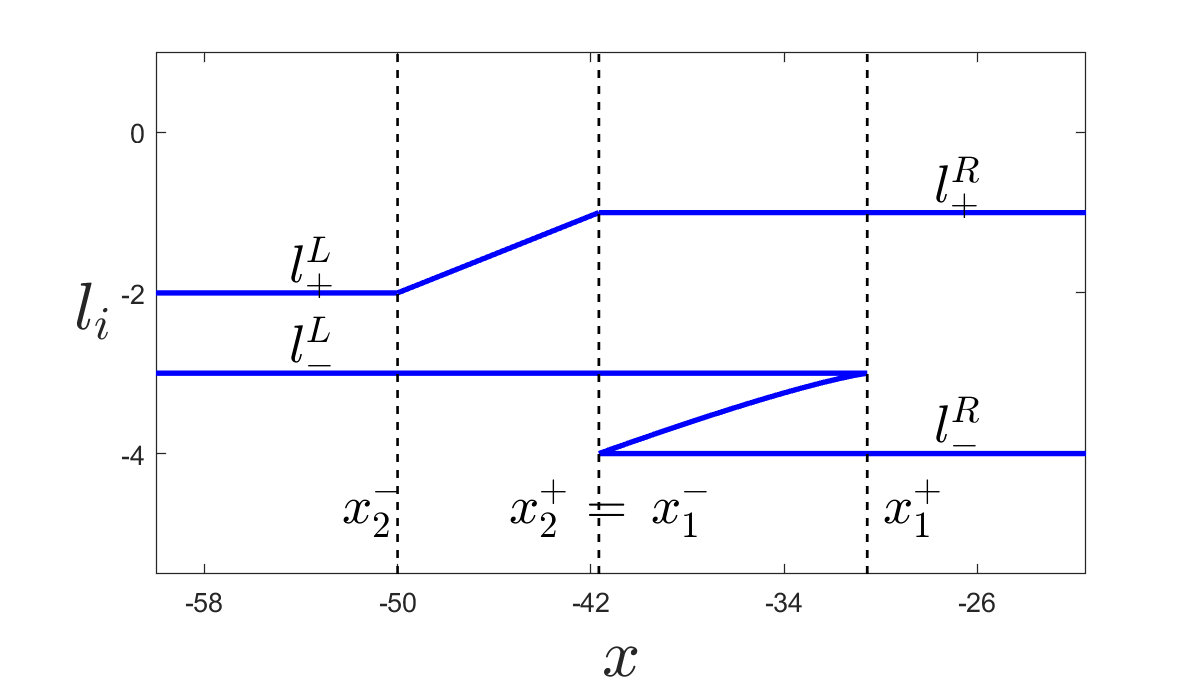}}\hfill
\subfigure[]{\includegraphics[width=0.33\linewidth]{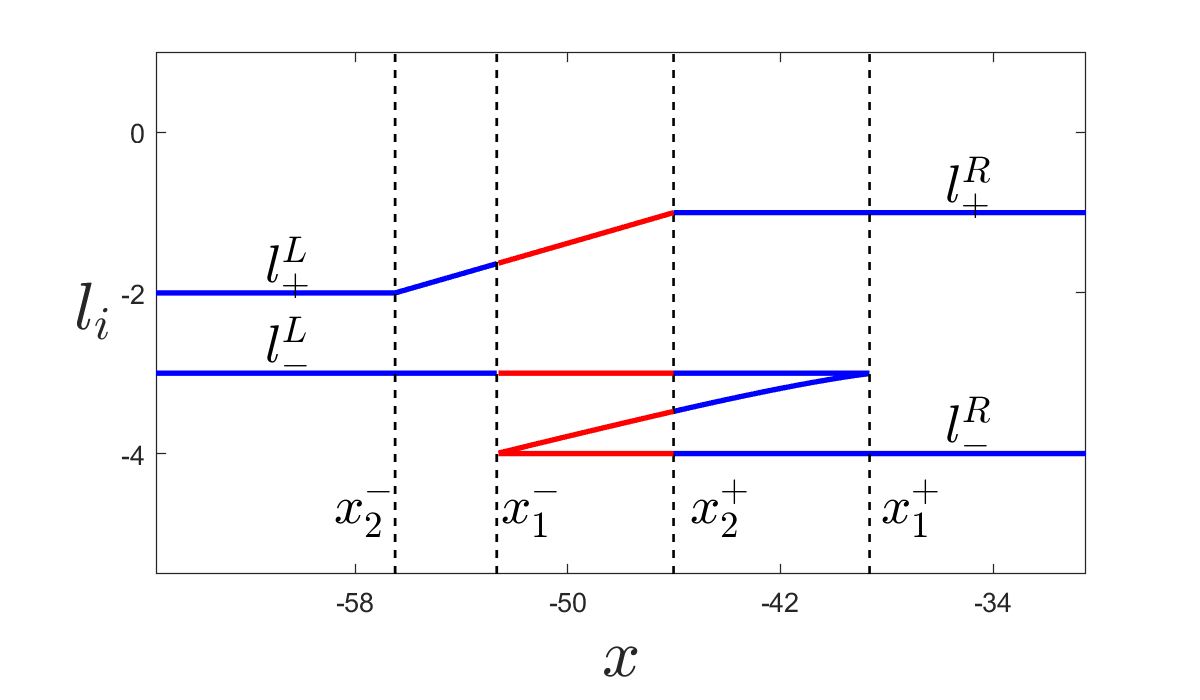}}\hfill
\subfigure[]{\includegraphics[width=0.33\linewidth]{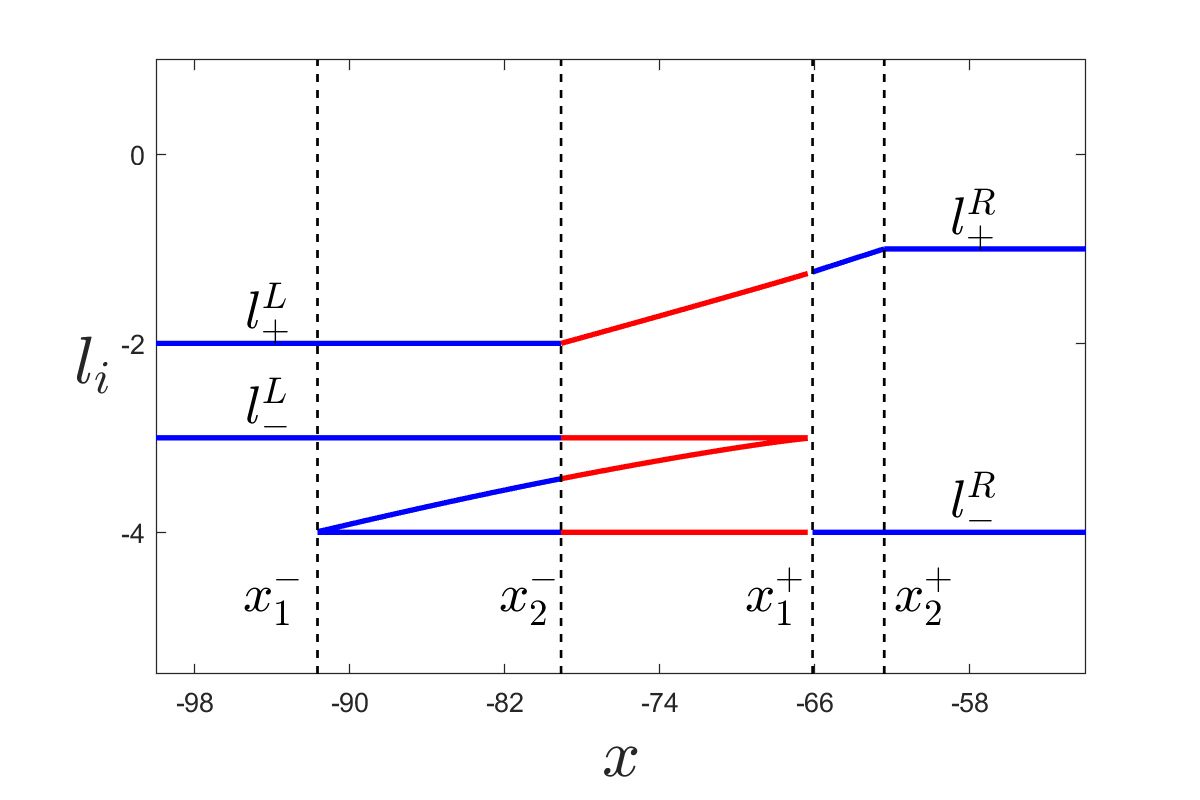}}\hfill
\subfigure[]{\includegraphics[width=0.33\linewidth]{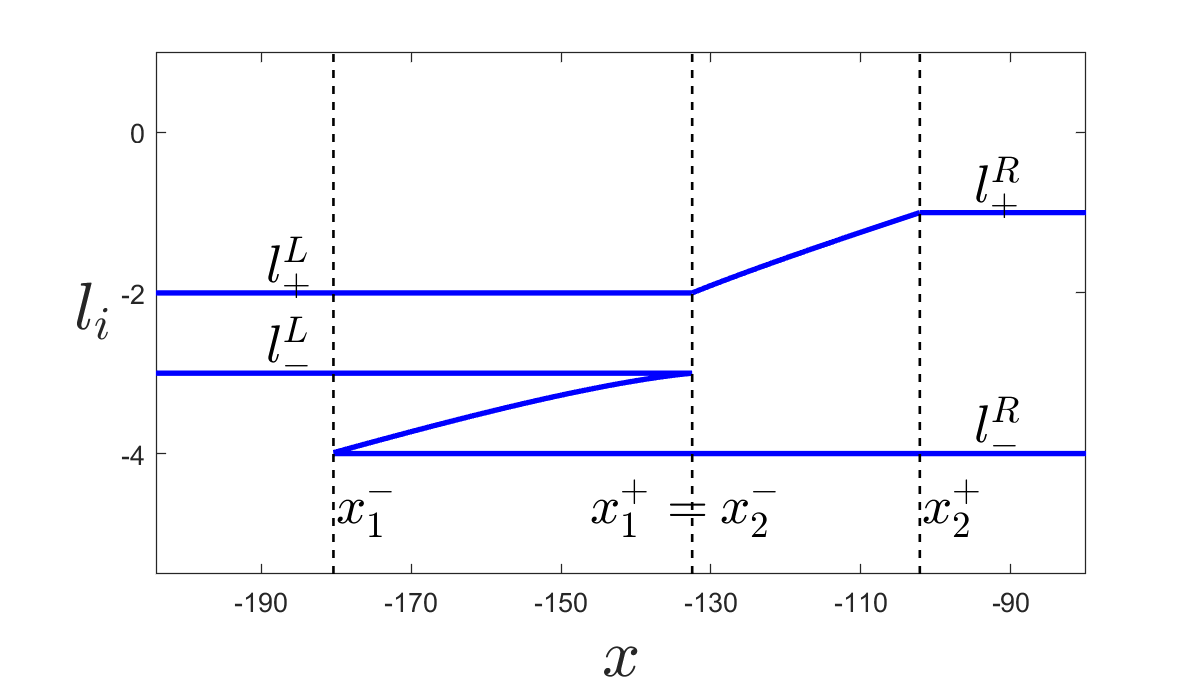}}\hfill
\subfigure[]{\includegraphics[width=0.33\linewidth]{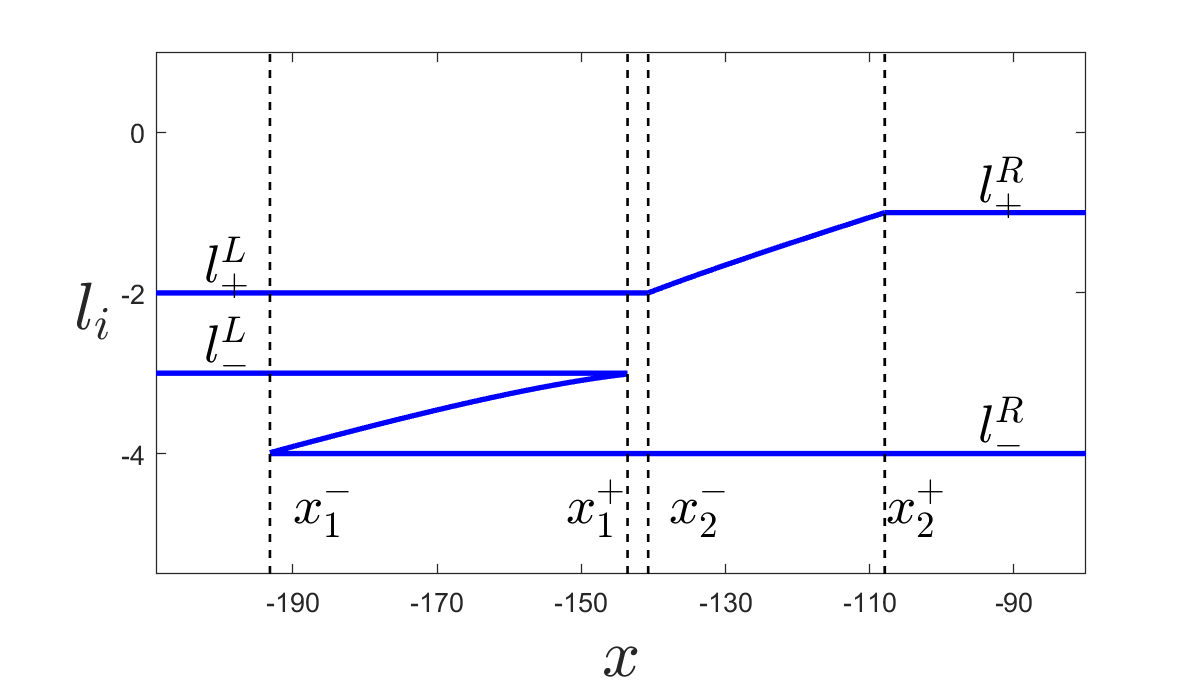}}\hfill
\flushleft{\footnotesize
\textbf{Fig.~$\bm{6}$.} (a) presents the boundaries of distinct wave regions throughout the evolution: blue lines indicate the boundaries of the DSW, green lines indicate the boundaries of the RW, and red lines indicate the boundaries of the interaction region. (b)-(g) illustrate the evolution of Riemann invariants, with solid red lines denoting the interaction region. }
\end{figure}
\begin{figure}[htbp]
\centering
\setcounter{subfigure}{0}
\subfigure[]{\includegraphics[width=0.33\linewidth]{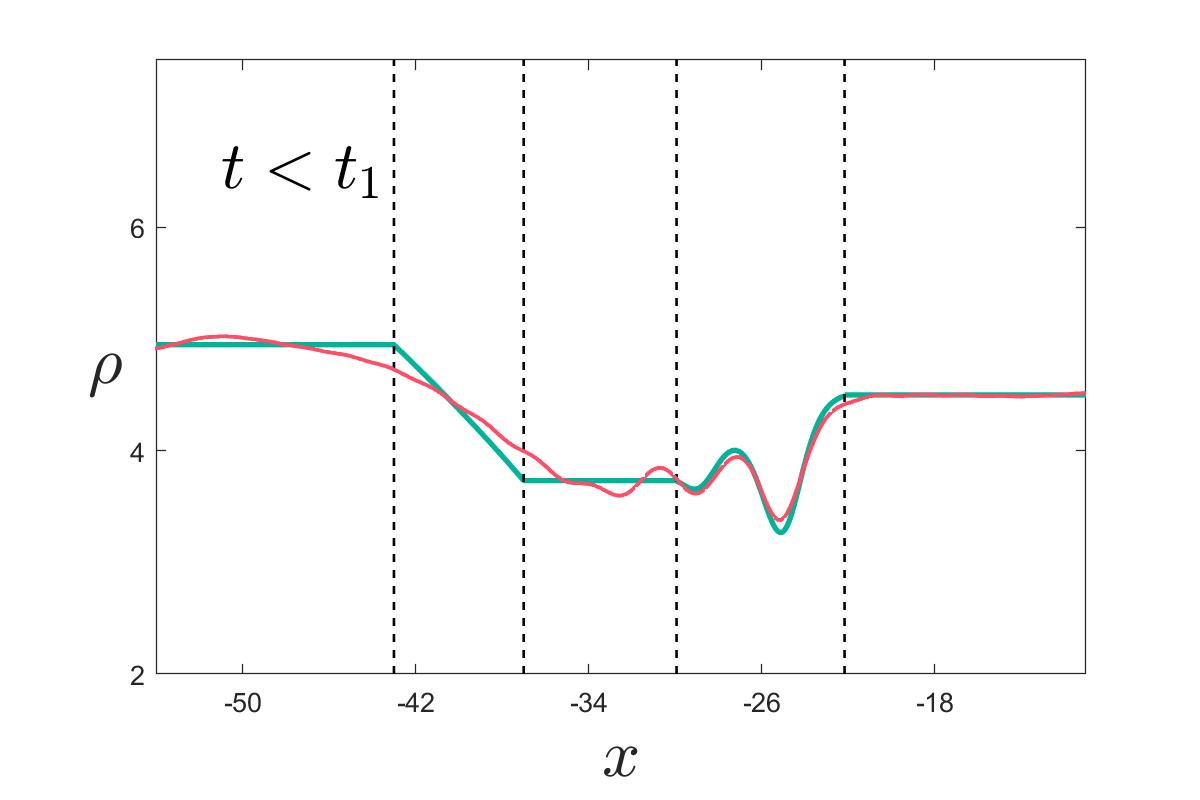}}\hfill
\subfigure[]{\includegraphics[width=0.33\linewidth]{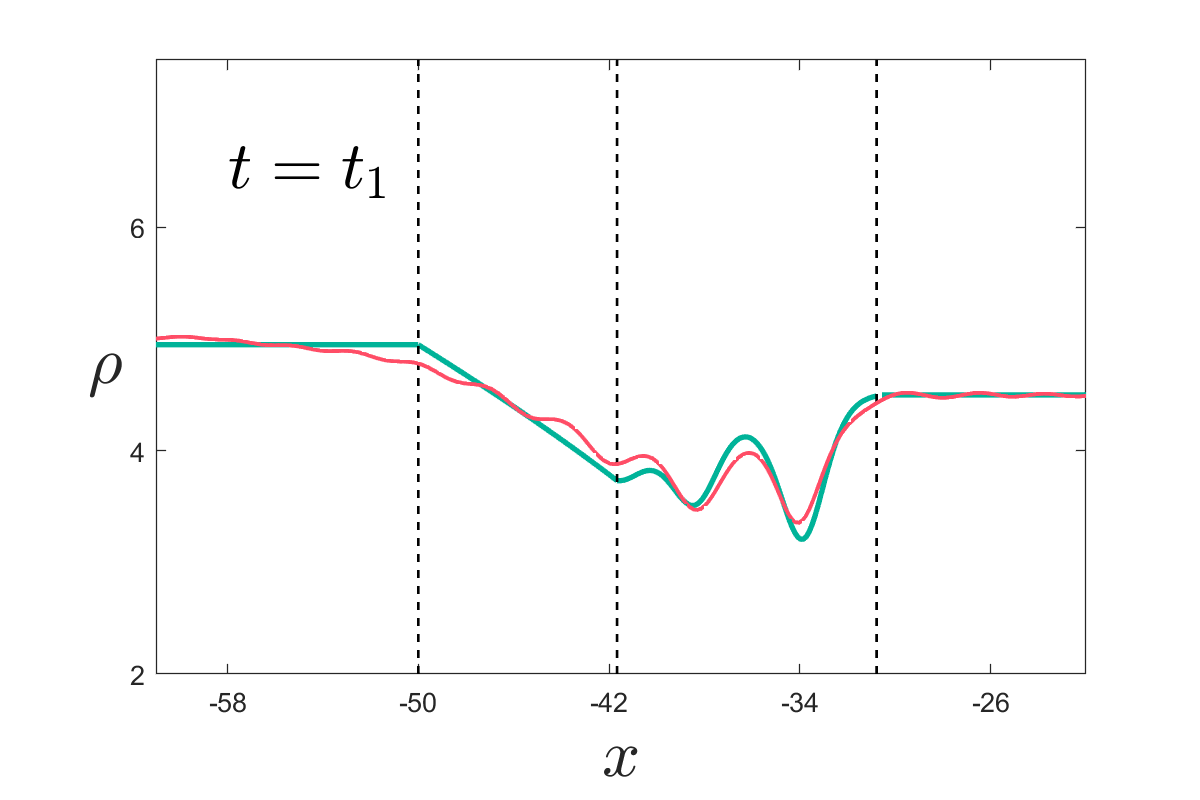}}\hfill
\subfigure[]{\includegraphics[width=0.33\linewidth]{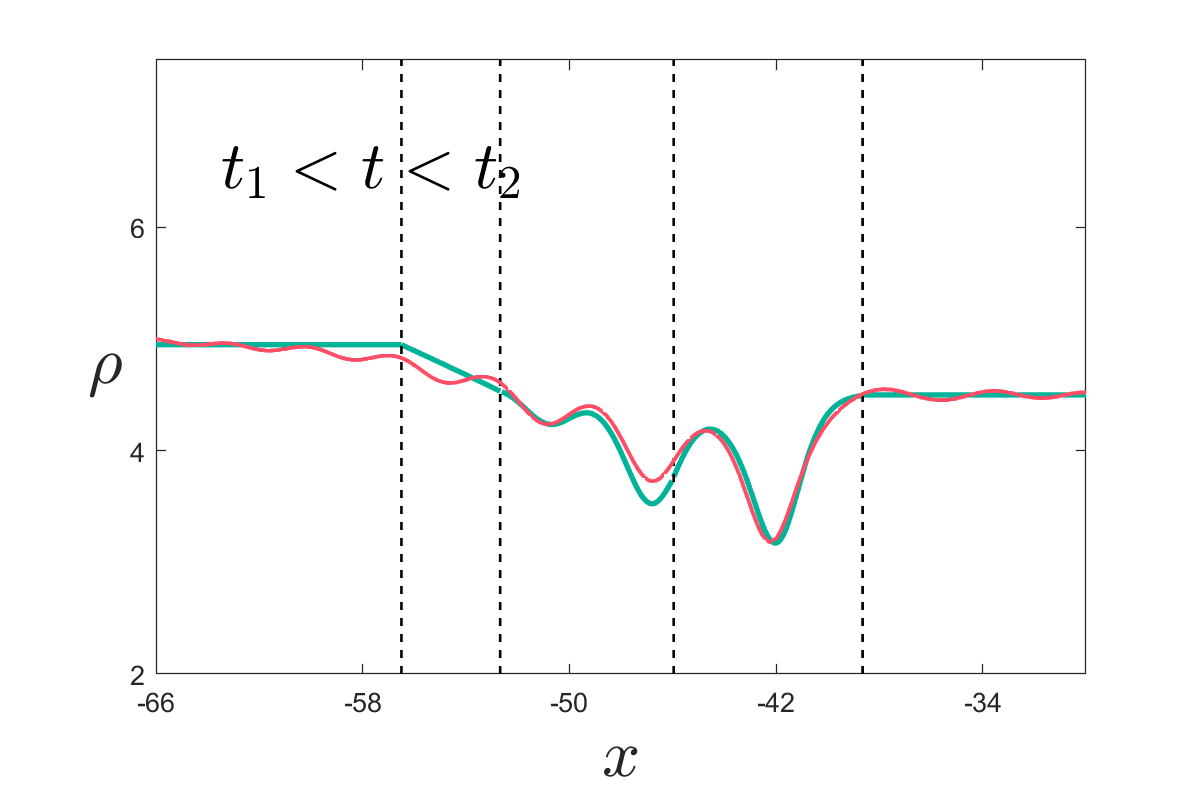}}\hfill
\subfigure[]{\includegraphics[width=0.33\linewidth]{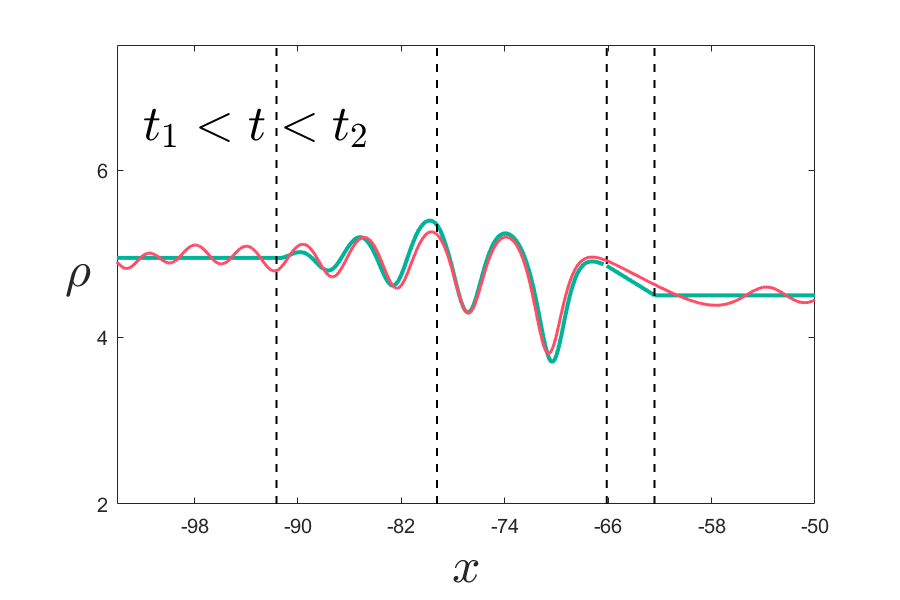}}\hfill
\subfigure[]{\includegraphics[width=0.33\linewidth]{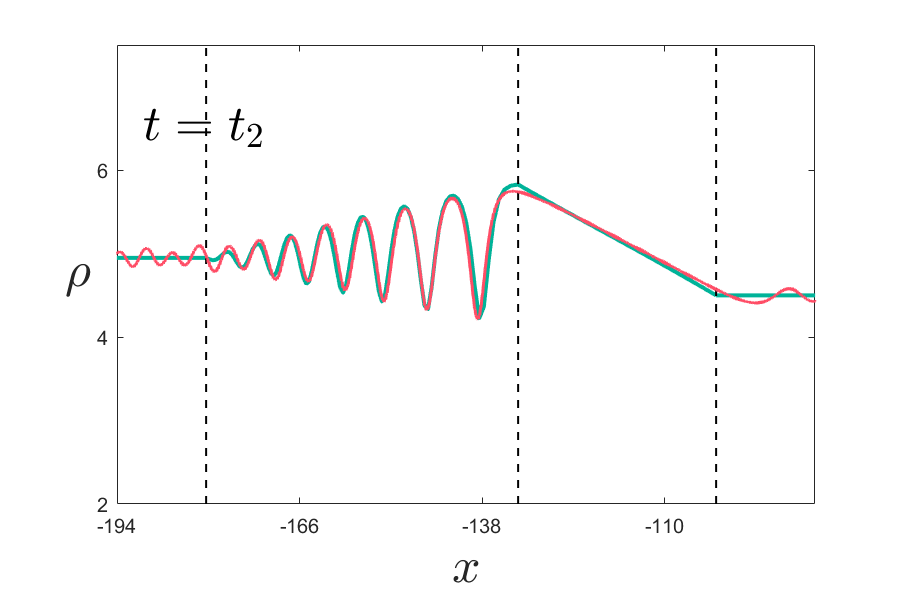}}\hfill
\subfigure[]{\includegraphics[width=0.33\linewidth]{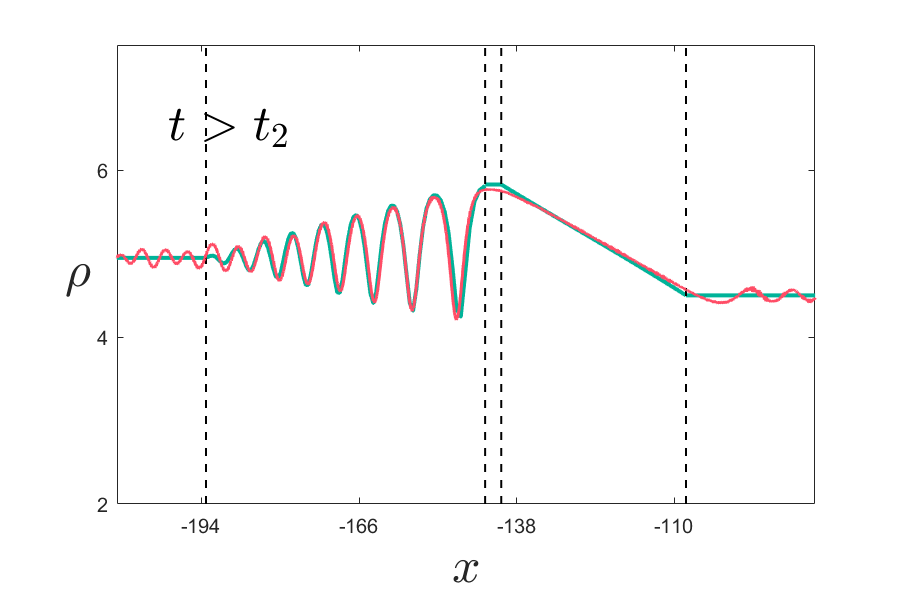}}
\flushleft{\footnotesize
\textbf{Fig.~$\bm{7}$.} The DSW analytical solutions constructed by mapping the Riemann invariants according to the relation given in Eq. (2.7) (green solid line) and the numerical simulation (red solid line) solutions on $x$.}
\end{figure}
\begin{figure}[htbp]
\centering
\setcounter{subfigure}{0}
\subfigure[]{\includegraphics[width=0.33\linewidth]{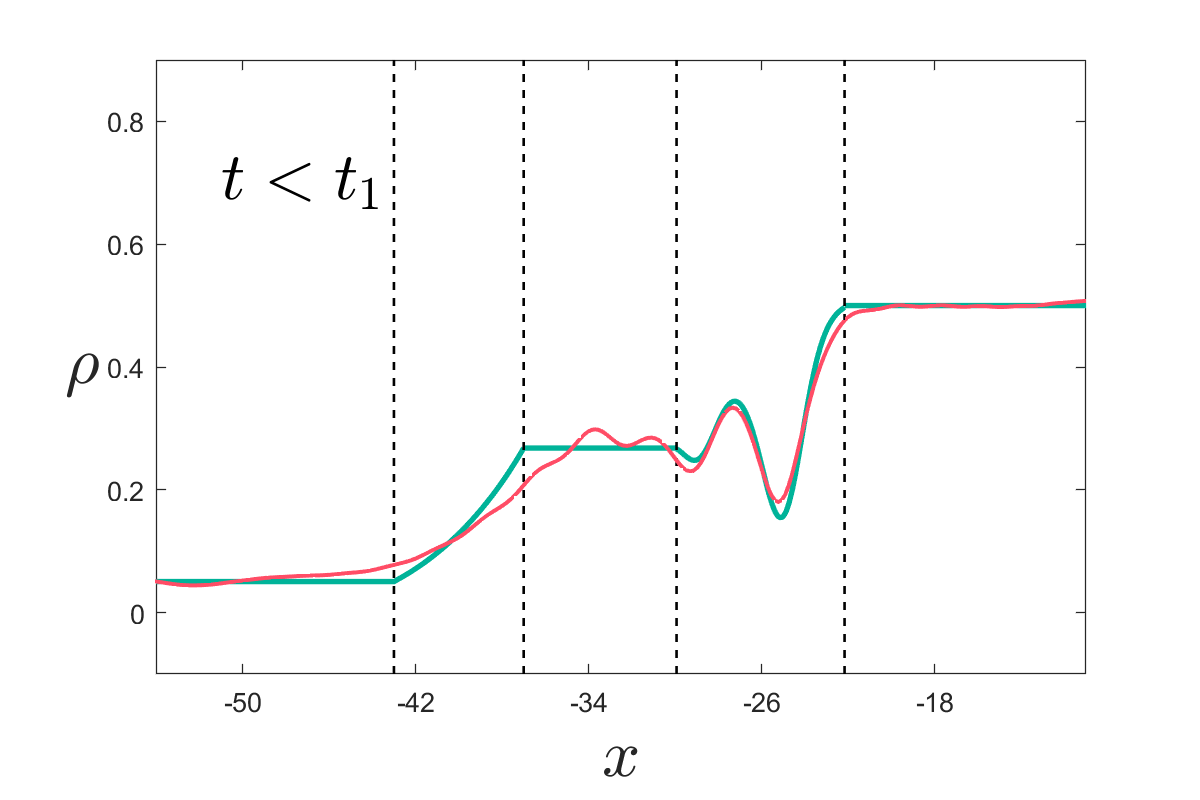}}\hfill
\subfigure[]{\includegraphics[width=0.33\linewidth]{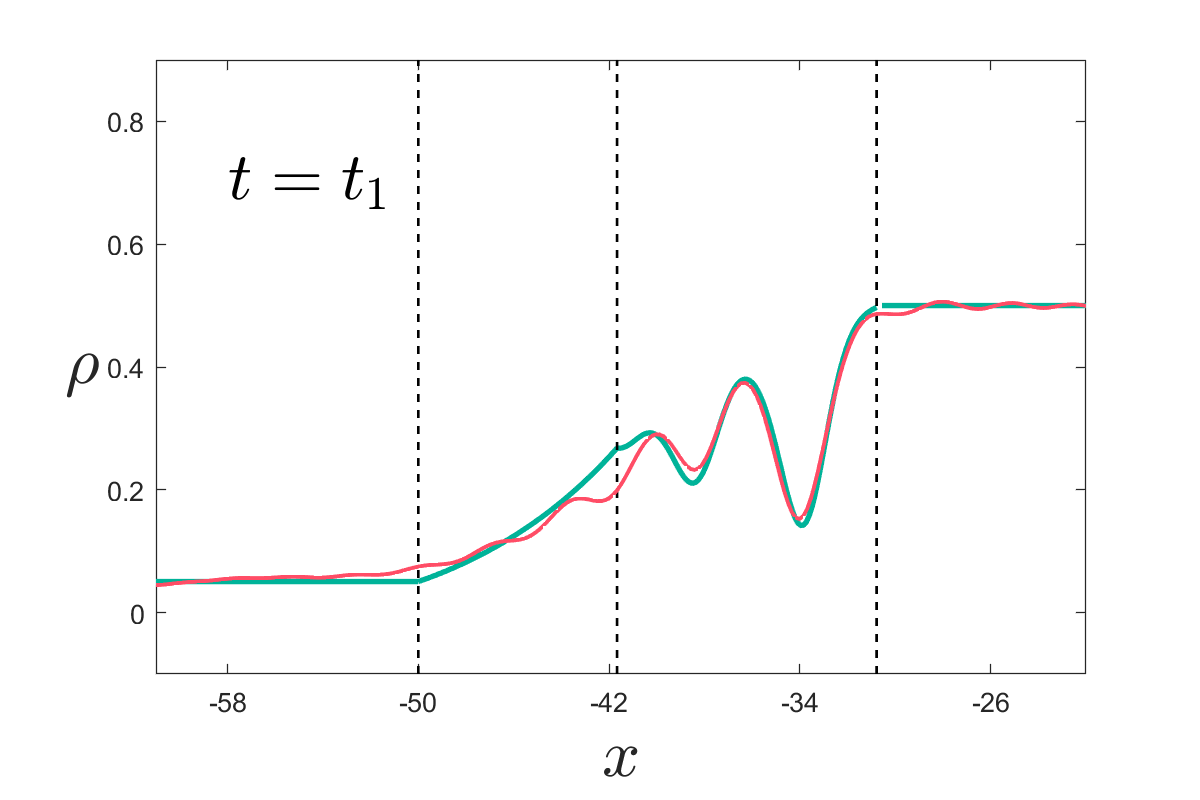}}\hfill
\subfigure[]{\includegraphics[width=0.33\linewidth]{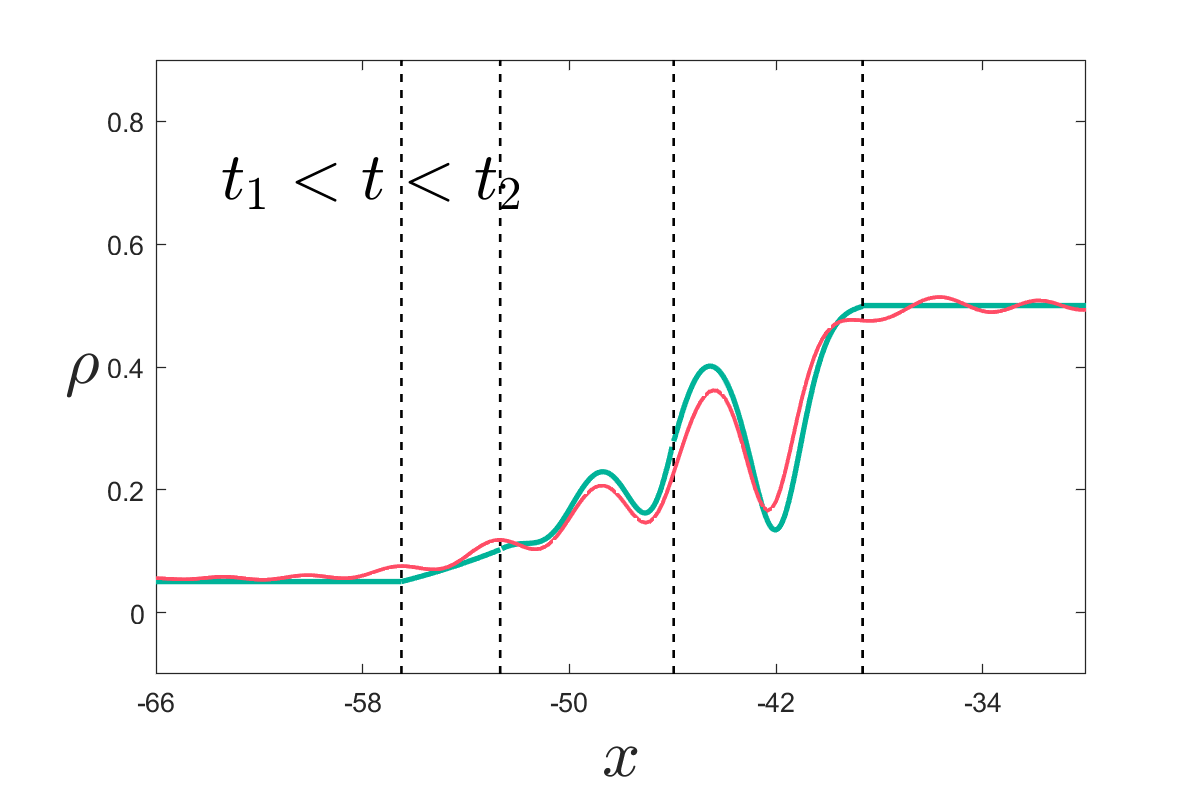}}\hfill
\subfigure[]{\includegraphics[width=0.33\linewidth]{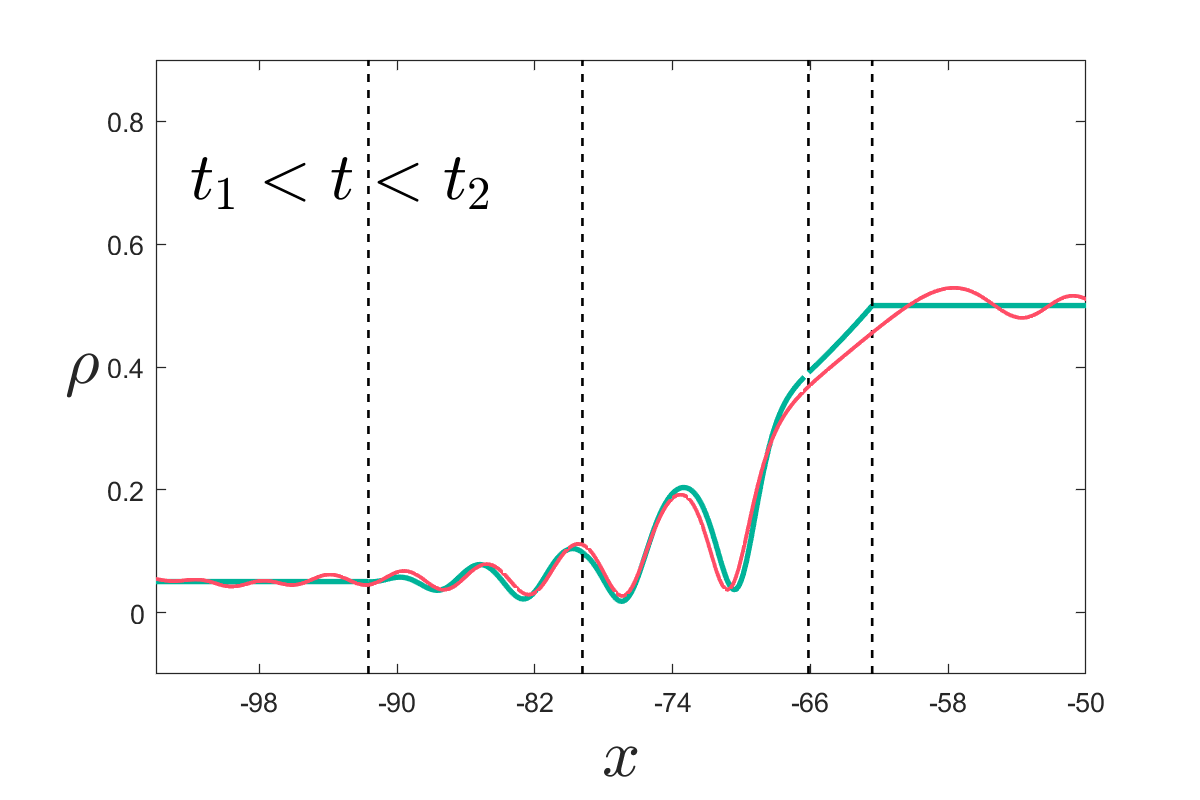}}\hfill
\subfigure[]{\includegraphics[width=0.33\linewidth]{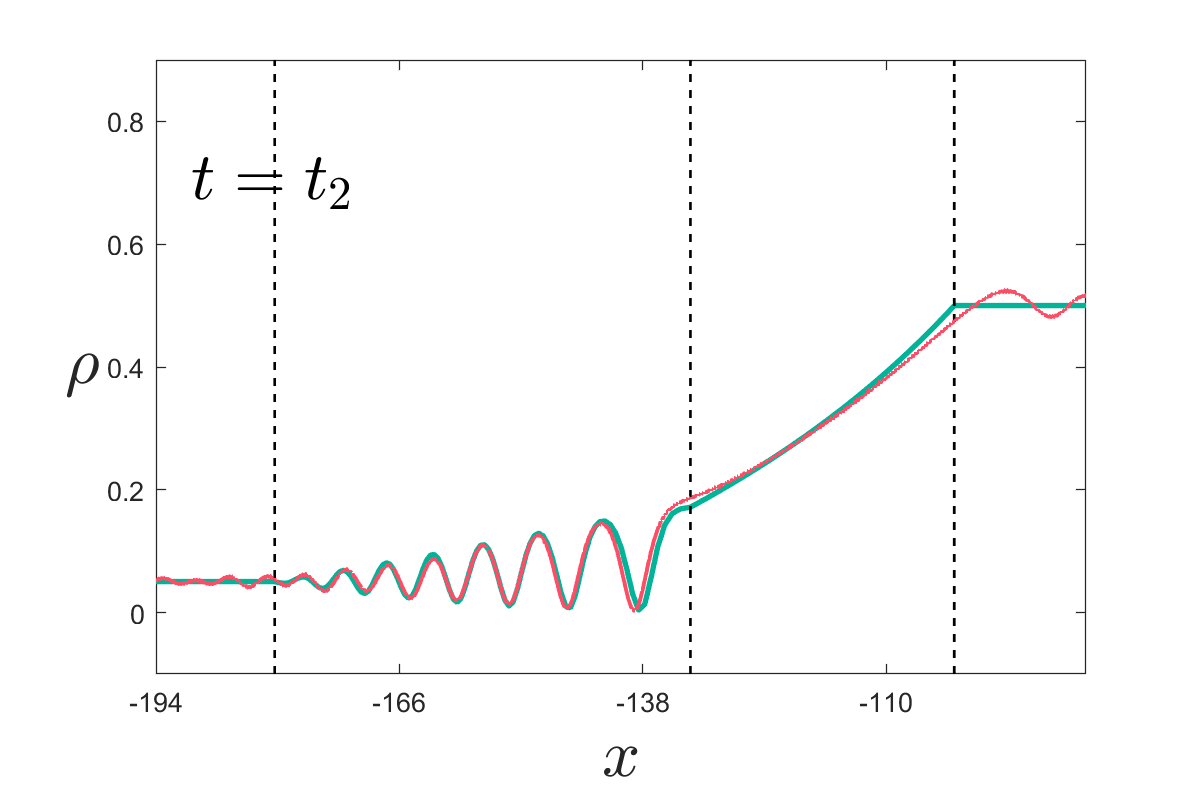}}\hfill
\subfigure[]{\includegraphics[width=0.33\linewidth]{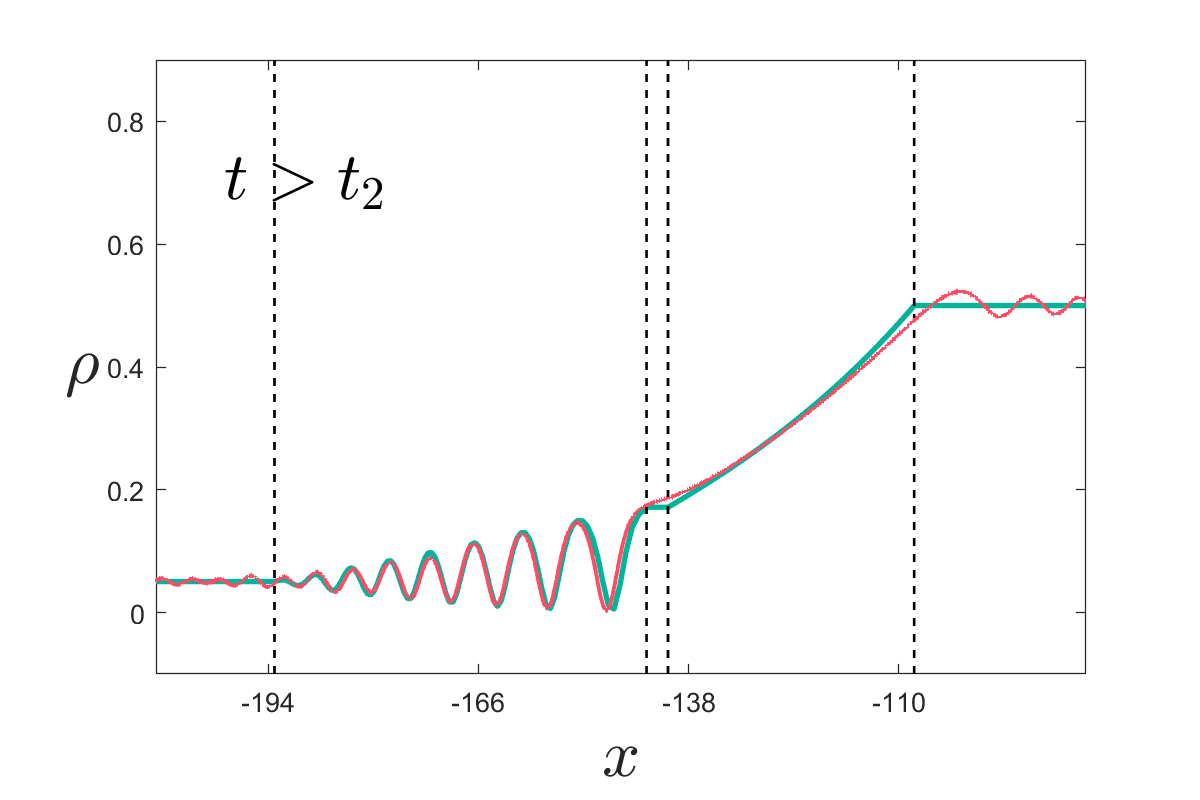}}
\flushleft{\footnotesize
\textbf{Fig.~$\bm{8}$.} The DSW analytical solutions constructed by mapping the Riemann invariants according to the relation given in Eq. (2.8) (green solid line) and the numerical simulation (red solid line) solutions on $x$.}
\end{figure}
$(i)$ Before interaction $(0<t<t_1)$:

The self-similar solution in the DSW region $x_1^-(t)<x(t)<x_1^+(t)$ is analogous to (3.4), with the Riemann invariant $l_2$ varying.
\begin{equation}
\begin{aligned}
 &l_1= l_-^R ,l_3 = l_-^L ,l_4 = l_+^R, \\
 V _ { 2 } (l_1 , l_2, l_3 , l_4 ) = \frac{1}{2} (l_-^R + l_2 +&~l_-^L + l_+^R) + \frac{(l_-^R - l_2) (l_2 - 
      l_-^L)\text K(m)} {(l_-^R - 
       l_-^L)\text E(m) + (-l_2 + 
       l_-^L)\text K(m)},
\end{aligned}
\end{equation}
where $m = \frac { (l_-^R-l_2) ( l_-^L - l_+^R ) } { ( l_-^R - l_-^L ) ( l_2 - l_+^R ) }.$
The boundaries of the DSW are then found as
\begin{equation}
x _ { 1 } ^{+} = \frac {(l_-^R + l_+^R + 2 l_-^R)} {2} t , ~x _ { 1 } ^{-} =  -\frac { -8l_-^{R2}+(l_-^L-l_+^R)^2+4l_-^R(l_+^R+l_-^{L})} {2(2l_-^R - l_-^L - l_+^R)} t .
\end{equation}
The RW centred at $x=-d$ can be described by Riemann invarients
\begin{equation}
\begin{aligned}
 &l_-= l_-^L, \\
 \frac{x+d}{t}=&V _ { + } (l_+ ,l_- ) = \frac{3l_++l_-}{2},
\end{aligned}
\end{equation}
and the boundaries $x_{2}^{\pm}$ are obtained by (3.36)
\begin{equation}
x _ { 2 } ^{-} = -d + \frac {  l_-^L + 3l_+^L } { 2 }t ,\   x _ { 2 } ^{+} = -d + \frac { l_-^L + 3l_+^R } { 2 } t .
\end{equation}
The time $t_1$ and position $x_1$ at which the leading edge of the DSW overtakes the trailing edge of the RW can be expressed as
\begin{equation}
t _ { 1 } =- \frac { d(l_-^L-2l_-^R+l_+^R)} { (l_-^R-l_+^R)(l_+^R-4l_-^R+3l_-^L) } , ~x _ { 1 } = d \frac {l_-^{L2}+4l_-^Ll_-^R-8l_-^{R2}-2l_-^Ll_+^R+4l_-^Rl_+^R+l_+^{R2}} { 2(l_-^R-l_+^R)(-l_+^R+4l_-^R-3l_-^L)} .
\end{equation}

$(ii)$ Being interaction $(t_1<t<t_2)$:

The modulated solutions in the non-interaction region remain consistent with Eqs. (3.34) and (3.36). While the modulated solution within the interaction region is governed by the Whitham systems (2.10) and satisfy the boundary conditions given below
\begin{equation}
\begin{cases} 
l_ { 1 } = l_-^R ,~ l_ { 3 } = l_-^L,~ l_ { 4 } = l_+^R,~
{x}= V _ { 2 } (l_1 ,l_2 , l_3 , l_4 )t,~at~x = x_{2}^{+}(t),\\
l_ { 1 } = l_ { 2 } = l_-^R, ~l_ { 3 } = l_-^L ,~
x = V _- (l_3 ,l_4 )t-d,~at~x = x_{1}^{-}(t).
\end{cases} 
\end{equation}
It is observed that $l_1$ and $l_3$ are constant and two Whitham equations remain. Carrying out calculations analogous to those in the previous section, we derive the solution for the function $f( l_ { 2 } , l_ { 4 } )$ in the EDP equation
\begin{equation}
\begin{aligned}
& f ( l_ { 2 } , l_ { 4 } ) = \frac {- 2 d ( l_+^R -l_-^R ) } { \pi \sqrt { ( l_+^R - l_ { 2 } ) ( l _ { 4 }-l_-^R) } } ( \Pi _ { 1 } ( s , z ) -\text K( z ) ) ,
 \end{aligned}
\end{equation}
where $z = \frac { ( l_-^R- l_ { 2 } ) ( l_ { 4 } -l_+^R ) } { ( l_+^R- l_ { 2 } ) ( l_ { 4 } -l_-^R ) } ,~ s = - \frac { l_ { 4 } -l_+^R } { l_-^R - l_ { 4 } } .$
Thus, the modulation solution for the interaction region between RW and DSW given
\begin{equation}
\begin{aligned} 
&l_ 1= l_-^R ,~l_ { 3 } = l_-^L,~ 
x - V _ { 2 , 4 } ( l_-^R,~l_2,~l_-^L ,~l_4 ) t= W_{2, 4}(l _ { 2 } , l _ { 4 } ),~
\end{aligned}
\end{equation}
where
\begin{equation}
\begin{aligned} 
W_{2, 4}(l _ { 2 } , l _ { 4 } )=\left( 1 - \frac { \mathfrak { L } } { \partial _ { 2 , 4 } \mathfrak { L } } \partial _ { 2 , 4 } \right) f ( l _ { 2 } , l _ { 4 } ).
\end{aligned}
\end{equation}
At $t=t_2$, the DSW wave fully overtakes the RW, i.e., $ x_{1}^{+}(t_2) = x_{2}^{-}(t_2)$, we give
\begin{equation}
t _{2} = \frac { 2 d(l_-^L-l_+^R)\text E(r)  } { \pi (l_-^L - l_+^L ) \sqrt { (l_+^R - l_-^R
)(l_+^L-l_-^R) } }, 
\end{equation}
where $r=\frac { (l_-^L-l_-^R)(l_+^L-l_+^R)  } { (l_+^L - l_-^L)(l_-^L-l_+^R) }$. And the corresponding $x_2 = x_{1}^{+}(t_2) = x_{2}^{-}(t_2)$ is given by

\begin{equation}
x_2 = \frac{l_-^R + l_+^L + 2l_-^L} {2} t_2 +\frac{2d(l_-^R-l_+^R)(-\text K(r)+\Pi_1(\frac{ l_+^L -l_+^R } { l_+^L - l_-^R},r))}{\pi \sqrt{(l_-^L -l_+^R)(l_-^R -l_+^L)}}.
\end{equation}

$(iii)$ After interaction $(t>t_2)$:

The separated DSW is described by the modulated solution below 
\begin{equation}
l_ { 1 } = l_-^R,~l_ { 3 } =l_-^L,~l_ { 4 } =l_+^L,\\
x=V_{2}t+W_2(l_2,l_+^L),
\end{equation}
where
\begin{equation}
\begin{aligned}
&W_2(l_2,l_+^L)=\frac{2d}{\pi \sqrt{(l_-^R -l_+^L)(l_2-l_+^R)}}((l_-^R-l_+^R)\Pi_1(\frac{-l_+^L+l_+^R}{l_-^R-l_+^L}
,{q})+\\ &\frac{(l_2-l_-^L)(l_2-l_+^R)(l_-^R-l_+^L)\text{E}(q)+(-l_2+l_+^L)
(l_-^R-l_+^R)(-l_-^L+l_-^R)\mu(y)\text{E}(q)}{(-l_2+l_+^L)((-l_+^L+l_+^R)
\mu(y)+(l_2-l_+^R))}),
 \end{aligned}
\end{equation}
and
\begin{equation}
q =\frac{(l_2-l_-^R)(l_+^L+l_+^R)}{(l_2-l_+^R)(l_+^L-l_-^R)}, ~y = \frac { (l_2-l_-^R) (l_-^L-l_+^L) } {(l_2-l_+^L)(l_-^L-l_-^R)} .
\end{equation}
The boundaries of the DSW are given
\begin{equation}
\begin{aligned}
&~~~~~~~x_1 ^+=\frac{l_-^R + l_+^R + 2l_-^L} {2} t +W_2(l_-^R,l_+^L),  \\
x _ { 1 } ^{-} =  &\frac { l_-^{L2}+4l_-^Ll_-^R-8l_-^{R2}-2l_-^Ll_+^L+4l_-^Rl_+^L+l_+^{L2}} {2(l_-^L - 2l_-^R + l_+^L)} t+W_2(l_-^L,l_+^L). \end{aligned}
\end{equation}
The solution for the separated RW gives 
\begin{equation}
l_-=l_-^R,~ x = \frac {l_{-}^R + 3l_+} {2} t + W_4 ( l_ {-}^R,l_+ ),
\end{equation}
and the function $W_4( l_ {-}^R,l_+ )$ has the form
\begin{equation}
W_4( l_ {-}^R,l_+ ) = \frac { 2d\left( (l_-^R-l_+^R) ( \Pi _ { 1 } (p, r ) - \text K(r)) + (l_-^L-l_+^R) \text E (r) \right) } {\pi \sqrt {(l_ + - l_-^R)(l_-^L-l_+^R)}}, 
\end{equation}
where $r= \frac { ( l_+-l_+^R) ( l_-^L-l_-^R) } {(l_+-l_-^R)( l_-^L-l_+^R) } ,~p = - \frac { l_+-l_+^R  } { l_-^R-l_+ } .$
The boundaries of the RW are given
$$
x _ { 2 } ^{-} = \frac { l_-^R+ 3l_+^L} { 2 } t + W _ { 4 } ( l_ {-}^R,l_+^L ) ,~ x _ { 2 } ^{+} = \frac { l_-^R+ 3l_+^R} { 2 } t + W _ { 4 } ( l_ {-}^R,l_+^R ) .
$$
The three evolutionary stages derived from the foregoing results are presented in Figs.~6-8.

\vspace{5mm}\noindent\textbf{3.3 RW overtakes CDSW}
\hspace*{\parindent}\\

The hydrodynamics equations of Eq.~(1.1) investigated in this paper is non-convex, which fundamentally differentiates its wave evolution behaviors from those of classical convex equations. Systems with convexity can only generate two fundamental wave structures, namely the RWs and DSWs. In contrast, the non-convexity of systems induces a novel wave structure that does not exist in convex equations, known as the CDSW. It is worth noting that in the separation regime, our calculations reveal that the velocity of the CDSW is always lower than that of the RW. Consequently, only the scenario in which the RW overtakes the CDSW is admissible and considered in this work.

Configuration: Suppose the CDSW and RW are generated at 
$(0,0)$ and $(d,0)$ on the $(x,t)$ plane at $t=0$, respectively. Their respective expanding regions given by $x_1^-(t)<x(t)<x_1^+(t)$ and $x _2^-(t)<x(t)<x_2^+(t)$. We have the following initial data:
\begin{equation}
\begin{aligned}
l_+(x,0)=
\begin{cases} 
l_+^L, & x<0,\\
 l_+^R, & x>0, 
\end{cases}
\quad \text{and} \quad
l_-(x,0)=
\begin{cases} 
l_-^L, & x<d, \\
l_-^R, & x>d,
\end{cases}
\end{aligned}
\end{equation}
where $l_-^L<l_-^L<l_+^L=l_+^R$ and $d>0$. The configuration at the origin constitutes a distinctive prerequisite for the generation of CDSW. Although the Riemann invariants at this location remain continuous, the corresponding density exhibits a distinct discontinuity. The density $\rho$ and velocity $v$ profiles corresponding to this construction are illustrated in the Fig. 9.

$(i)$ Before interaction $(0<t<t_1)$:

In the CDSW region $x_1^-(t)<x(t)<x_1^+(t)$, the two largest Riemann invariants are identical. Along this
solution, we have $m = 0$, which yields
\begin{equation}
\begin{aligned}
&l_1= l_-^L ,~l_2 = l_+^L,~l_3=l_4,\\
V _ { 4 } =&~V_3= \frac{l_-^{L2}-2l_-^Ll_+^L+4l_-^Ll_4-8l_4^2}{2(l_-^L+l_+^L-2l_4)}.
\end{aligned}
\end{equation}
The boundaries of the CDSW are determined by setting $l_{4} = 0$ for the trailing edge $x_{1}^{+}$ and $l_{4} = l_2= l_+^L$ for the leading edge $x_{1}^{-}$:
\begin{equation}
x _ { 1 } ^{-} = \frac {l_-^{L2} + 2 l_-^Ll_+^L - 3 l_+^{L2}} {2(l_-^L-l_+^L)} t , ~x _ { 1 } ^{+} =  \frac { l_-^{L2}-2l_-^Ll_+^L+l_+^{L2}} {2(l_-^L -l_+^L)} t .
\end{equation}

The left-propagating RW is asymptotically described by
\begin{equation}
\begin{aligned}
 &l_+= l_+^R , \\
 \frac{x-d}{t}=&V _ { - } (l_+ ,l_- ) = \frac{3l_-+l_+}{2},
\end{aligned}
\end{equation}
and the boundaries $x_{2}^{\pm}$ are given
\begin{equation}
x _ { 2 } ^{-} = d + \frac { 3 l_-^L + l_+^R } { 2 }t ,\   x _ { 2 } ^{+} = d + \frac { 3 l_-^R + l_+^R } { 2 } t .
\end{equation}
\begin{figure}[htbp]
\centering
\setcounter{subfigure}{0}
\subfigure[]{\includegraphics[width=0.333\linewidth]{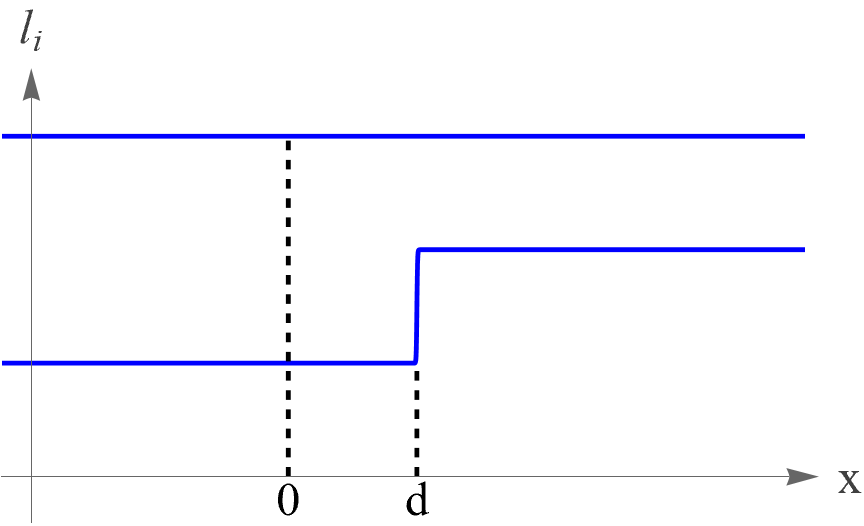}}\hfill
\subfigure[]{\includegraphics[width=0.333\linewidth]{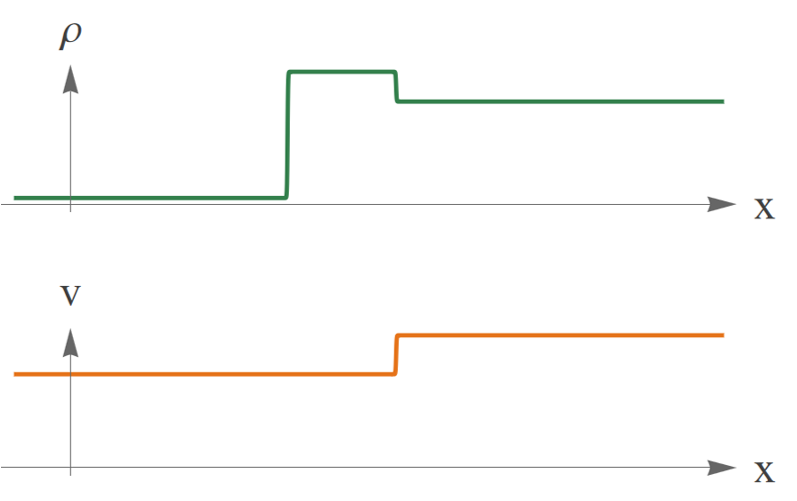}}\hfill
\subfigure[]{\includegraphics[width=0.333\linewidth]{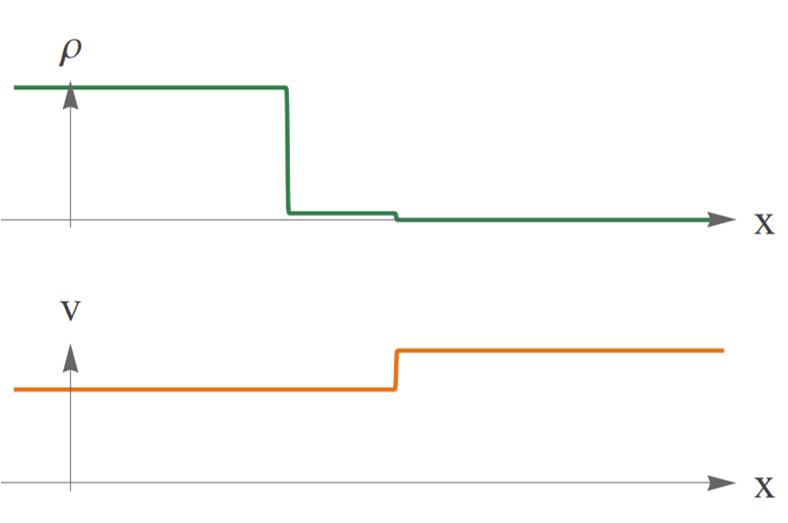}}\hfill
\flushleft{\footnotesize
\textbf{Fig.~$\bm{9}$.} The initial configuration (3.51) of the Riemann invariants, along with two sets of initial density and velocity corresponding to the constant states of the Riemann problem.}
\end{figure}
\begin{figure}[htbp]
\centering
\setcounter{subfigure}{0}
\subfigure[]{\includegraphics[width=0.71\linewidth]{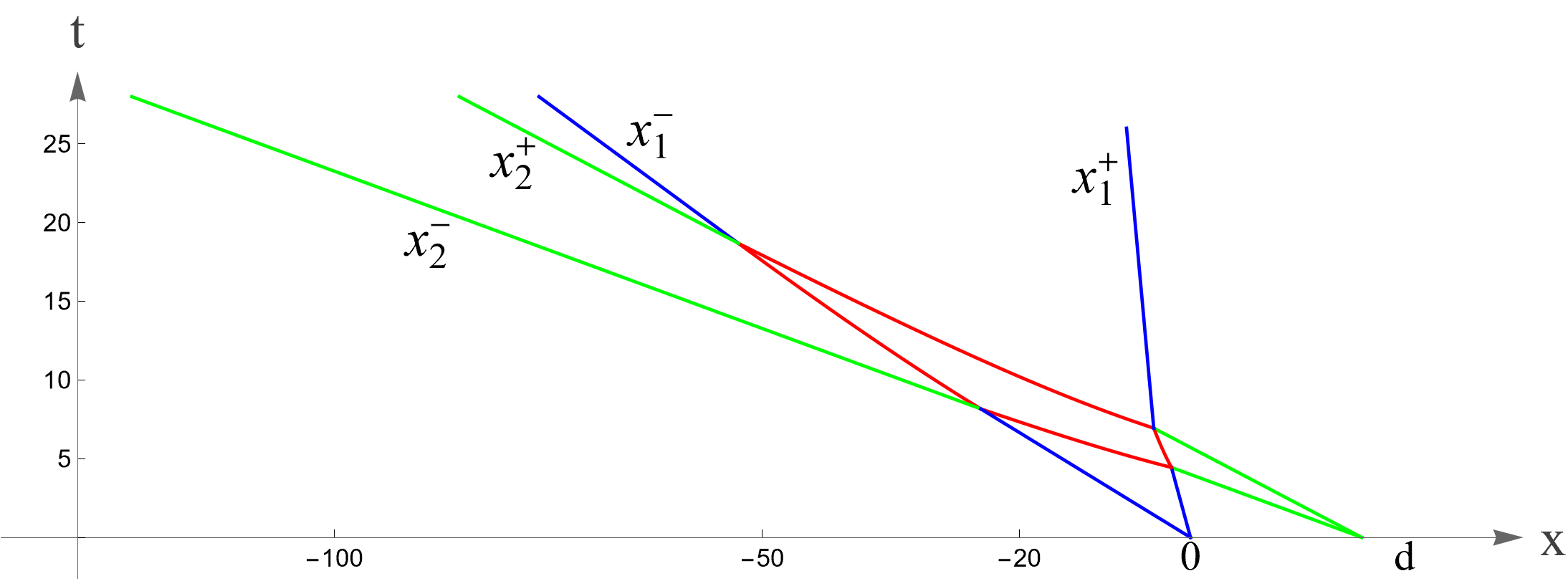}}\hfill
\subfigure[]{\includegraphics[width=0.333\linewidth]{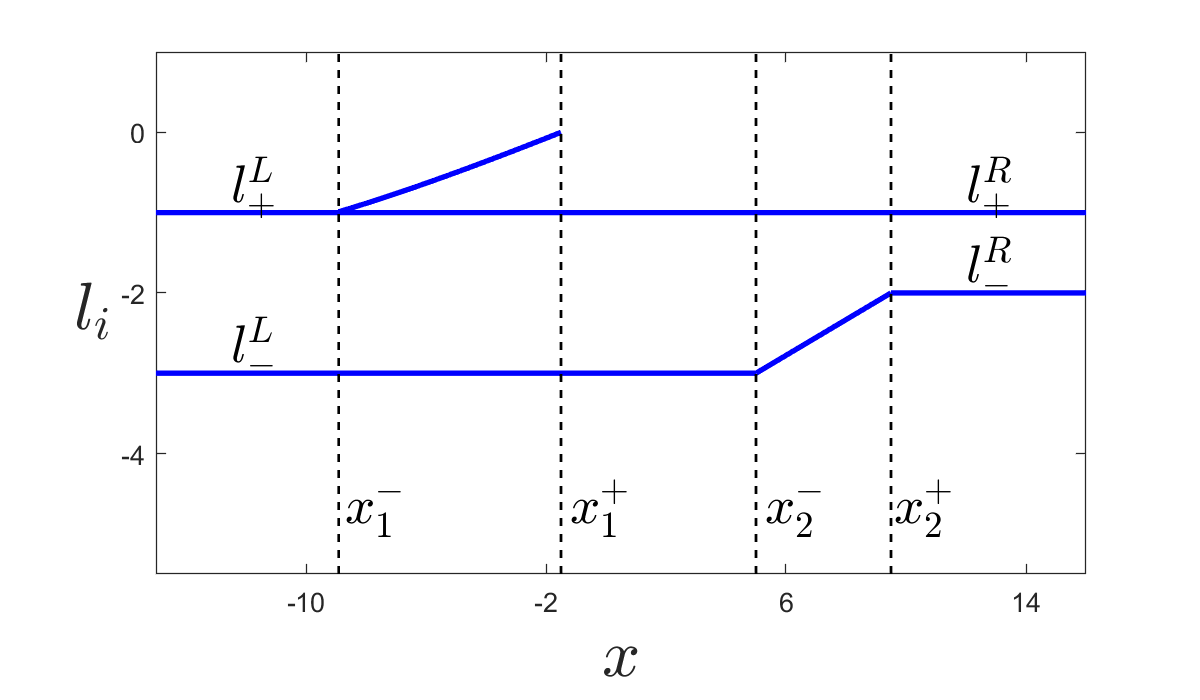}}\hfill
\subfigure[]{\includegraphics[width=0.333\linewidth]{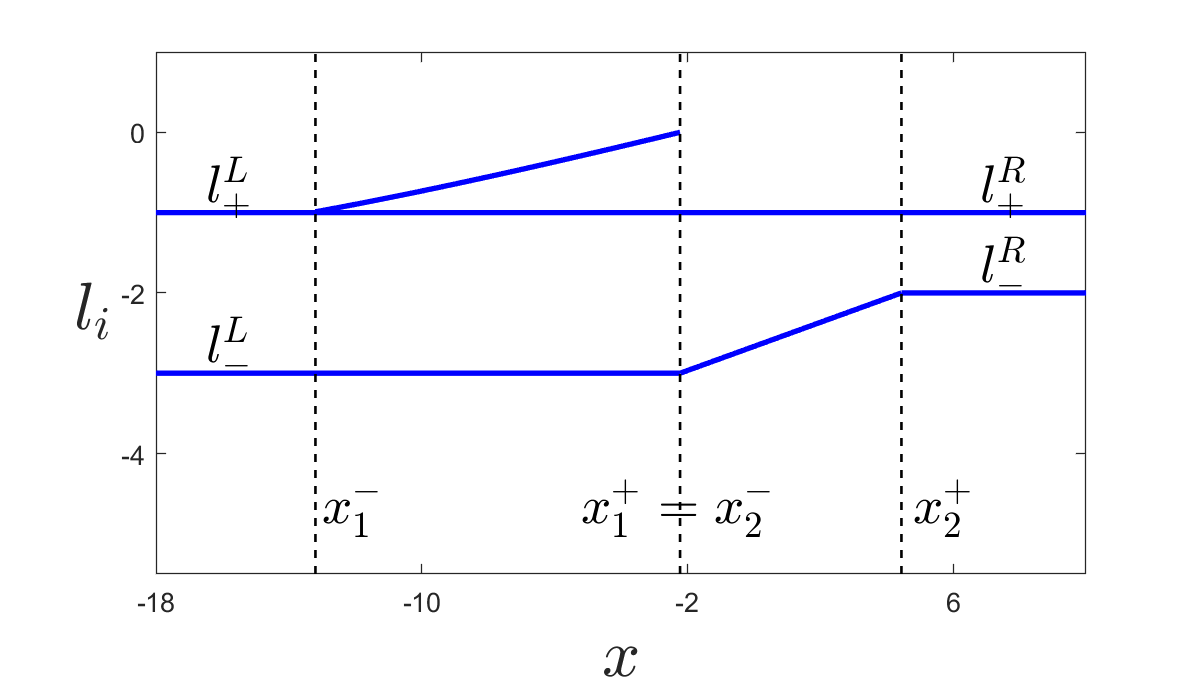}}\hfill
\subfigure[]{\includegraphics[width=0.333\linewidth]{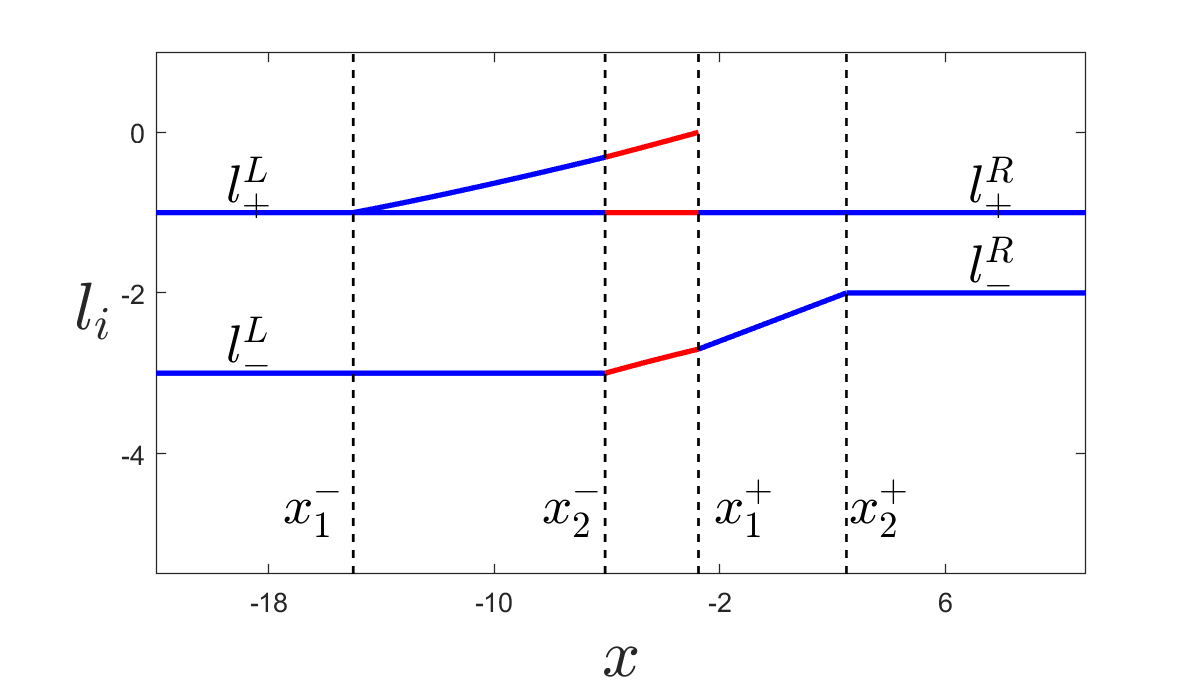}}\hfill
\subfigure[]{\includegraphics[width=0.333\linewidth]{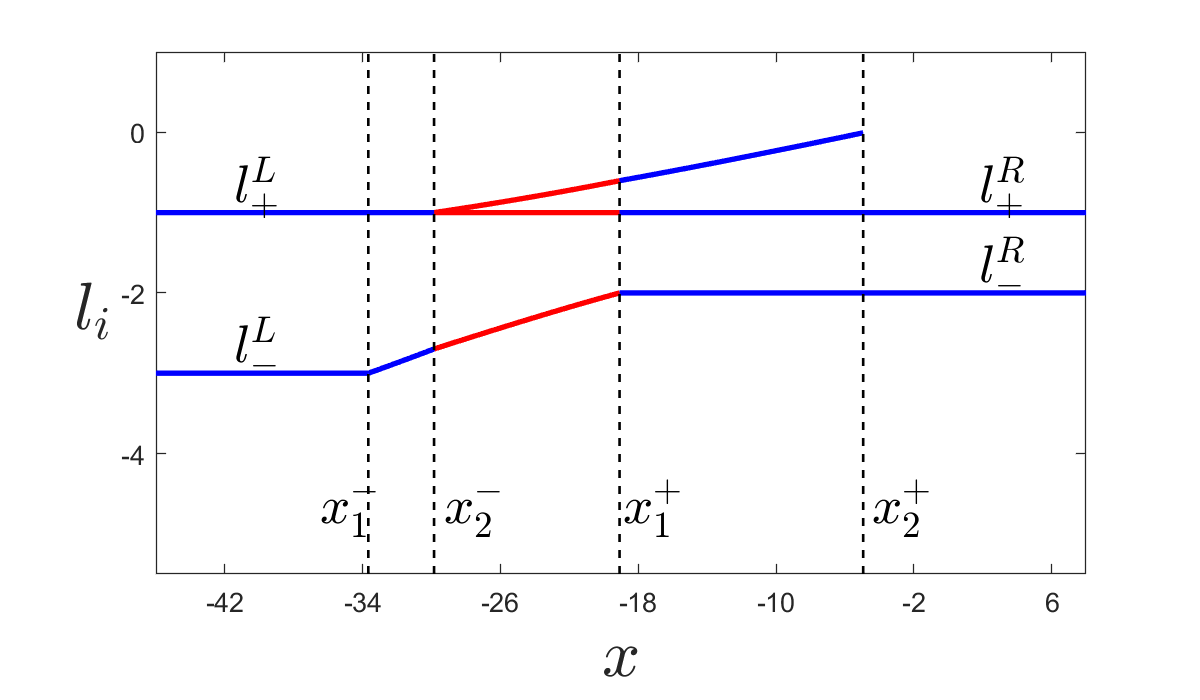}}\hfill
\subfigure[]{\includegraphics[width=0.333\linewidth]{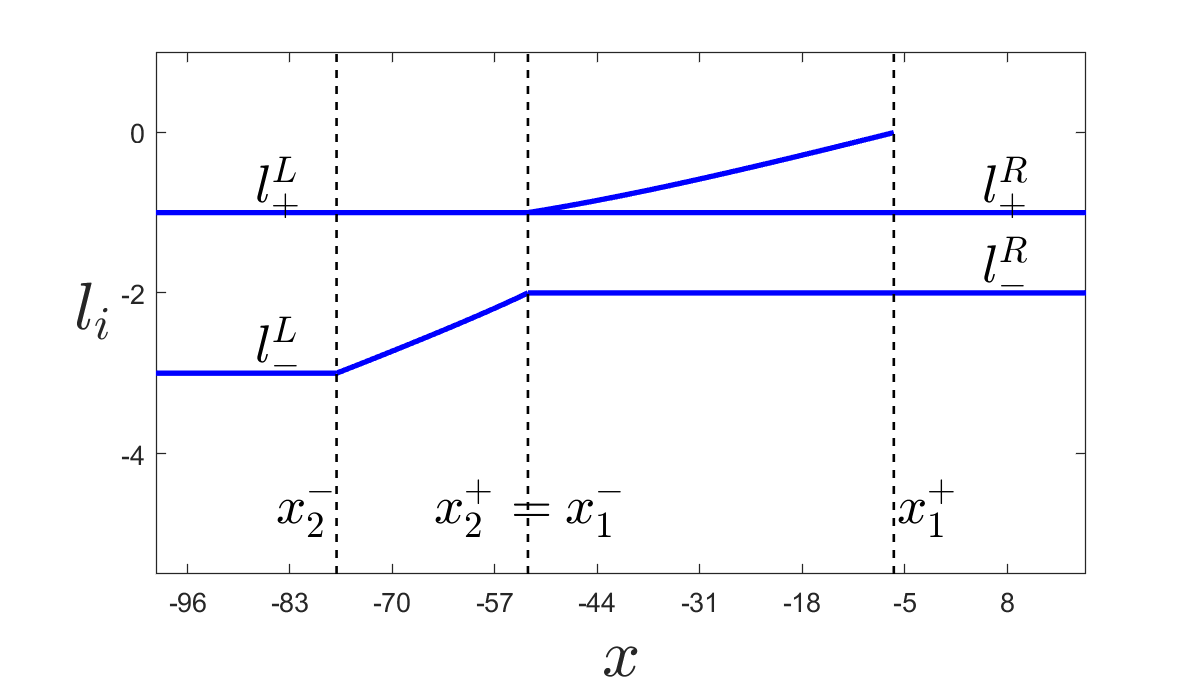}}\hfill
\subfigure[]{\includegraphics[width=0.333\linewidth]{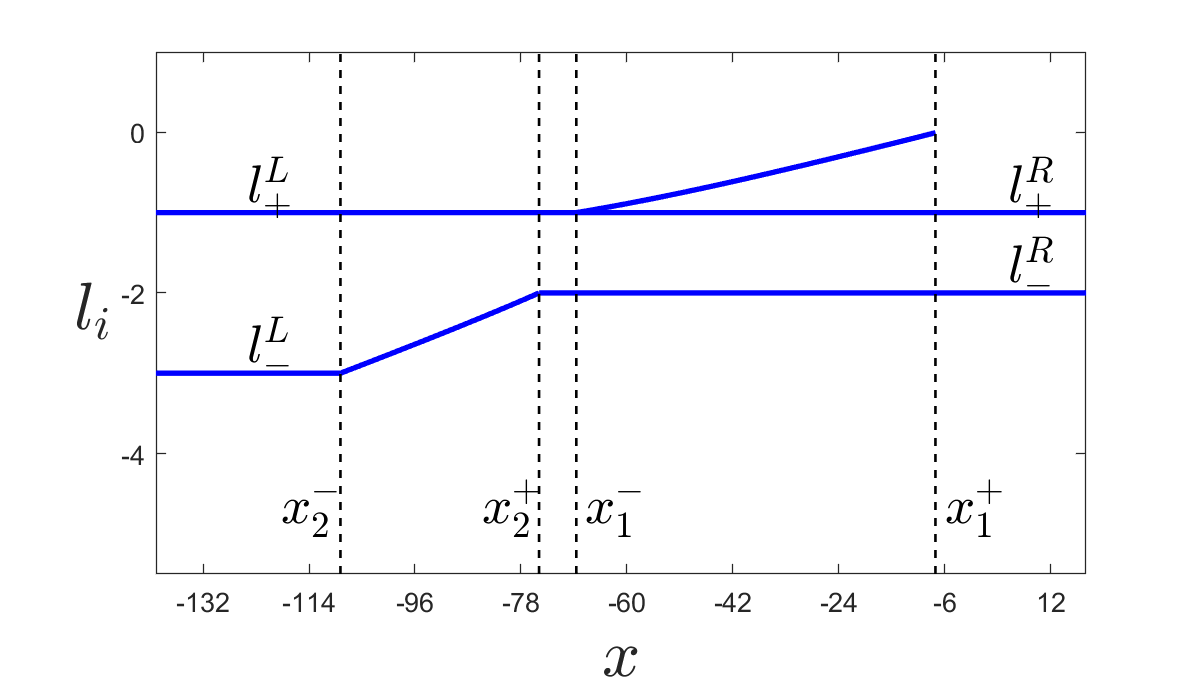}}
\flushleft{\footnotesize
\textbf{Fig.~$\bm{10}$.} (a) presents the boundaries of distinct wave regions throughout the evolution: blue lines indicate the boundaries of the CDSW, green lines indicate the boundaries of the RW, and red lines indicate the boundaries of the interaction region. (b)-(g) illustrate the evolution of Riemann invariants, with solid red lines denoting the interaction region.}
\end{figure}

\begin{figure}[htbp]
\centering
\setcounter{subfigure}{0}
\subfigure[]{\includegraphics[width=0.333\linewidth]{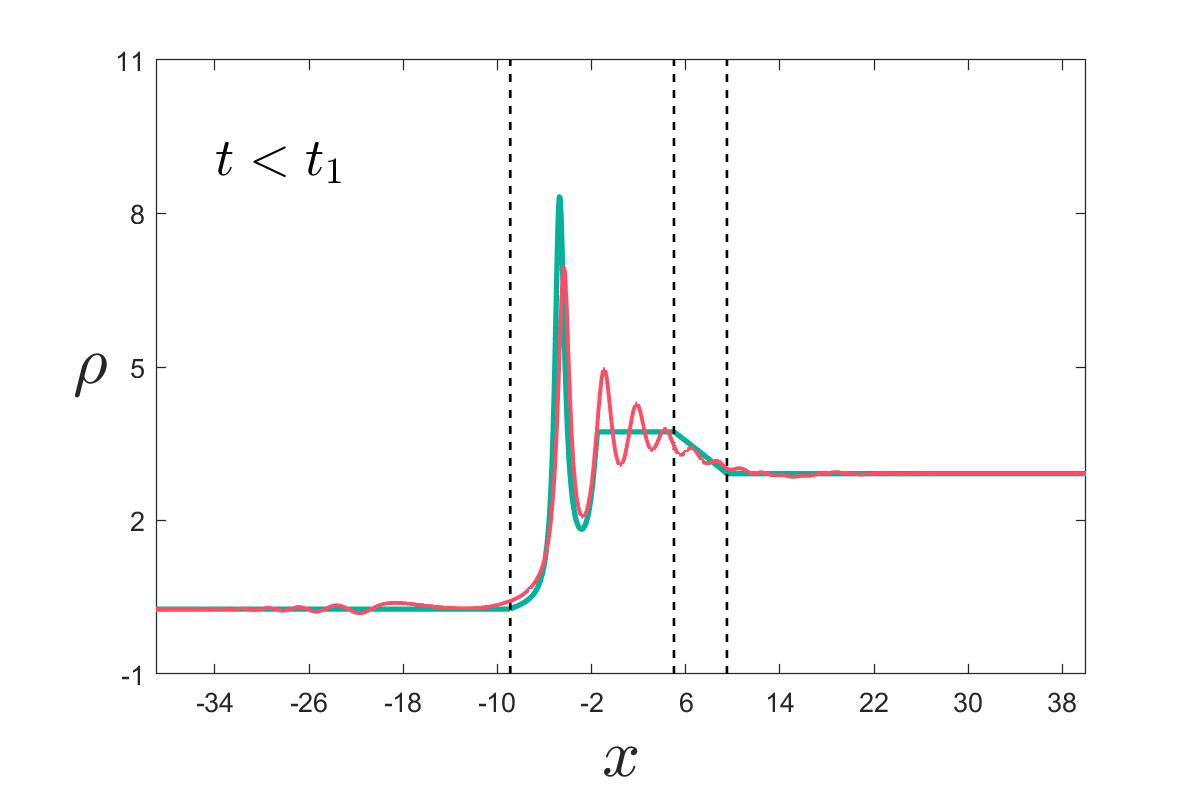}}\hfill
\subfigure[]{\includegraphics[width=0.333\linewidth]{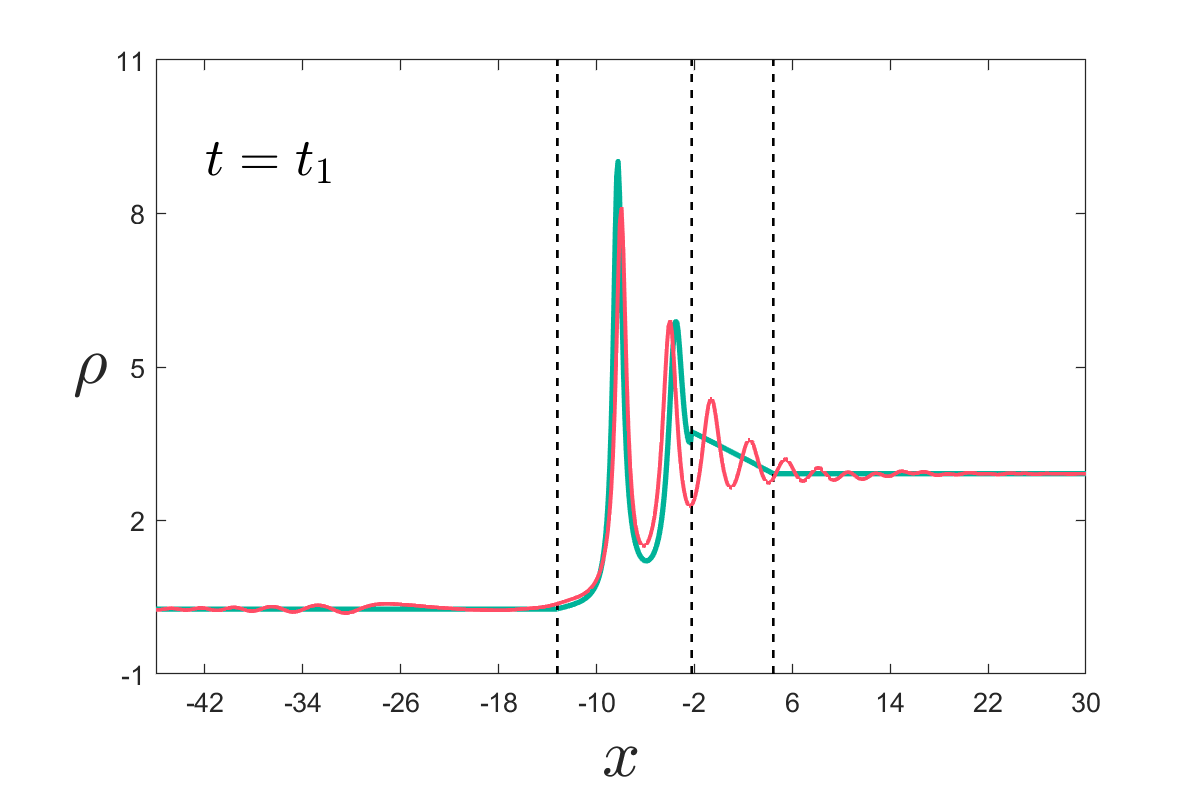}}\hfill
\subfigure[]{\includegraphics[width=0.333\linewidth]{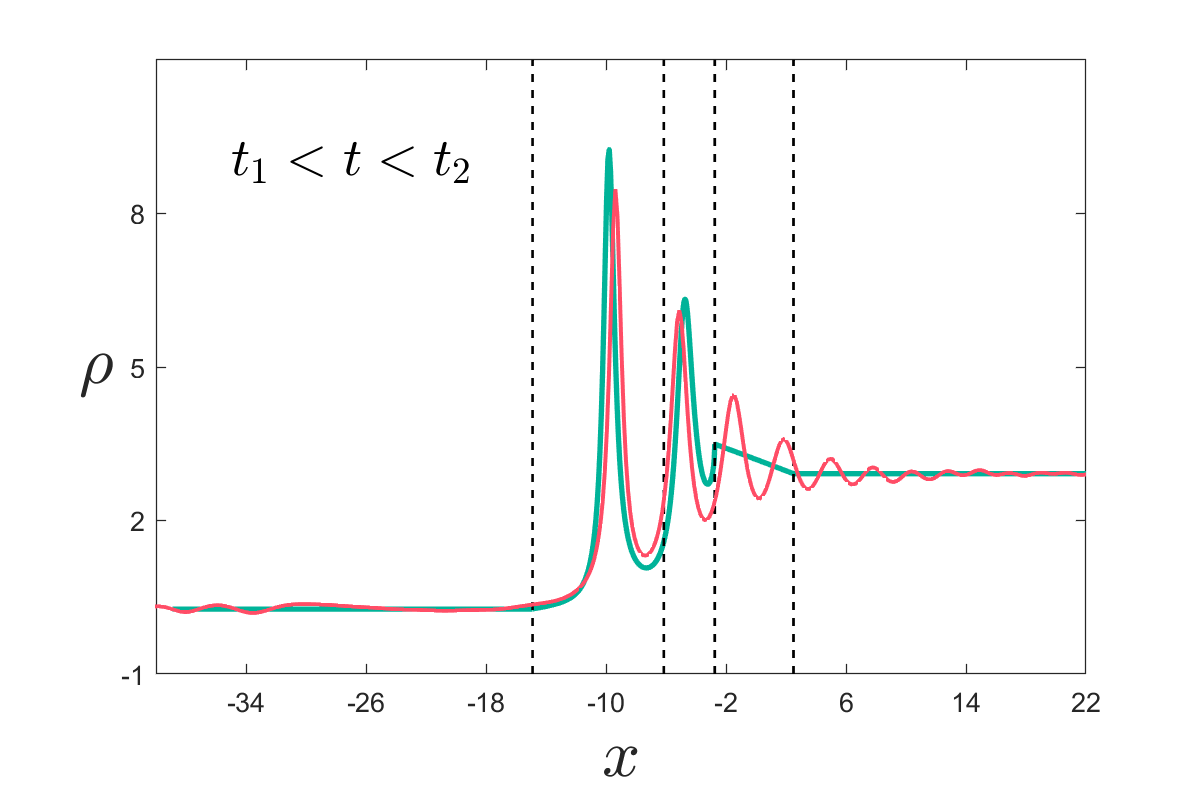}}\hfill
\subfigure[]{\includegraphics[width=0.333\linewidth]{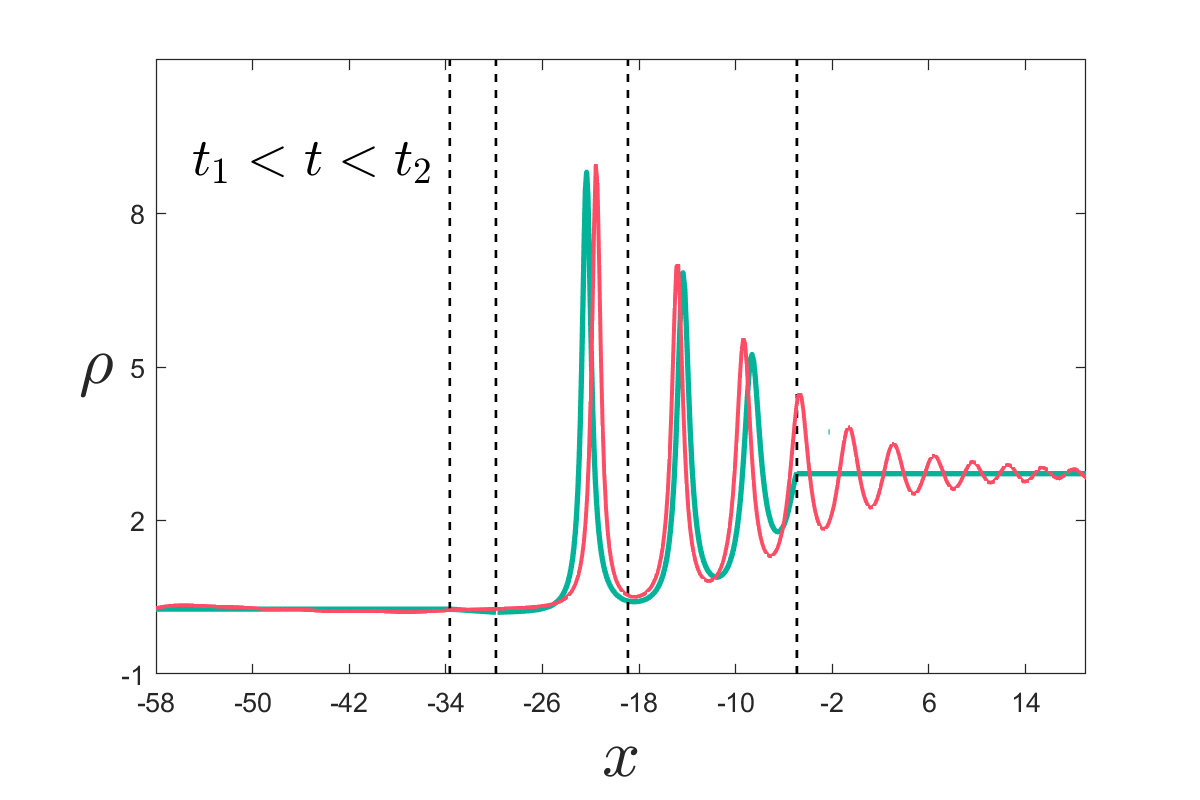}}\hfill
\subfigure[]{\includegraphics[width=0.333\linewidth]{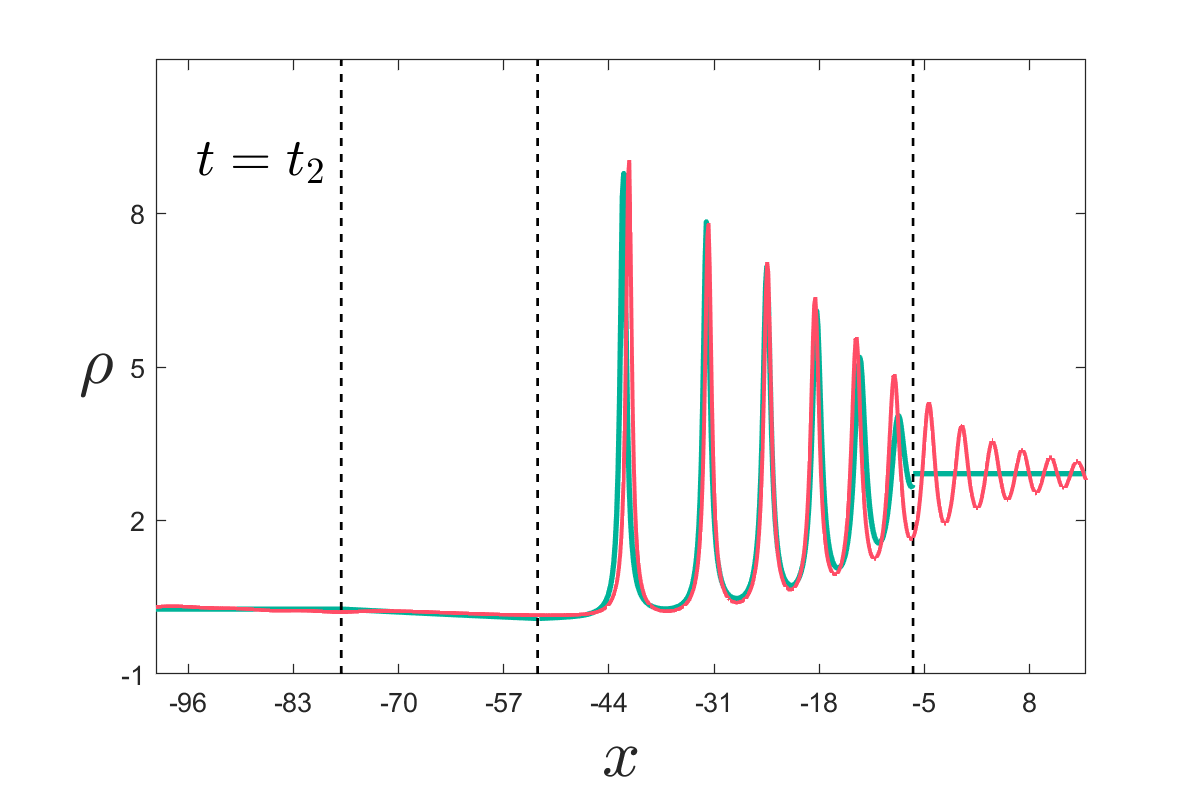}}\hfill
\subfigure[]{\includegraphics[width=0.333\linewidth]{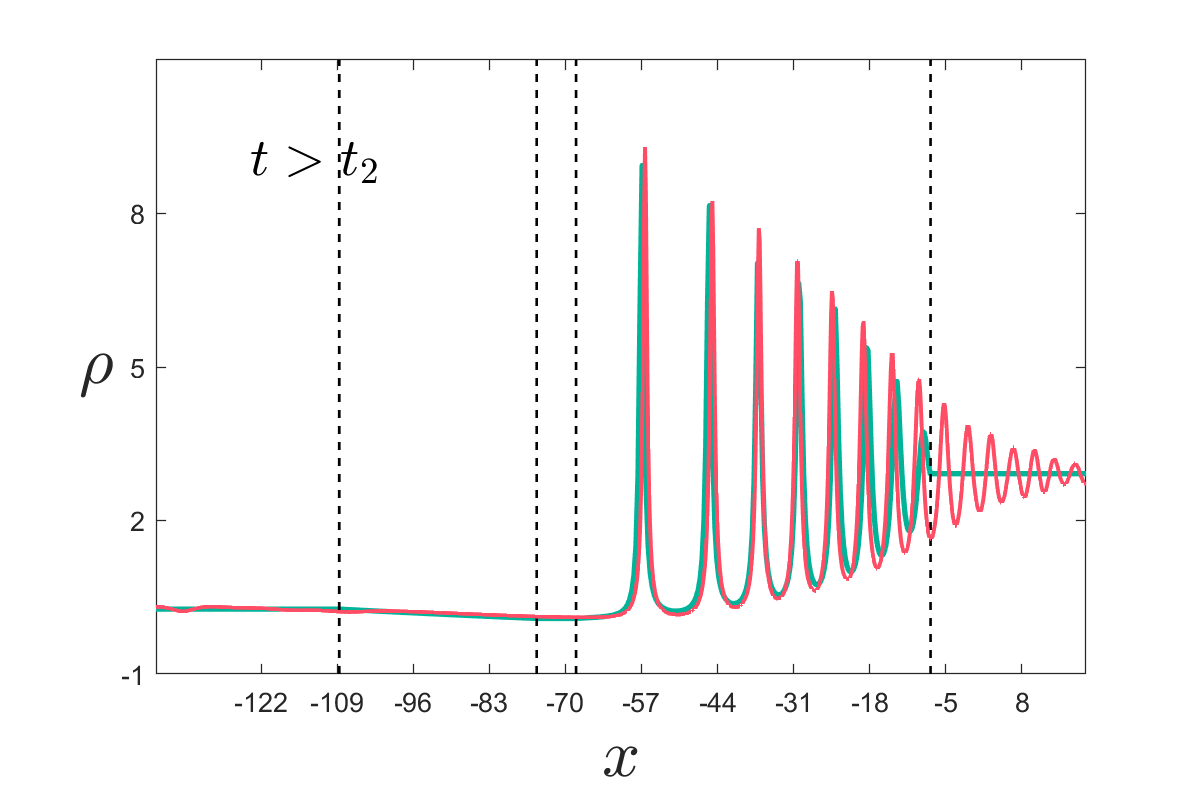}}
\flushleft{\footnotesize
\textbf{Fig.~$\bm{11}$.} The CDSW analytical solutions constructed by mapping the Riemann invariants according to the relation given in Eq. (2.7) (green solid line) and the numerical simulation (red solid line) solutions on $x$.}
\end{figure}
\begin{figure}[htbp]
\centering
\setcounter{subfigure}{0}
\subfigure[]{\includegraphics[width=0.333\linewidth]{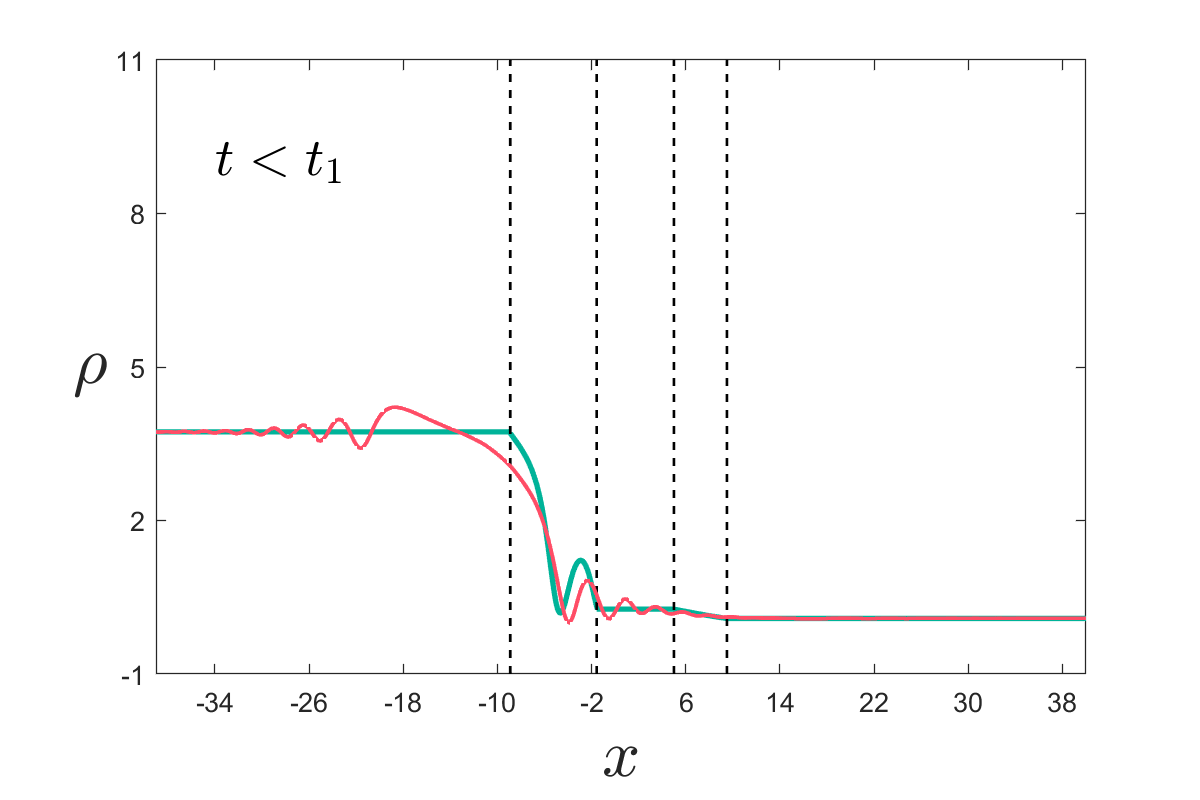}}\hfill
\subfigure[]{\includegraphics[width=0.333\linewidth]{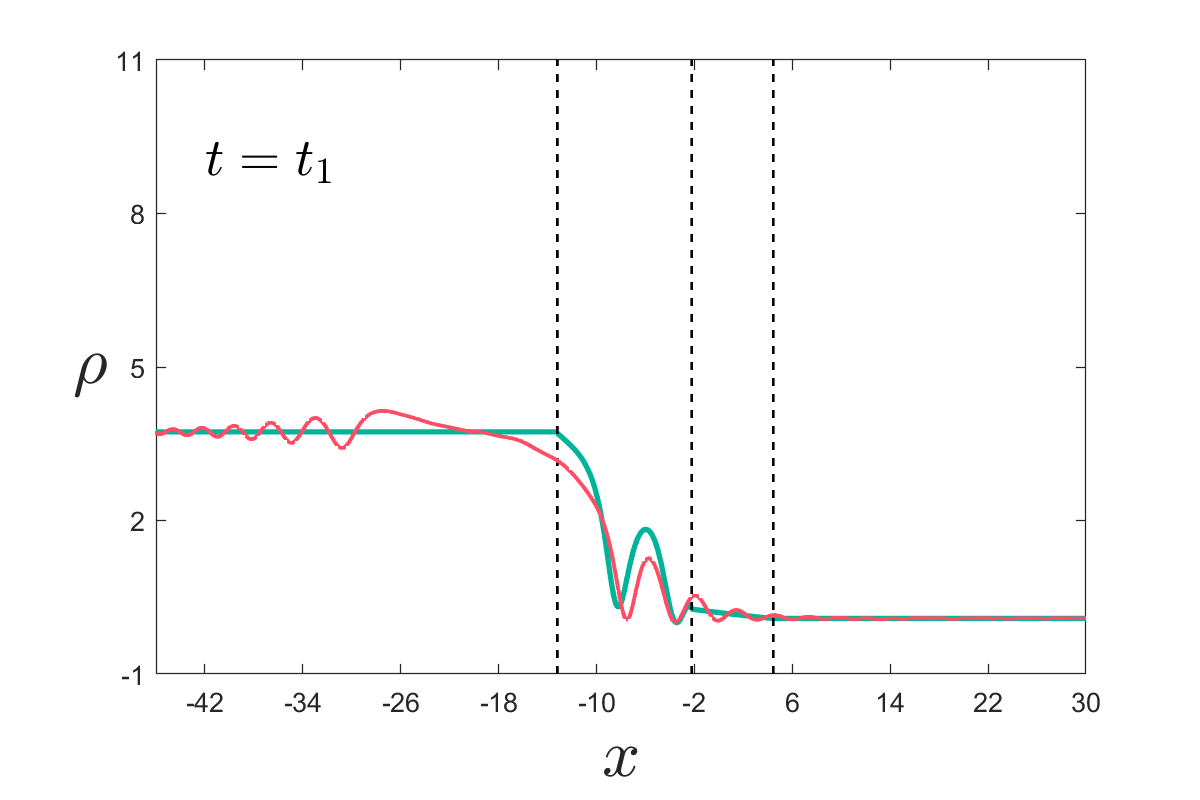}}\hfill
\subfigure[]{\includegraphics[width=0.333\linewidth]{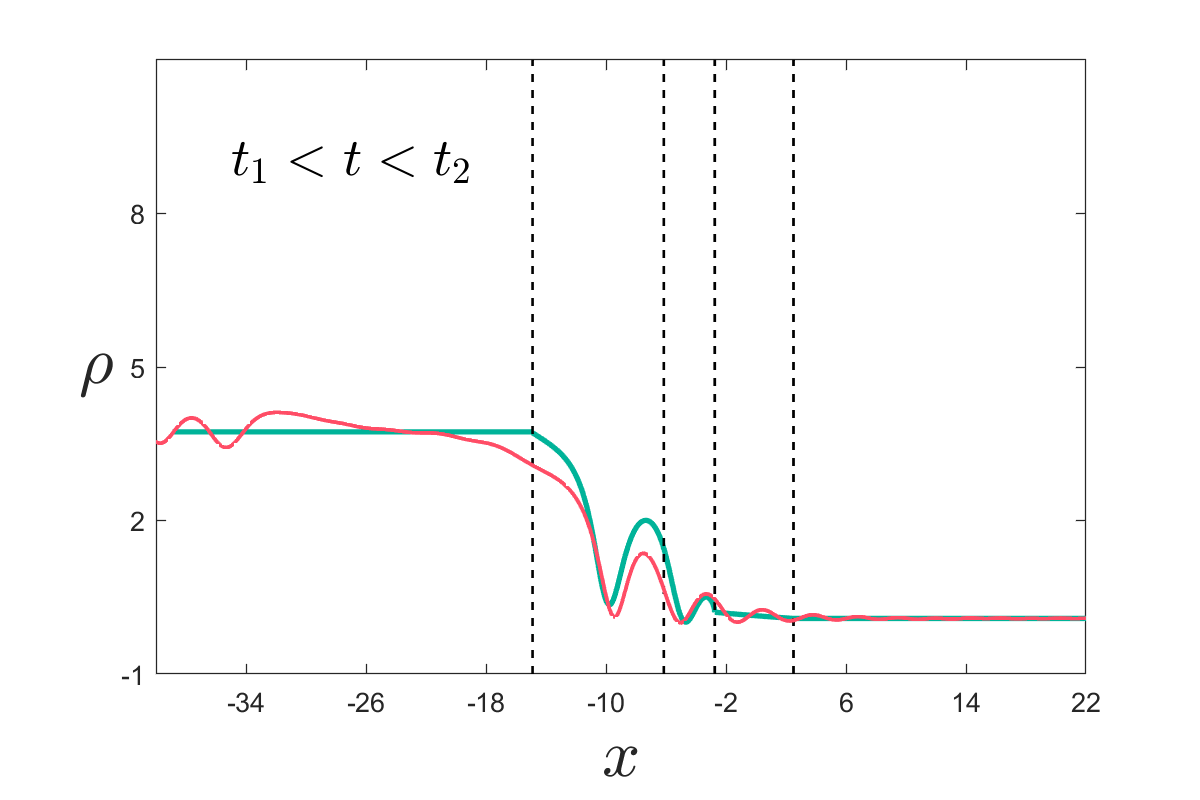}}\hfill
\subfigure[]{\includegraphics[width=0.333\linewidth]{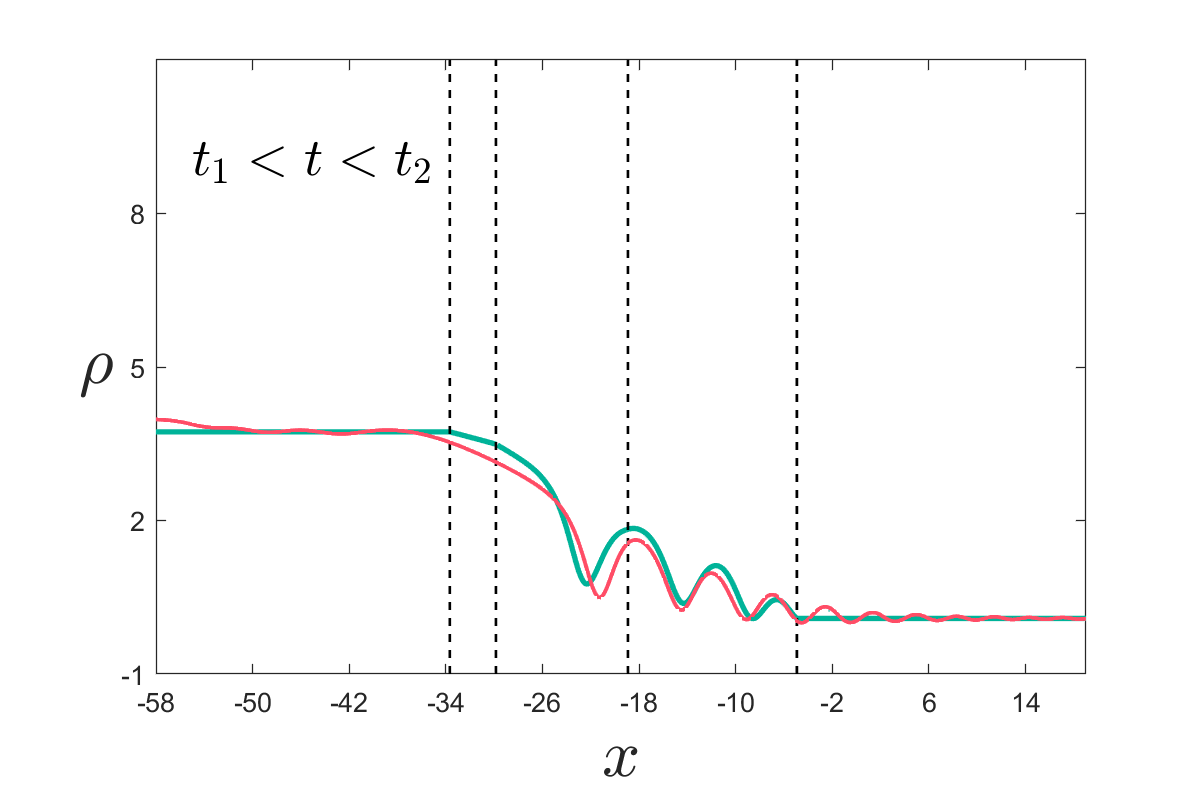}}\hfill
\subfigure[]{\includegraphics[width=0.333\linewidth]{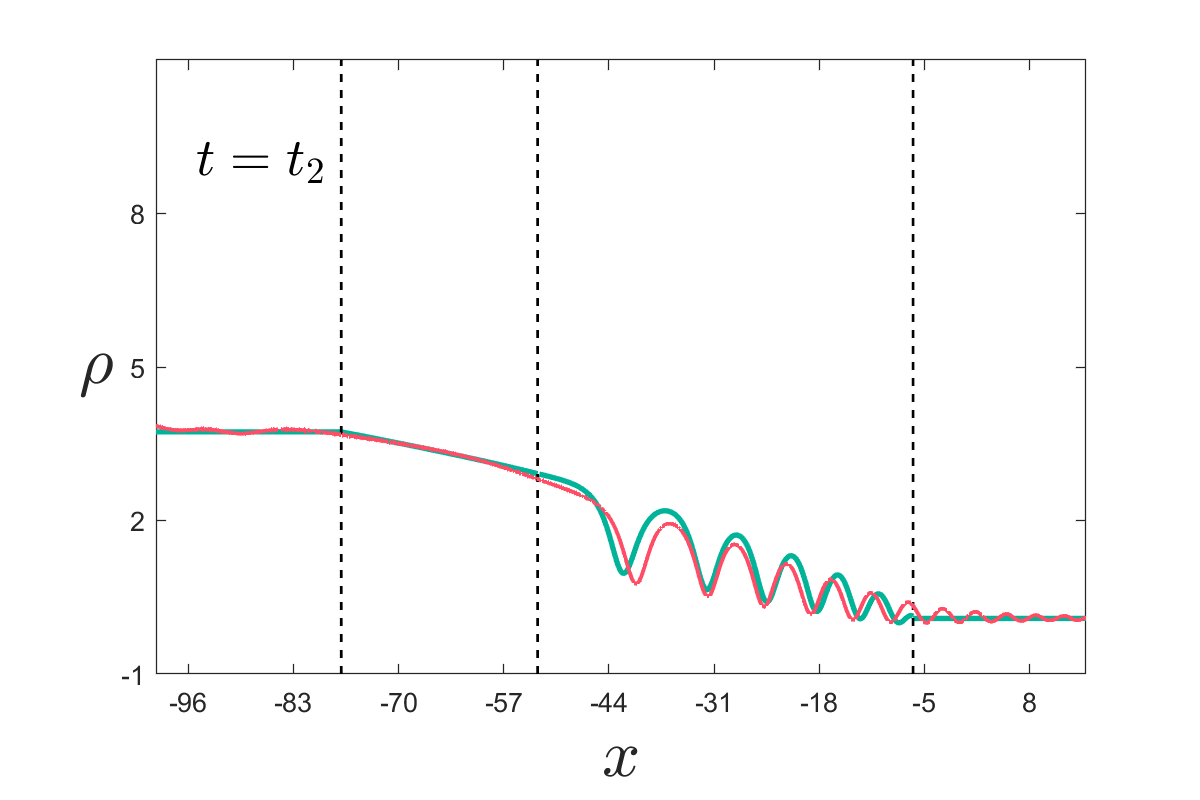}}\hfill
\subfigure[]{\includegraphics[width=0.333\linewidth]{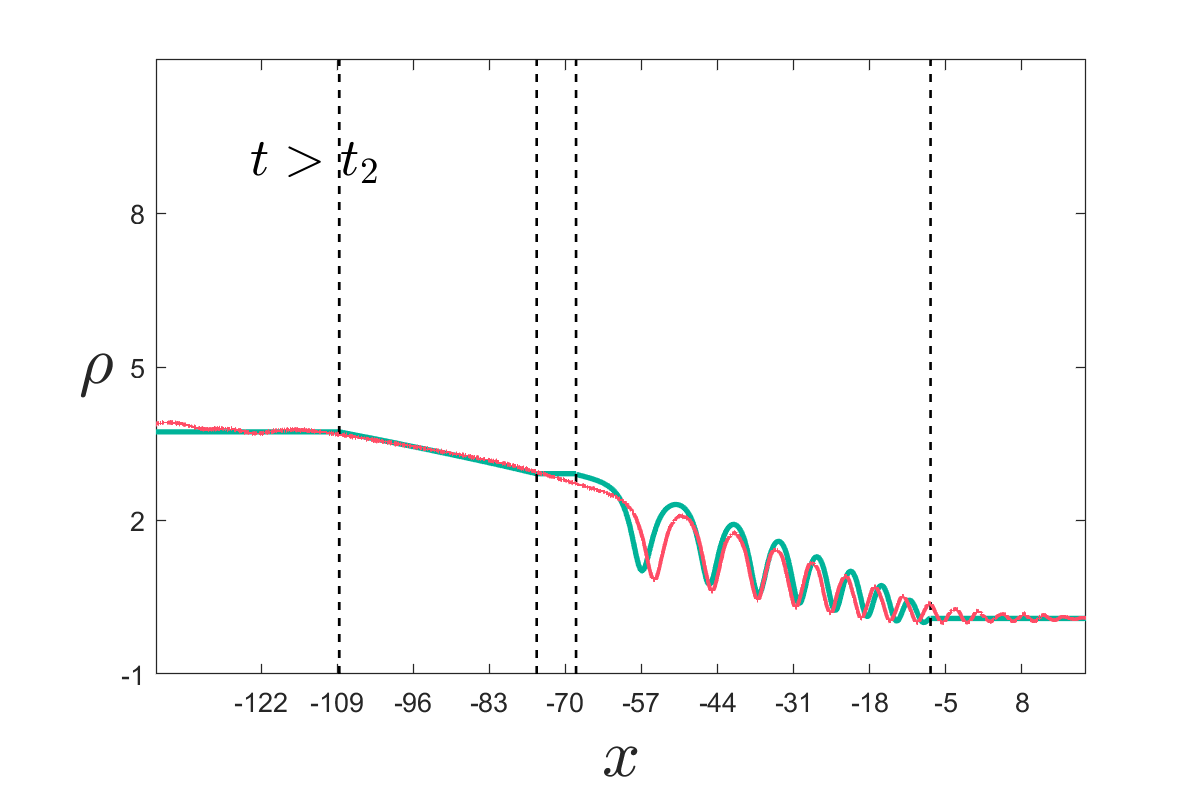}}
\flushleft{\footnotesize
\textbf{Fig.~$\bm{12}$.} The CDSW analytical solutions constructed by mapping the Riemann invariants according to the relation given in Eq. (2.8) (green solid line) and the numerical simulation (red solid line) solutions on $x$.}
\end{figure}
At the moment $t=t_1$, the RW begins to interact with the trailing boundary of the CDSW. Simple calculation we obtain
\begin{equation}
t _ { 1 } =- \frac {d(l_-^L+l_+^L)} { l_-^L(3l_+^L+l_-^L) } , ~x _ { 1 } = -d \frac {(l_-^{L}-l_+^{L})^2} {2l_-^L(3l_+^L+l_-^L)}.
\end{equation}

$(ii)$ Being interaction $(t_1<t<t_2)$:

The modulated solutions for the non-interacting domains still comply with Eqs. (3.52) and (3.54), whereas the solution within the interacting domain are solved with the aid of boundary conditions
\begin{equation}
\begin{cases} 
l_ { 1 } = l_-^L,~l_ { 2 } = l_+^L ,~ l_ { 3 } = l_ { 4 },~
{x}= V _ { 4 } (l_1 ,l_2 , l_3 , l_4 )t,~at~x = x_{2}^{-}(t),\\
l_ { 3 } = l_ { 4 } = 0, ~l_ { 2 } = l_+^L ,~
x = V _- (l_1 ,l_2 )t+d,~at~x = x_{1}^{+}(t).
\end{cases} 
\end{equation}
Here, $l_1$ and $l _4$ are varying Riemann invariants, so we adopt the hodograph transformation once more to solve the EDP equation and obtain its solution $f(l_1,l_4)$.
\begin{equation}
\begin{aligned}
f ( l_ { 1 } , l_ { 4 } ) = \frac {- 2 d~l_-^L } { \pi \sqrt { - l_ { 1 } ( l _ { 4 }-l_-^L) } } ( \Pi _ { 1 } ( s , z ) -\text K( z ) ) ,
 \end{aligned}
\end{equation}
where 
$$
z = \frac {l_ { 4 } ( l_ { 1 } -l_-^L ) } { l_ { 1 } ( l_ { 4 } -l_-^L ) } ,~ s = \frac { l_ { 1 } -l_-^L } {l_ { 1 } } .
$$
Accordingly, we derive the modulated solution for the interaction region between the RW and the CDSW as follows
\begin{equation}
\begin{aligned} 
&l_ 2= l_+^L ,~l_ { 3 } = l_4,  \\ 
x - V _ { 1 , 4 } ( l_ { 1 } ,~l_+^L & ,~l_ {4} ,~l_4 ) t= \left( 1 - \frac { \mathfrak { L } } { \partial _ { 1 ,4 } \mathfrak { L } } \partial _ { 1 , 4 } \right) f ( l _ { 1 } , l _ {4} ) .
\end{aligned}
\end{equation}

Furthermore, at $t=t_2$, the RW completely overtakes the CDSW. After straightforward calculations, we obtain that
\begin{equation}
t_{2} = \frac { 2 d\sqrt{l_-^R(l_-^L-l_+^R)}\text E(r)} { \pi l_-^R(l_-^R - l_+^L)}, 
\end{equation}
where $r=\frac { (l_-^R-l_-^L)l_+^L } { l_-^R(-l_-^L+ l_+^L)}$. The corresponding coordinate $x_{2} = d+x_{2}^{-}(t_2) = x_{1}^{+}(t_2)$ is given by
\begin{equation}
x_2 = \frac{d\sqrt{l_-^R(l_-^L-l_+^L)}(l_-^L+3l_+^L)\text E(r)}{\pi l_-^R(l_-^R -l_+^L)}.
\end{equation}

$(iii)$ After interaction $(t>t_2)$:

The modulated solution for the separated CDSW is given by:
\begin{equation}
l_ { 1 } = l_-^R,~l_ { 2 } =l_+^L,~l_ {3} =l_4,\\
x=V_{4}t+W_4(l_-^R,l_4).
\end{equation}
the function $W_4(l_-^R,l_4)$ is found as
\begin{equation}
\begin{aligned}
&W_4(l_-^R,l_4)=\frac{2d}{\pi \sqrt{l_-^R (l_-^L-l_4)}}(l_-^L~\Pi_1(\frac{-l_-^L+l_-^R}{l_-^R}
,{q})+\\ &\frac{l_-^R(l_4-l_+^R)(-l_4+l_-^L)\text{E}(q)+l_-^L(-l_4+l_-^R)
(l_4-l_+^R)\text{K}(q)}{(-2l_4+l_-^R+l_+^R)l_4)}),
 \end{aligned}
\end{equation}
with $q =\frac{l_4(l_-^R-l_-^L)}{l_-^R(l_4-l_-^L)}.$
The boundaries $x_{1}^{-}$ and $x_{1}^{+}$ are given respectively
\begin{equation}
\begin{aligned}
&x_1 ^-=  \frac { l_-^{R2}+2l_-^Rl_+^L-3l_+^{L2}} {2(l_-^R - l_+^L)} t+W_4(l_-^R,l_+^L), \\
&x _ { 1 } ^{+} =  \frac { l_-^{R2}-2l_-^Ll_+^L+l_+^{L2}} {2(l_-^R + l_+^L)} t+W_4(l_-^R,0). 
\end{aligned}
\end{equation}
\begin{figure}[htbp]
\centering
\setcounter{subfigure}{0}
\subfigure[]{\includegraphics[width=0.333\linewidth]{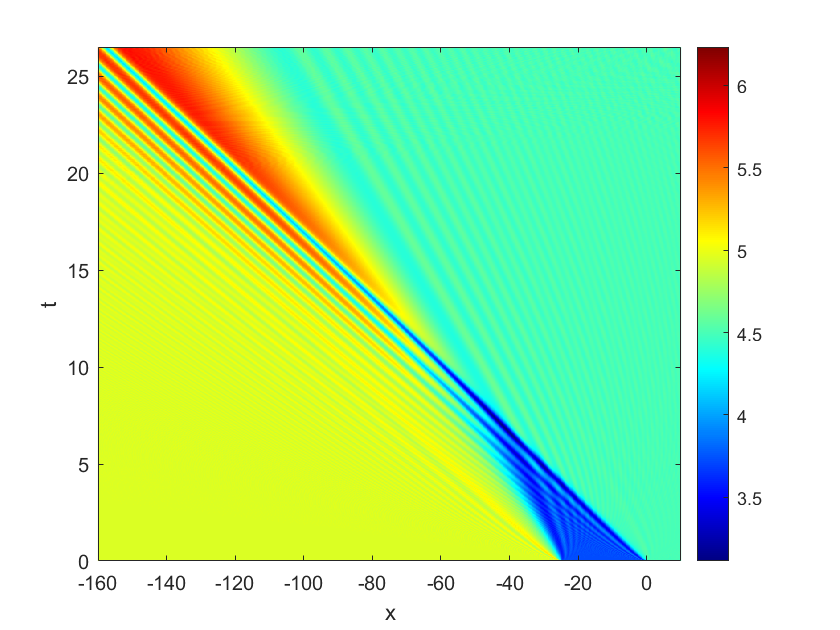}}\hfill
\subfigure[]{\includegraphics[width=0.333\linewidth]{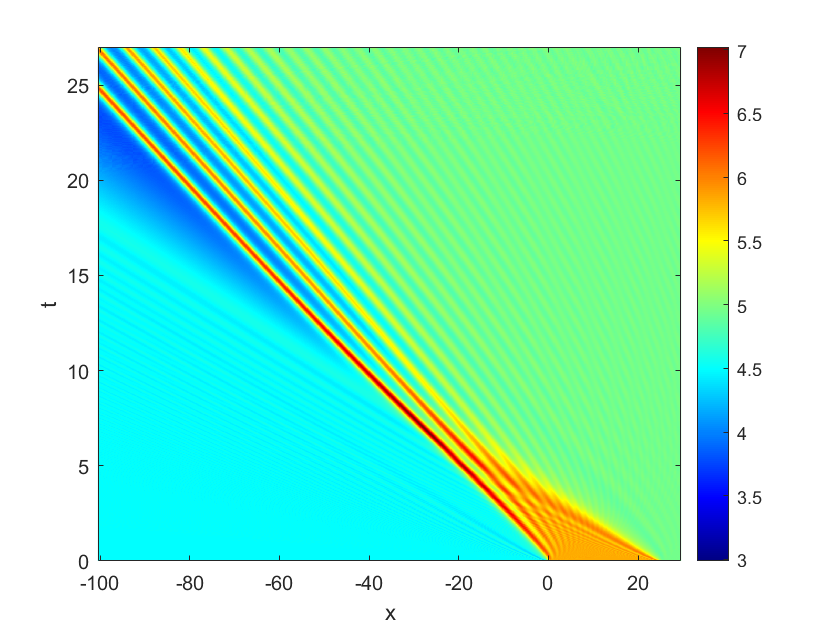}}\hfill
\subfigure[]{\includegraphics[width=0.333\linewidth]{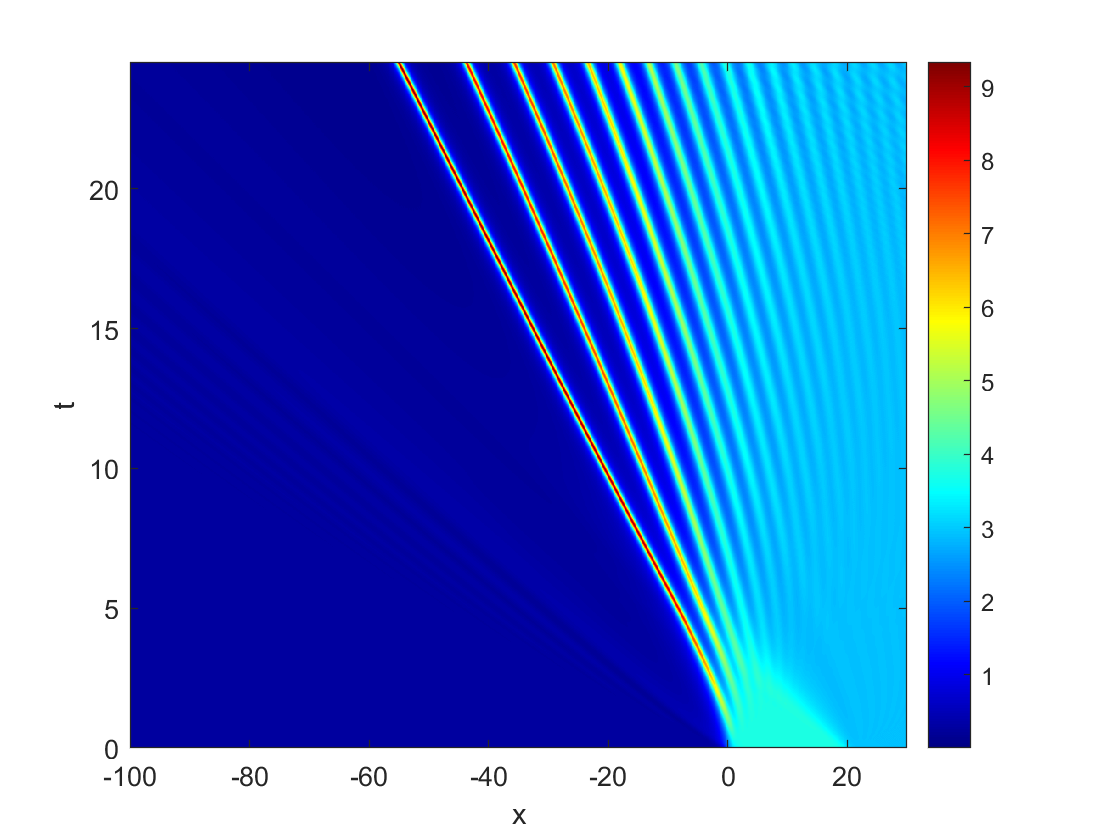}}\hfill
\subfigure[]{\includegraphics[width=0.333\linewidth]{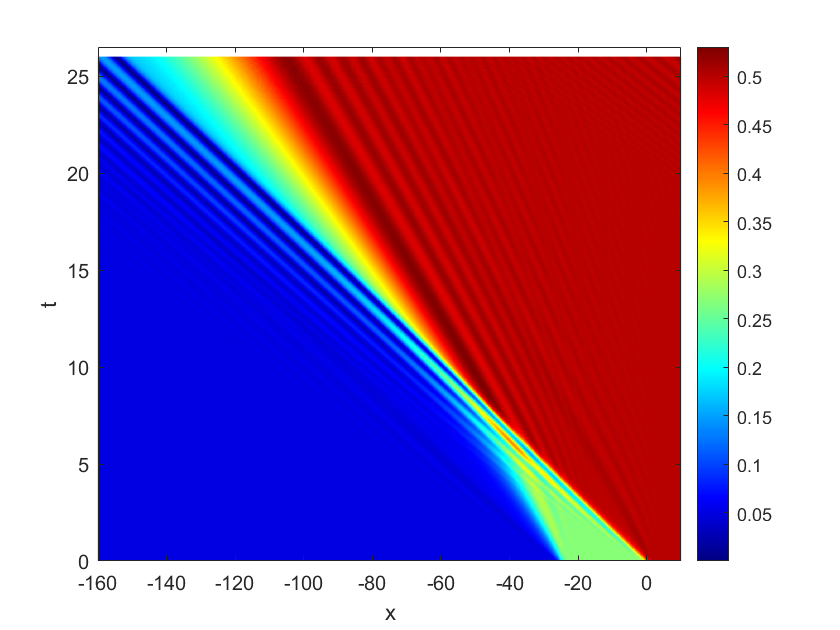}}\hfill
\subfigure[]{\includegraphics[width=0.333\linewidth]{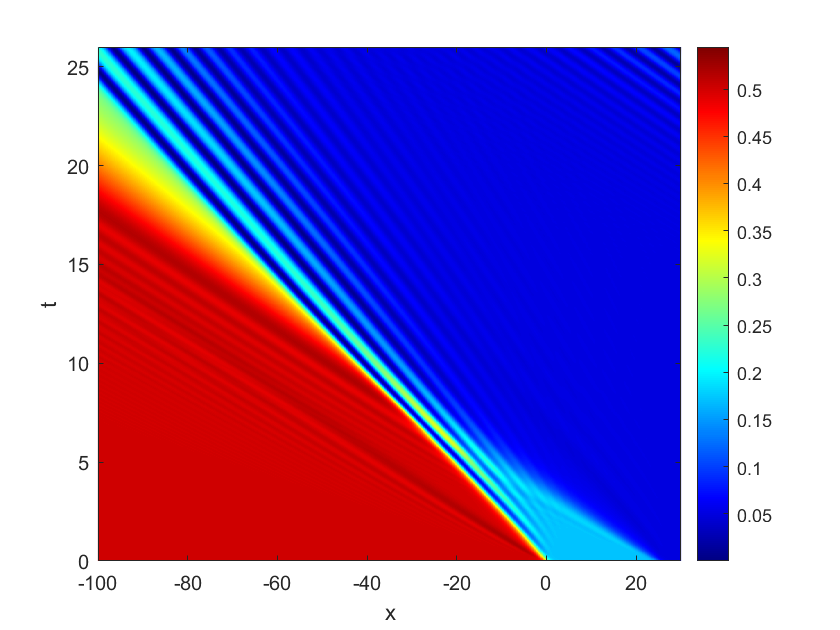}}\hfill
\subfigure[]{\includegraphics[width=0.333\linewidth]{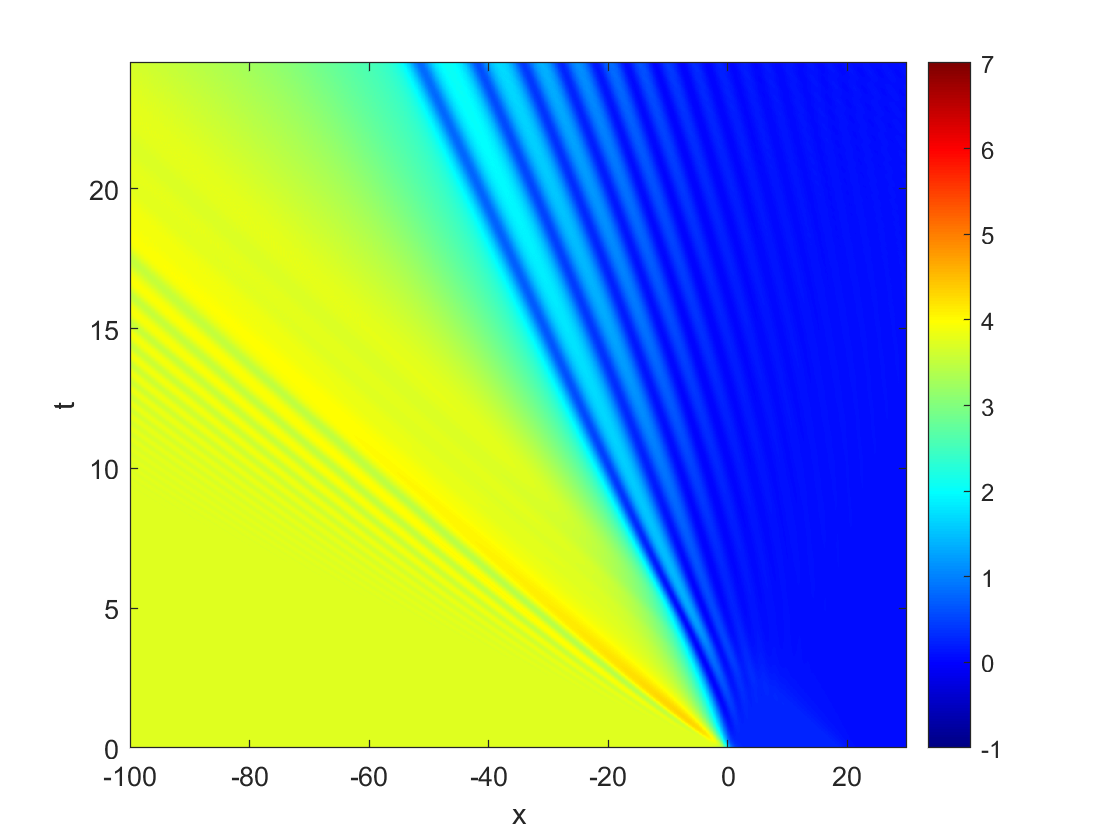}}
\flushleft{\footnotesize
\textbf{Fig.~$\bm{13}$.} The density figure of the two-wave separation regime after interaction.}
\end{figure}

The solution for the RW is given by:
\begin{equation}
x = \frac { 3 l_ { - } + l_+^L} { 2 } t + W_1 ( l_ { - },l_+^L ),
\end{equation}
where the function $W_1( l_ { - },l_+^L )$ has the form
\begin{equation}
W _ { 1 } ( l_ {-},l_+^L ) = \frac { 2d\left(-l_-^L ( \Pi _ { 1 } (p, r ) - \text K(r)) - (l_-^L-l_+^L) \text E (r) \right) } {\pi \sqrt {l_ - (l_-^L-l_+^L)}}  , 
\end{equation}
where $r= \frac { ( l_--l_-^L)l_+^L} {l_-( l_+^L-l_-^L) } ,~p = \frac { l_--l_-^L  } { l_- } .$
The boundaries of RW are given by the expressions
$$
x _ { 2 } ^{-} = \frac { 3 l_-^L+ l_+^L} { 2 } t + W _ { 1 } ( l_ {-}^L,l_+^L ) ,~ x _ { 2 } ^{+} = \frac { 3 l_-^R+ l_+^L} { 2 } t + W _ { 1 } ( l_ {-}^R,l_+^L ) .
$$
It is worth noting that the mapping corresponding to the Riemann invariants changes after the interaction between the RW and the CDSW. That is, the CDSW modifies the branches in the physical variables of the RW, a phenomenon that does not occur for convex equations. On the basis of the foregoing analysis, we present the detailed evolutionary process in Figs. $10–12$.

Having concluded our analysis of the two-wave separation regime after interaction, we present the density profiles for three distinct cases in Fig. 13.

\vspace{7mm}\noindent\textbf{4  Overtaking followed by merging}
\hspace*{\parindent}\\
\renewcommand{\theequation}{4.\arabic{equation}}\setcounter{equation}{0}

In another scenario, when a RW interacts with a DSW, the two do not separate completely but remain in a state of persistent interaction. In what follows, we present a detailed discussion.

\vspace{5mm}\noindent\textbf{4.1 DSW overtakes RW}
\hspace*{\parindent}\\

Configuration: Suppose the DSW and RW are generated at 
$(0,0)$ and $(-d,0)$ on the $(x,t)$ plane at $t=0$, respectively. In contrast to the previously constructed configurations, both the RW and DSW generated in this case are induced by the jump of the same Riemann invariant $l_+$, whose jump profile takes a form of box, with their respective expanding regions given by $x_1^-(t)<x(t)<x_1^+(t)$ and $x _2^-(t)<x(t)<x_2^+(t)$. We construct the following initial data (see Fig. 13(a)):
\begin{equation}
\begin{aligned}
l_-(x,0)=l_-^0,
\quad \text{and} \quad
l_+(x,0)=
\begin{cases} 
l_+^0, & x<-d, \\
l_+^C, & -d<x<0, \\
l_+^0, & x>0,
\end{cases}
\end{aligned}
\end{equation}
where $l_+^0<l_+^C$.
\begin{figure}[htbp]
\centering
\setcounter{subfigure}{0}
\subfigure[]{\includegraphics[width=0.33\linewidth]{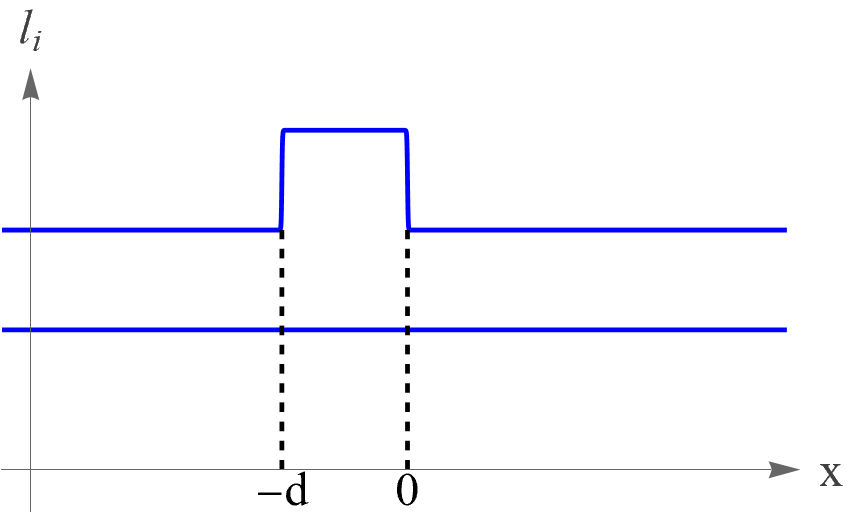}}
\quad
\subfigure[]{\includegraphics[width=0.33\linewidth]{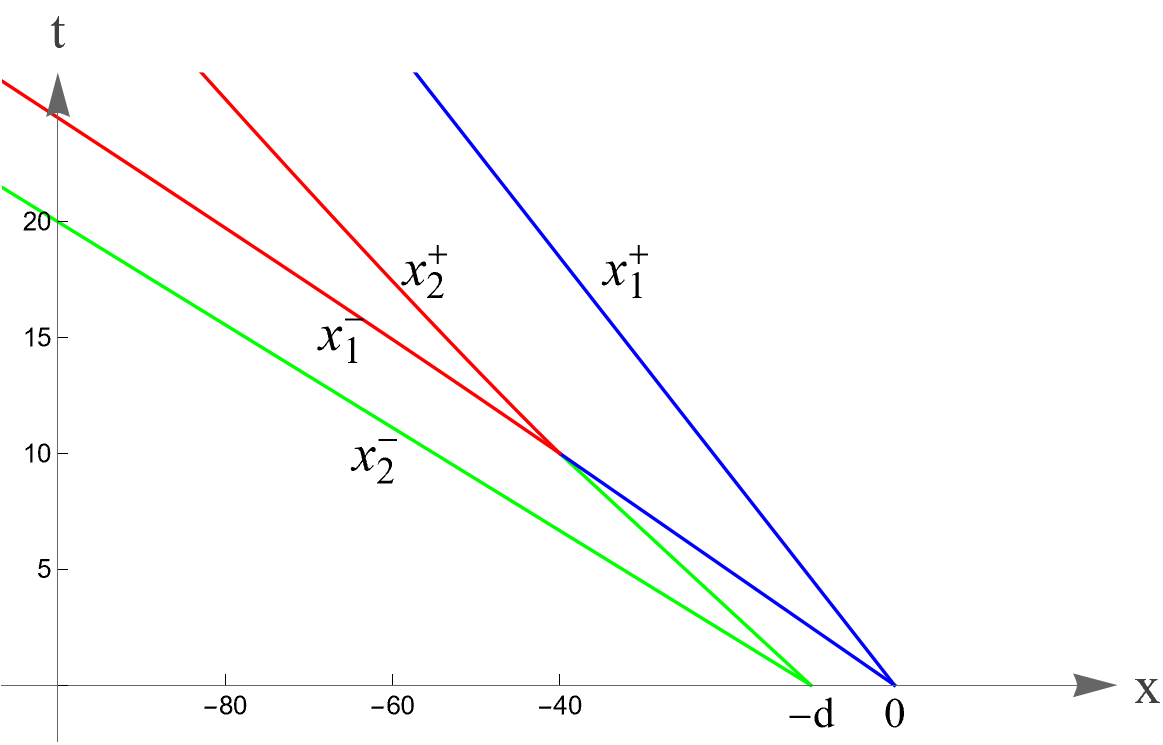}}
\quad
\flushleft{\footnotesize
\textbf{Fig.~$\bm{14}$.} (a) The initial configuration (4.1) of the Riemann invariants. (b) presents the boundaries of distinct wave regions throughout the evolution.}
\end{figure}
\begin{figure}[htbp]
\centering
\setcounter{subfigure}{0}
\subfigure[]{\includegraphics[width=0.33\linewidth]{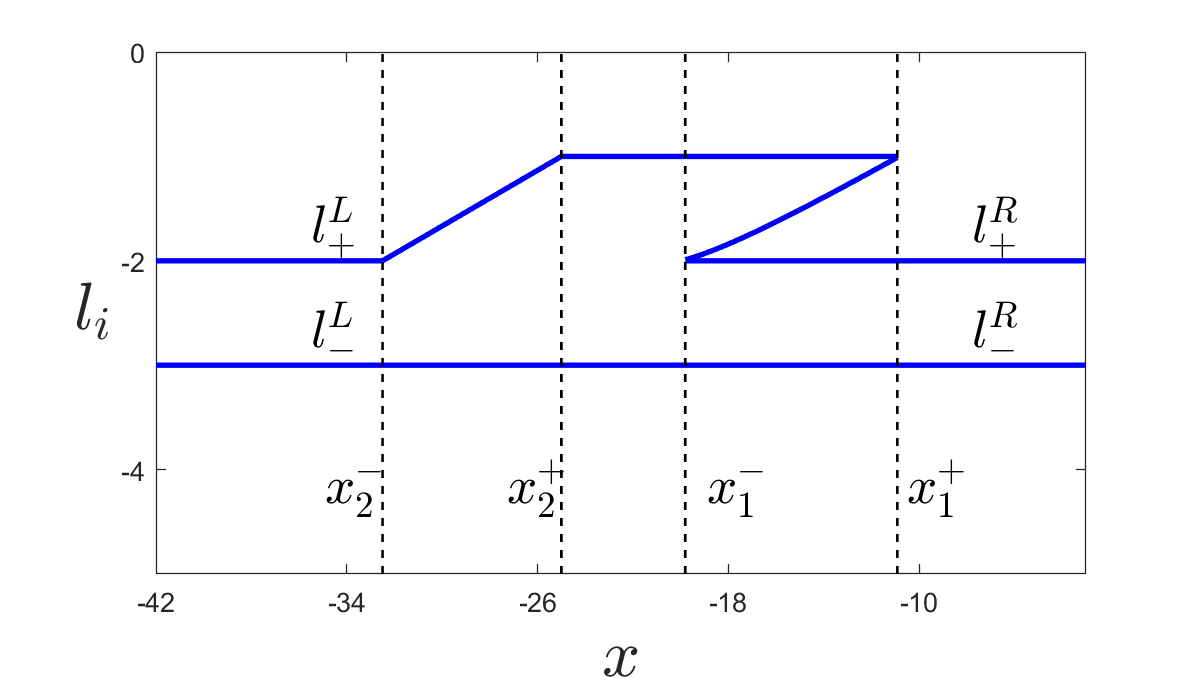}}\hfill
\subfigure[]{\includegraphics[width=0.33\linewidth]{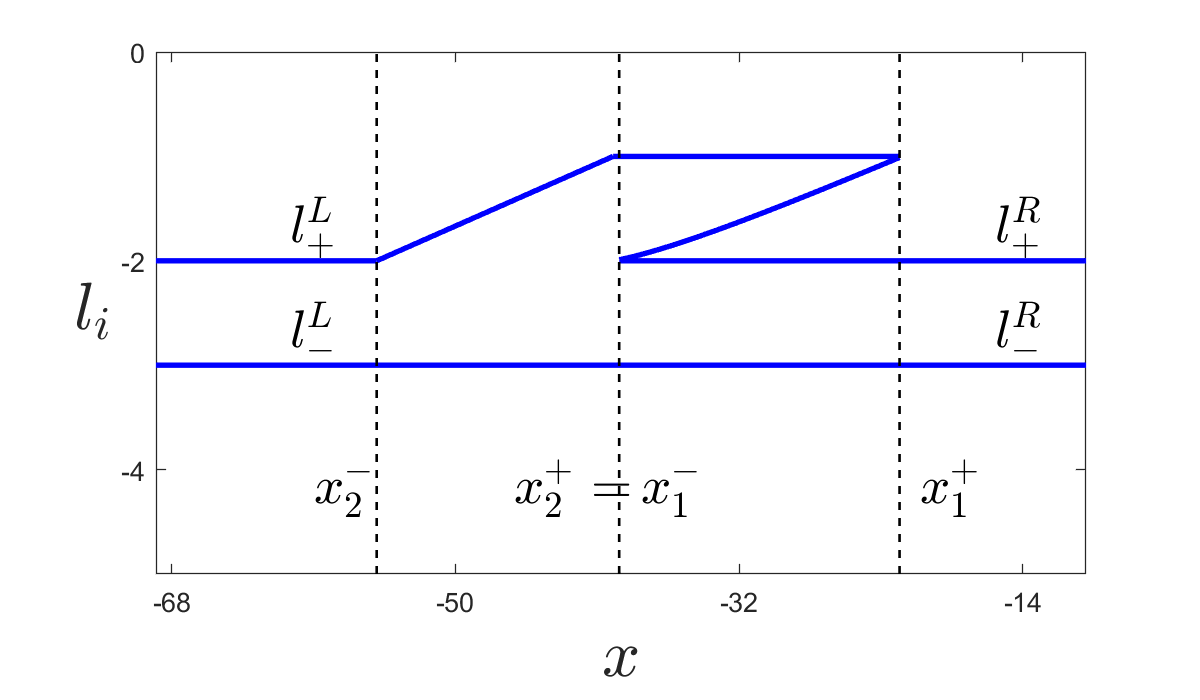}}\hfill
\subfigure[]{\includegraphics[width=0.33\linewidth]{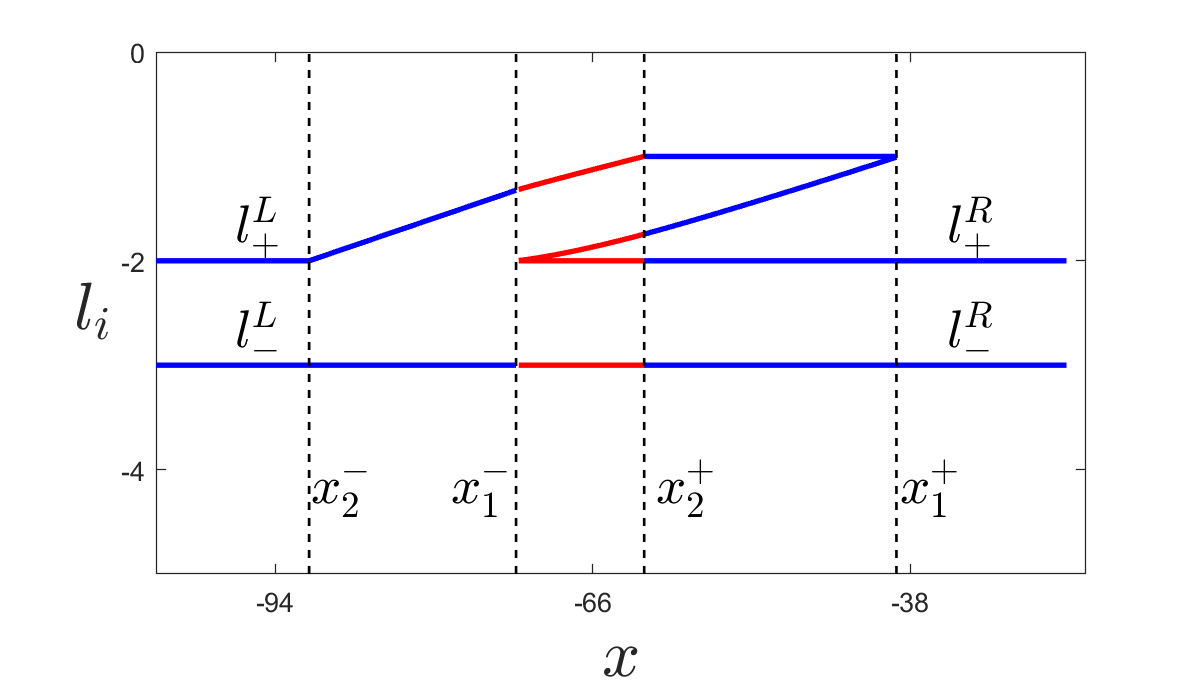}}
\flushleft{\footnotesize
\textbf{Fig.~$\bm{15}$.} The evolution of Riemann invariants, with solid red lines denoting the interaction region.}
\end{figure}
\begin{figure}[htbp]
\centering
\setcounter{subfigure}{0}
\subfigure[]{\includegraphics[width=0.333\linewidth]{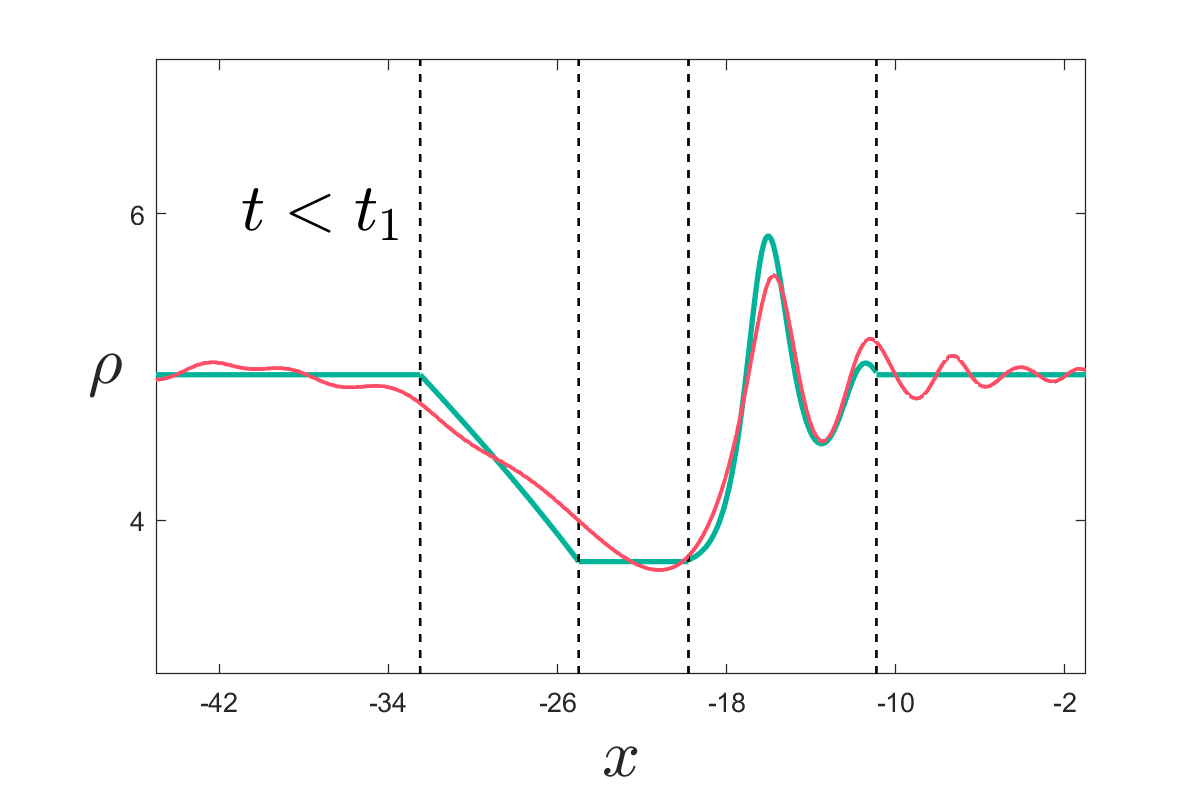}}\hfill
\subfigure[]{\includegraphics[width=0.333\linewidth]{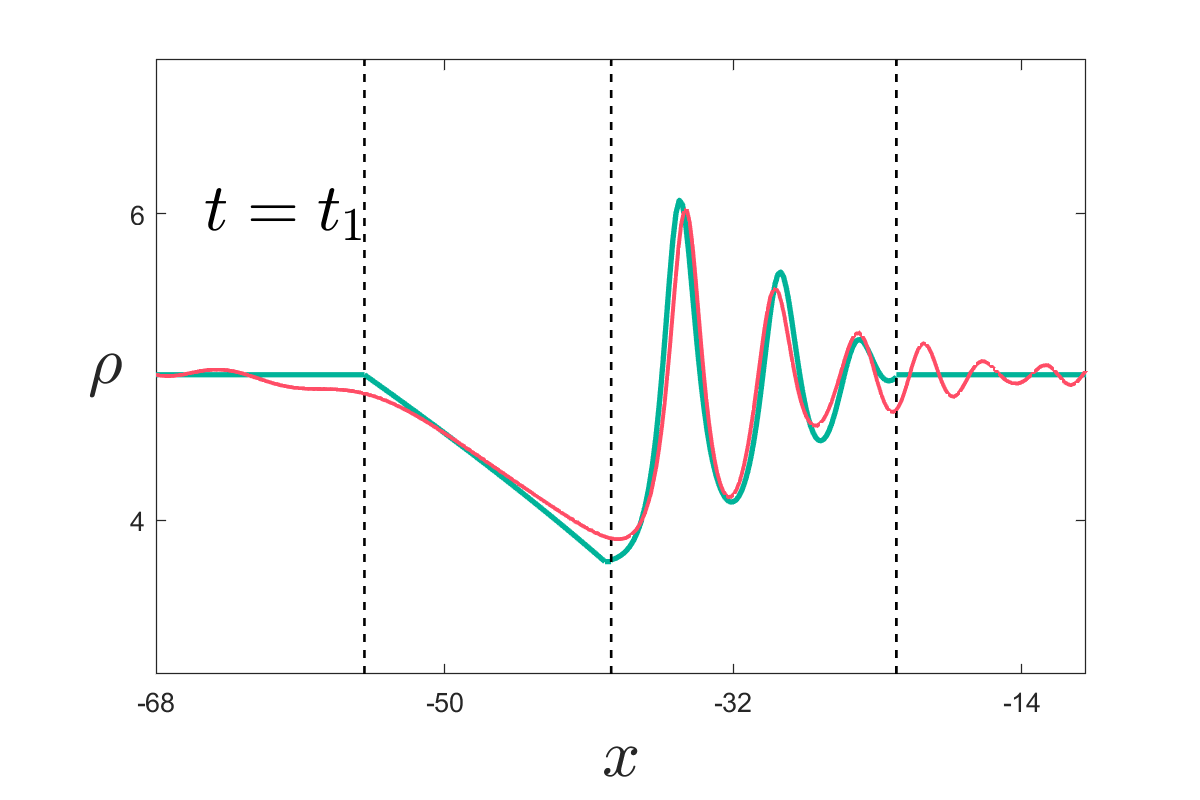}}\hfill
\subfigure[]{\includegraphics[width=0.333\linewidth]{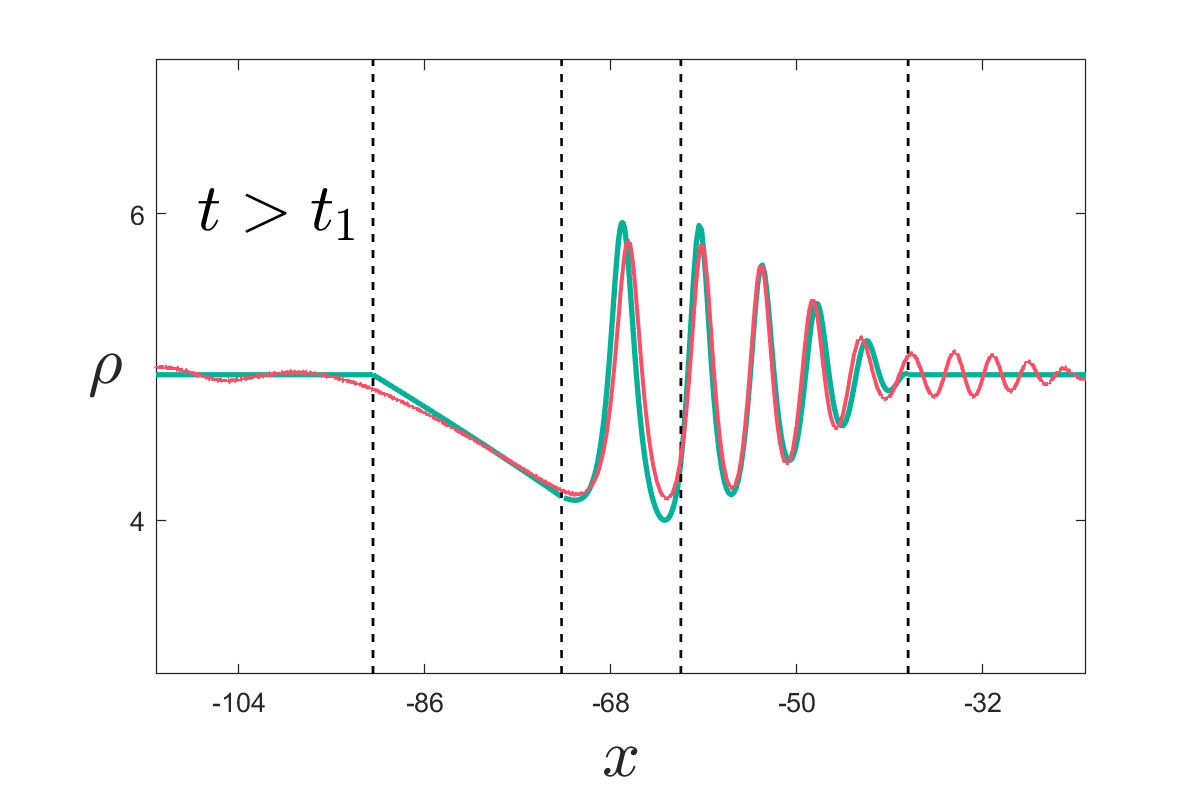}}\hfill
\subfigure[]{\includegraphics[width=0.333\linewidth]{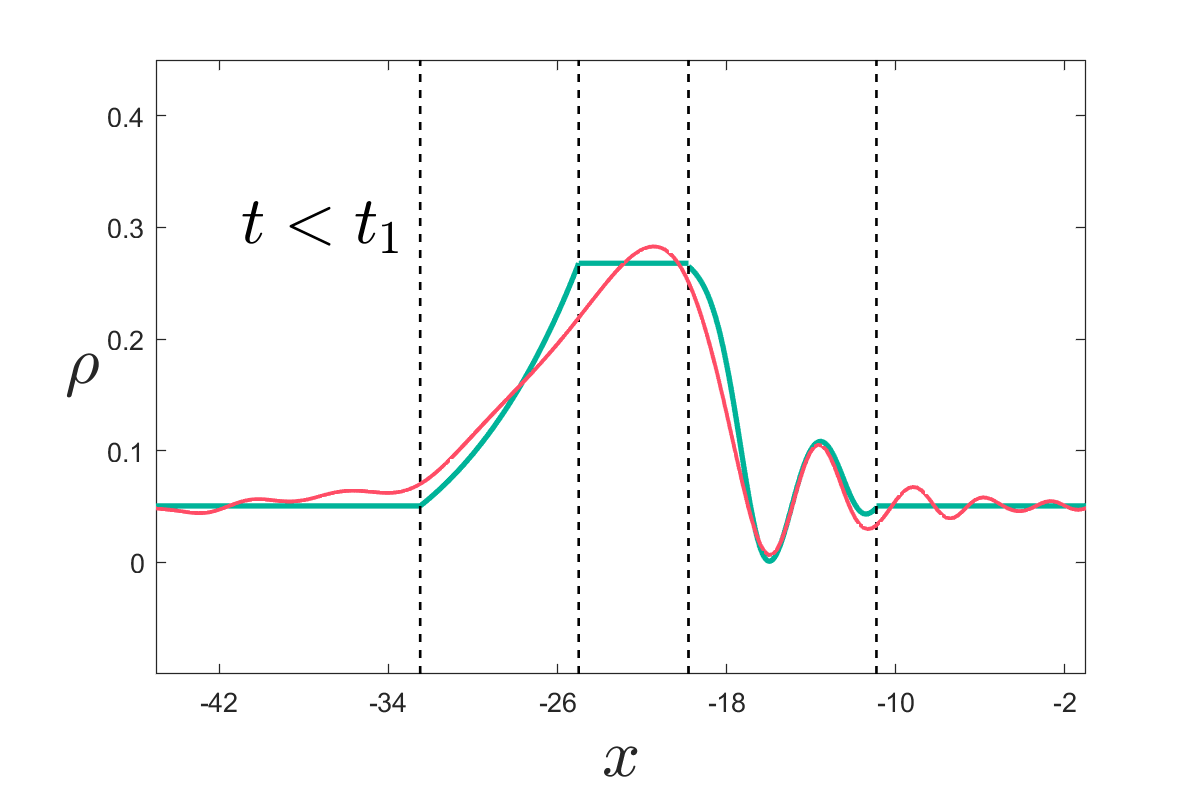}}\hfill
\subfigure[]{\includegraphics[width=0.333\linewidth]{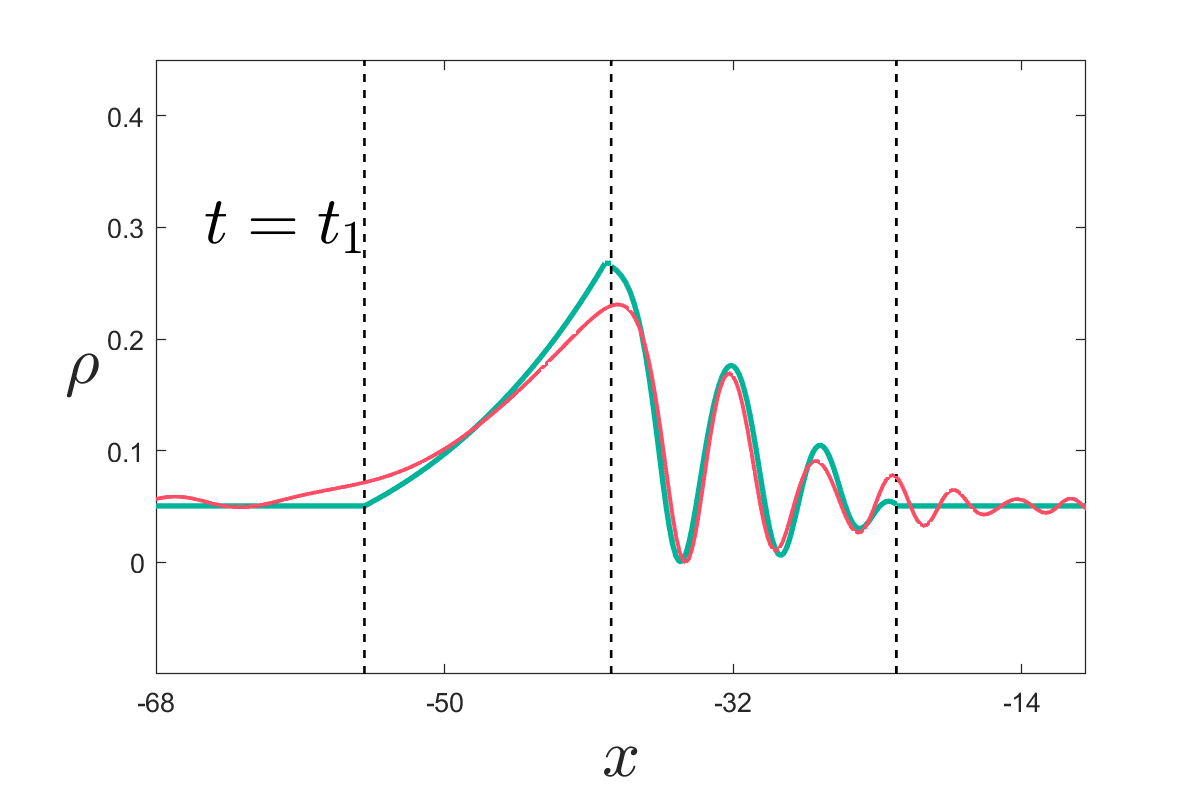}}\hfill
\subfigure[]{\includegraphics[width=0.333\linewidth]{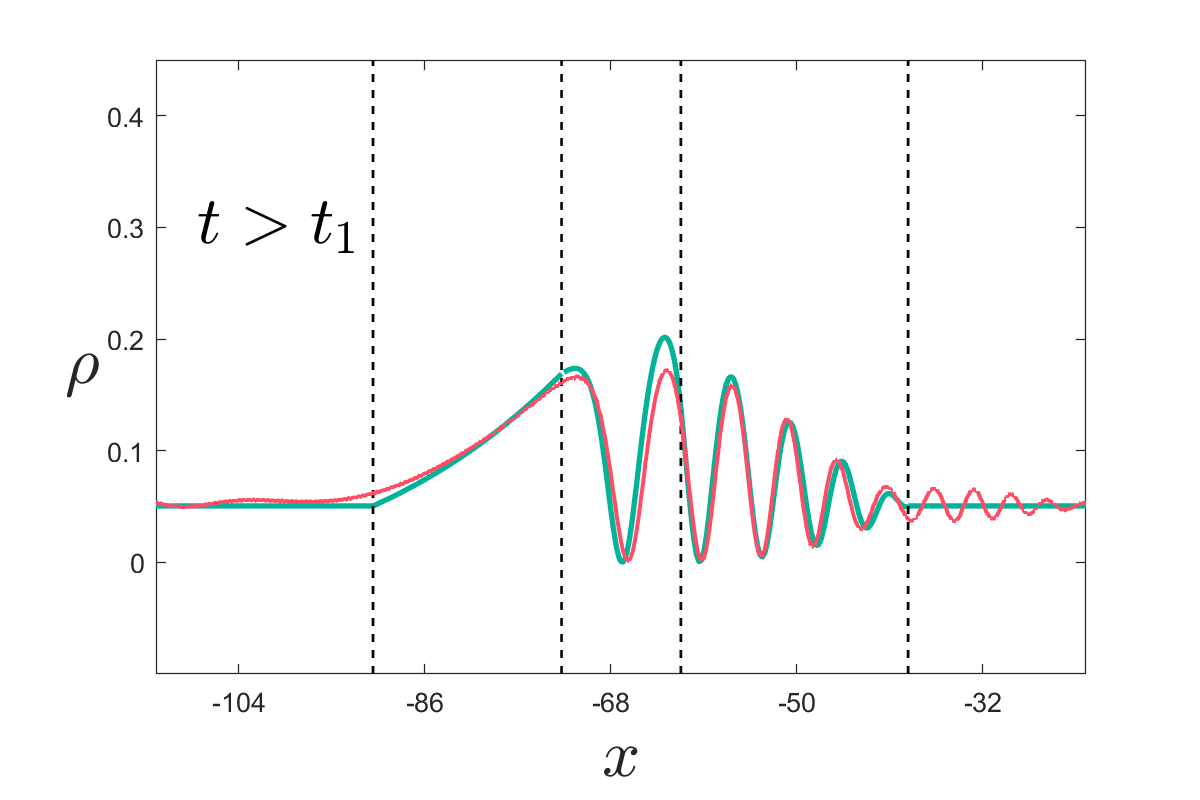}}
\flushleft{\footnotesize
\textbf{Fig.~$\bm{16}$.}  The DSW analytical solutions constructed by mapping the Riemann invariants according to the relation given in Eqs. (2.7) and (2.8) (green solid line) and the numerical simulation (red solid line) solutions on x.}
\end{figure}

$(i)$ Before interaction $(t<t_1)$:

The DSW generated at the origin is described by the following modulation solution
\begin{equation}
\begin{aligned}
 &l_1= l_-^0 ,l_2 = l_+^0 ,l_4 = l_+^C, \\
 \frac{x}{t}=V _ { 3 } = \frac{1}{2} (l_-^0 +l_+^0 +& l_3 + l_+^C) + \frac{(l_+^0 - l_3) (l_3 - 
      l_+^C) \text K(m)} {(l_+^0 - 
       l_+^0) \text E(m) + (l_+^0- 
       l_3) \text K(m)},
\end{aligned}
\end{equation}
where $m = \frac { (l_-^0-l_+^0) ( l_3 - l_+^C ) } { ( l_-^0 - l_3 ) ( l_+^0 - l_+^C ) }, $
The expanding boundaries are respectively given by
\begin{equation}
x _ { 1 } ^{-} = \frac {(l_-^0 + l_+^0 + 2 l_+^C)} {2} t , ~x _ { 1 } ^{+} =  \frac { (l_-^0-l_+^0)^2-8l_+^{C2}+4l_+^C(l_-^0+l_+^{0})} {2(-2l_+^C+ l_-^0 + l_+^0)} t .
\end{equation}

The solution of the RW centered at $x = -d$ is expressed as
\begin{equation}
\begin{aligned}
 &l_-= l_-^0, \\
 \frac{x+d}{t}=&V _ { + } (l_+ ,l_- ) = \frac{3l_++l_-^0}{2},
\end{aligned}
\end{equation}
and the boundaries $x_{2}^{\pm}$ are given
\begin{equation}
x _ { 2 } ^{-} = -d + \frac {  3l_+^0 + l_-^0 } { 2 }t ,\   x _ { 2 } ^{+} = -d + \frac { 3l_+^C + l_-^0 } { 2 } t .
\end{equation}
When the left boundary of the RW catches up with the right boundary of the DSW, the two begin to interact, and the corresponding $t_1$ and $x_1$ can be obtained
\begin{equation}
t _ { 1 } =\frac { d} { l_+^C-l_+^0} , ~x _ { 1 } = d \frac {l_-^{0}+2l_+^0+l_+^C} { 2(l_+^C-l_+^0)}.
\end{equation}

$(ii)$ Being interaction $(t>t_1)$:

Again, the solution in the non-interaction region remains unchanged, and the boundary conditions in the interaction region satisfy
\begin{equation}
\begin{cases} 
l_ { 1 } = l_-^0,~l_ { 2 } = l_+^0 ,~ l_ { 4 }=l_+^C,~
{x}= V _ { 3 } (l_1 ,l_2 , l_3 , l_4 )t,~at~x = x_{1}^{-}(t),\\
l_ { 2 } = l_ { 3 } = l_+^0, ~l_ { 1 } = l_-^0 ,~
x = V _- (l_1 ,l_4 )t-d,~at~x = x_{2}^{+}(t).
\end{cases} 
\end{equation}
Applying them to the Whitham equations yields the solution for $f(l_3, l_4)$ in the EDP equation
\begin{equation}
\begin{aligned}
& f ( l_ { 3 } , l_ { 4 } ) = \frac {- 2 d ( l_+^C -l_+^0 ) } { \pi \sqrt { ( l_+^C - l_ { 3 } ) ( l _ { 4 }-l_+^0) } } ( \Pi _ { 1 } ( s , z ) -\text K( z ) ) ,
 \end{aligned}
\end{equation}
where $z = \frac { ( l_+^C- l_ { 4 } ) ( l_ { 3 } -l_+^0 ) } { ( l_+^C- l_ { 3 } ) ( l_ { 4 } -l_+^0 ) } ,~ s = - \frac { l_ { 4 } -l_+^C } { l_+^0 - l_ { 4 } } .$
Hence, the modulated solution within the interaction zone is expressed as
\begin{equation}
\begin{aligned} 
&l_ 1= l_-^0 ,~l_ { 2 } = l_+^0,  \\ 
x - V _ {3, 4} (l_-^0 ,~l_+^0 & ,~l_3,~l_4 ) t= \left( 1 - \frac { \mathfrak { L } } { \partial _ { 3 , 4 } \mathfrak { L } } \partial _ { 3 , 4 } \right) f ( l _ { 3} , l _ { 4 } ) , 
\end{aligned}
\end{equation}

Now given the solution we can find the boundaries of the interaction zone. The corresponding characteristics of Whitham system at trailing edge $x_1^-$ can be represented as
\begin{equation}
\frac {dx_1^-}{dt}=V_3=\frac{l_4+l_-^0+2l_+^0}{2}.
\end{equation}
Taking account of $l_4$ matches with the solution (4.4) we get 
\begin{equation}
x_1^-=\frac{3l_+^0+l_-^0}{2}t-d+Ct^\frac{1}{3}.
\end{equation}
From (4.6), we can obtain
\begin{equation}
x_1^-=-d+\frac{3}{2}d^2t^\frac{1}{3}(l_+^C-l_+^0)^\frac{1}{3}+\frac{3l_+^0+l_-^0}{2}t.
\end{equation}

The right edge $x_2^+$ of the interaction zone is represented as at $l_4=l_+^C$
\begin{equation}
x_2^+=V_4(l_3, l_+^C)t+W_4(l_3 ,l_+^C),
\end{equation}
where $W_4(l_3 ,l_+^C)$ has the following form
\begin{equation}
W_4(l_3 ,l_+^C)=\frac{d(l_3-l_+^C)(l_+^C-l_-)\text K(q)}{\sqrt{(l_3-l_+^C)(l_+^0-l_+^C)}((l_-^0-l_3)\text{E}(q)+(l_+^C-l_-^0)\text K(q))},
\end{equation}
with $q=\frac{(l_-^0-l_+^0)(l_3-l_+^C)}{(l_3-l_-^0)(l_+^C-l_+^0)}$. 
The boundary structure of the interaction region is illustrated in Fig. 14(b), while Figs. 15 and 16 present the temporal evolution of the Riemann invariants and the corresponding density respectively. The solution (4.9) shows that in the long time limit as $t\rightarrow +\infty $, the Riemann invariants has the asymptotic behaviour 
$l_3\rightarrow l_4$, with the modulus $m\rightarrow 0$,
$$
t\approx \frac{16d\left( l_4-l_{-}^{0} \right) \left( 2l_4-l_{-}^{0}-l_{+}^{0} \right) \sqrt{-\left( \left( l_4-l_{+}^{0} \right) \left( l_4-l_{+}^{C} \right) \right)}}{\pi \left( 8l_{4}^{2}+3l_{-}^{02}+2l_{-}^{0}l_{+}^{0}+3l_{+}^{02}-8l_4\left( l_{-}^{0}+l_{+}^{0} \right) \right) \left( l_3-l_4 \right) ^2}\gg 1.
$$
Correspondingly, the finite-gap periodic wave degenerates into small-amplitude harmonic oscillations shown in Fig. 17.
\begin{figure}[htbp]
\centering
\setcounter{subfigure}{0}
\subfigure[]{\includegraphics[width=0.5\textwidth, height=0.25\textwidth]{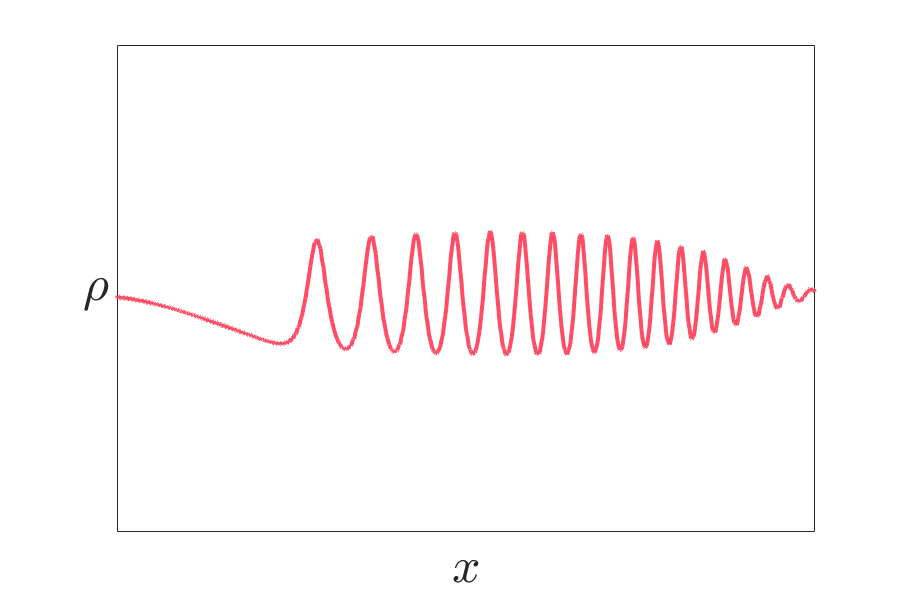}}\hfill
\subfigure[]{\includegraphics[width=0.5\textwidth, height=0.25\textwidth]{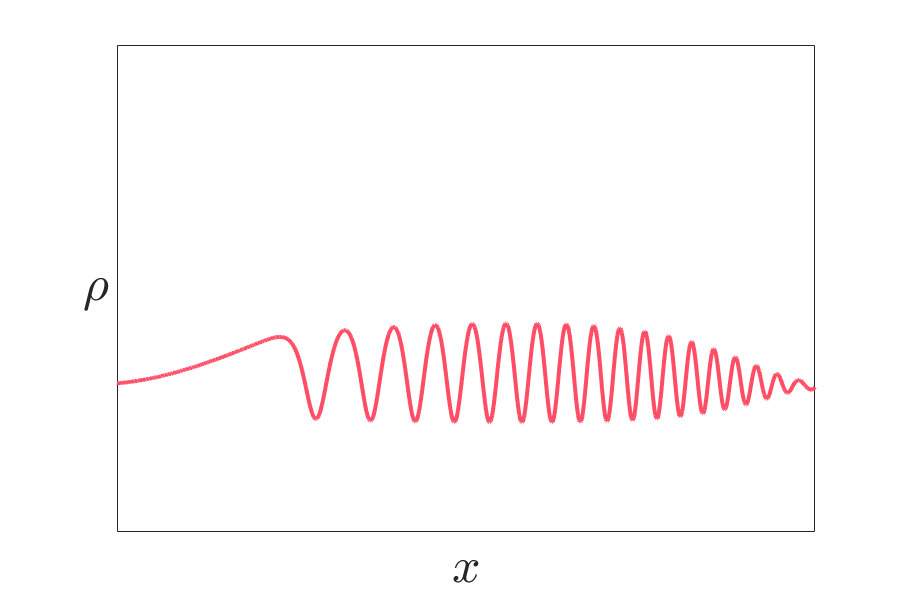}}
\flushleft{\footnotesize
\textbf{Fig.~$\bm{17}$.} The evolution of wave train in the long time limit as $t\rightarrow +\infty $.}
\end{figure}

\vspace{5mm}\noindent\textbf{4.2 RW overtakes DSW}
\hspace*{\parindent}\\

Configuration: Suppose the DSW and RW are generated at 
$(0,0)$ and $(d,0)$ on the $(x,t)$ plane at $t=0$, respectively. The initial condition for the Riemann invariant $l_+$ takes the form of a well, with their respective expanding regions given by $x_1^-(t)<x(t)<x_1^+(t)$ and $x _2^-(t)<x(t)<x_2^+(t)$. We construct the following initial data (see Fig. 18(a)):
\begin{equation}
\begin{aligned}
l_-(x,0)=l_-^0,
\quad \text{and} \quad
l_+(x,0)=
\begin{cases} 
l_+^0, & x<0, \\
l_+^C, & 0<x<d, \\
l_+^0, & x>d, 
\end{cases}
\end{aligned}
\end{equation}
where $l_+^C<l_+^0$.

$(i)$ Before interaction $(t<t_1)$:

Similarly, we obtain the solutions of the Whitham equations of the DSW structure
\begin{equation}
\begin{aligned}
 &l_1= l_-^0 ,l_2 = l_+^C ,l_4 = l_+^0, \\
 \frac{x}{t}=V _ { 3 } = \frac{1}{2} (l_-^0 +l_+^C +& l_3 + l_+^0) + \frac{(l_+^C - l_3) (l_3 - 
      l_+^0) \text K(m)} {(l_+^0 - 
       l_+^C) \text E(m) + (l_+^C- 
       l_3) \text K(m)},
\end{aligned}
\end{equation}
where $m = \frac { (l_3-l_+^0) ( l_-^0 - l_+^C) } { ( l_-^0 - l_3 ) ( l_+^C - l_+^0 ) }, $
The boundaries of the DSW are found as
\begin{equation}
x _ { 1 } ^{-} = \frac {(l_-^0 + 2l_+^C + l_+^0)} {2} t , ~x _ { 1 } ^{+} =  \frac { (l_-^0-l_+^C)^2-8l_+^{02}+4l_+^0(l_-^0+l_+^{C})} {2(-2l_+^0+ l_-^0 + l_+^C)} t .
\end{equation}
The RW solution takes the following form
\begin{equation}
\begin{aligned}
l_-= l_-^0 , ~\frac{x-d}{t}=&V _ { + } (l_+ ,l_- ) = \frac{3l_++l_-^0}{2}.
\end{aligned}
\end{equation}
Its two boundaries are given by
\begin{equation}
x _ { 2 } ^{-} = d+ \frac { 3 l_+^C + l_-^0 } { 2 }t ,\   x _ { 2 } ^{+} = d + \frac { 3 l_+^0 + l_-^0 } { 2 } t .
\end{equation}

The corresponding time $t_1$ and position $x_1$, where the left expansion front of the DSW overtakes the right expansion front of the RW, are expressed as
\begin{equation}
t _ { 1 } =\frac { d(l_-^0-2l_+^C+l_+^0)} { (l_+^C-l_+^0)(4l_+^0-l_+^C-3l_-^0) } , ~x _ { 1 } = d \frac {-8l_+^{C2}+(l_-^{0}-l_+^C)^2+4l_+^0(l_-^0+l_+^C)} { 2(l_+^C-l_+^0)(4l_+^0-l_+^C-3l_-^0)} .
\end{equation}

$(ii)$ Being interaction $(t>t_1)$:

The modulated solution within the interaction zone is expressed as
\begin{equation}
\begin{aligned} 
&l_ 1= l_-^0 ,~l_ { 4 } = l_+^0,  \\ 
x - V _ {2, 3}t&= \left( 1 - \frac { \mathfrak { L } } { \partial _ { 2 , 3 } \mathfrak { L } } \partial _ { 2 , 3 } \right) f ( l _ { 2 } , l _ { 3 } ) , 
\end{aligned}
\end{equation}
where
\begin{equation}
f ( l_ { 2 } , l_ { 3 } ) = \frac {- 2 d ( l_+^C -l_+^0 ) } { \pi \sqrt { ( l_+^0 - l_ { 2 } ) ( l _ {3 }-l_+^C) } } ( \Pi _ { 1 } ( s , z ) - K( z ) ) ,
\end{equation}
and
$$
z = \frac { ( l_+^0- l_ { 3 } ) ( l_ { 2 } -l_+^C ) } { ( l_+^0- l_ { 2 } ) ( l_ { 3 } -l_+^C ) } ,~ s = - \frac { l_ { 2 } -l_+^C } { l_+^0 - l_ { 2 } } .
$$
\begin{figure}[htbp]
\centering
\setcounter{subfigure}{0}
\subfigure[]{\includegraphics[width=0.35\linewidth]{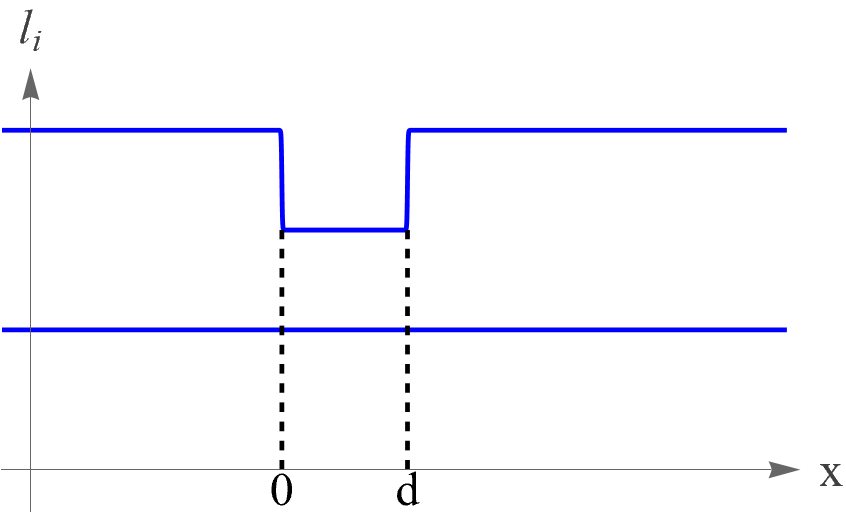}}
\subfigure[]{\includegraphics[width=0.35\linewidth]{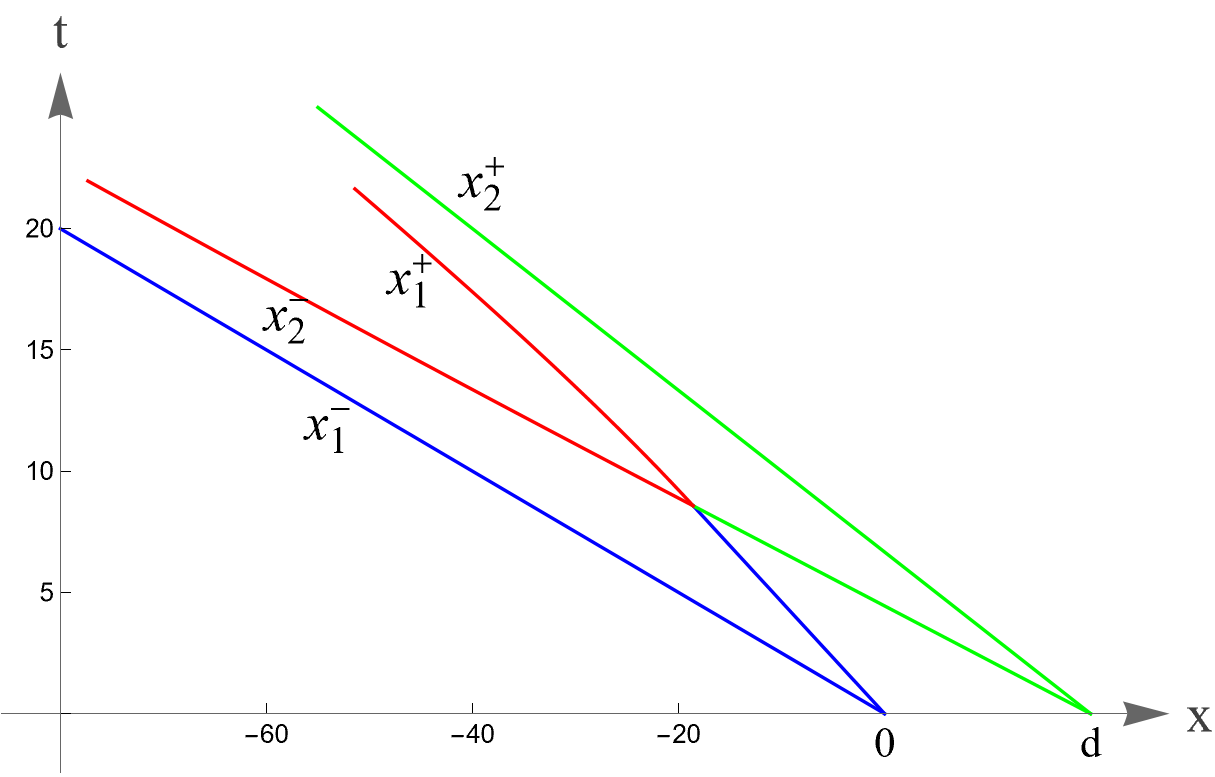}}
\flushleft{\footnotesize
\textbf{Fig.~$\bm{18}$.} (a) The initial configuration (4.1) of the Riemann invariants. (b) presents the boundaries of distinct wave regions throughout the evolution.}
\end{figure}
\begin{figure}[htbp]
\centering
\setcounter{subfigure}{0}
\subfigure[]{\includegraphics[width=0.333\linewidth]{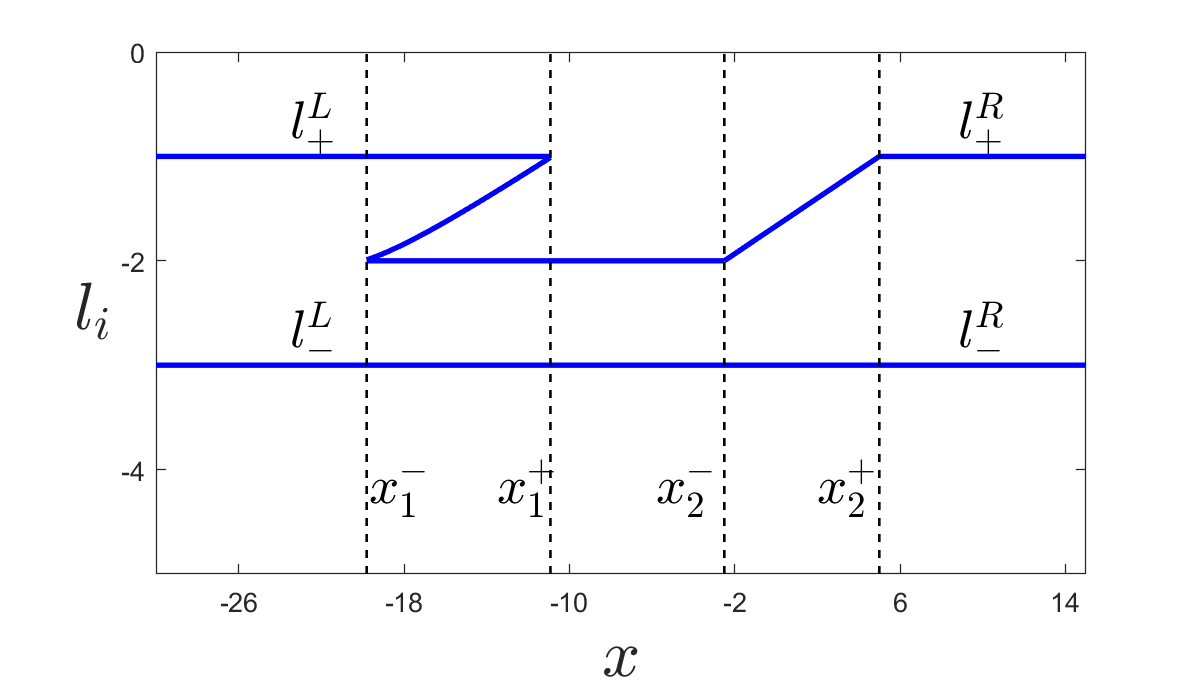}}\hfill
\subfigure[]{\includegraphics[width=0.333\linewidth]{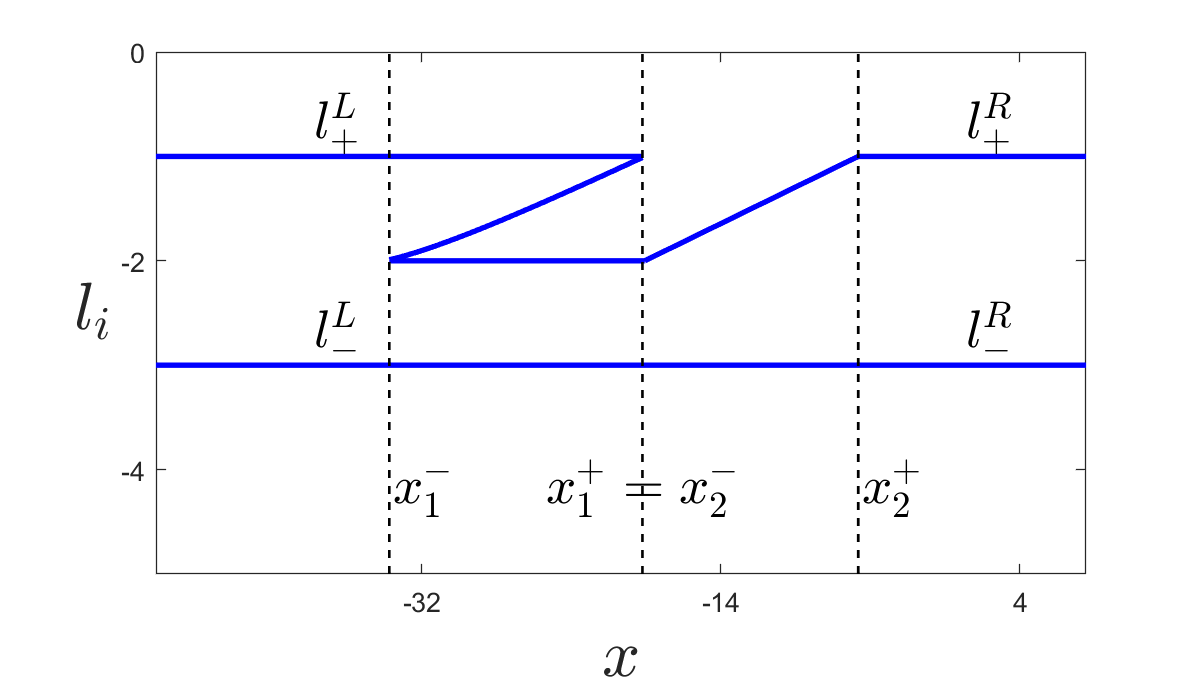}}\hfill
\subfigure[]{\includegraphics[width=0.333\linewidth]{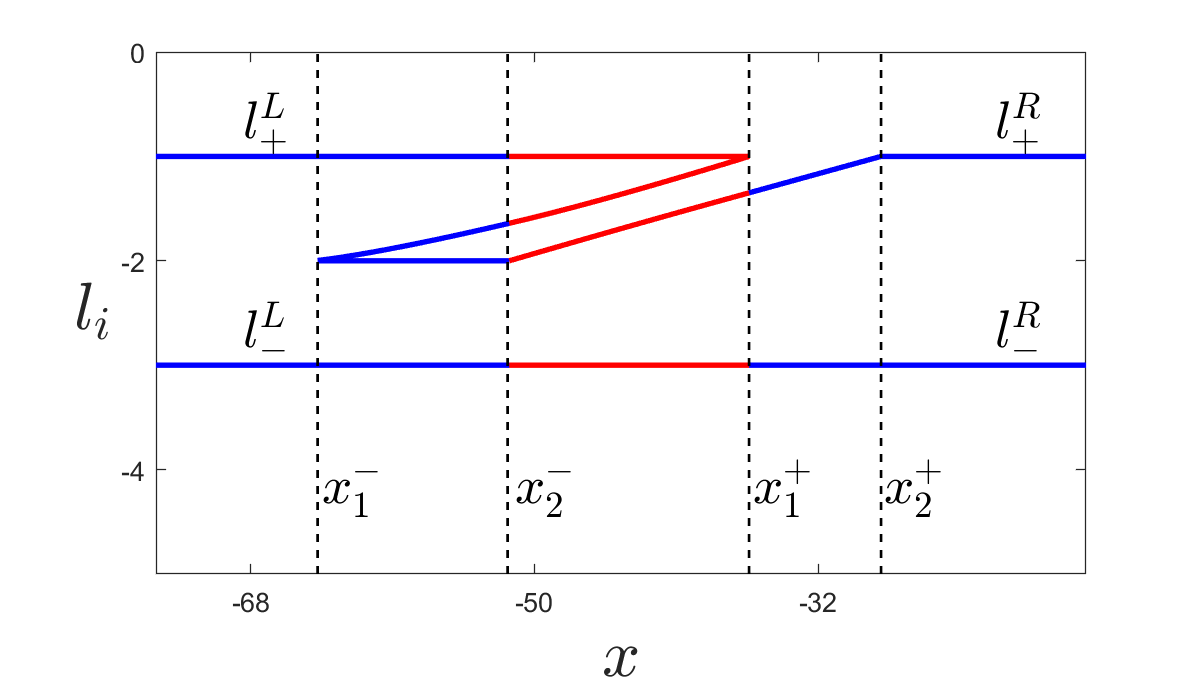}}
\flushleft{\footnotesize
\textbf{Fig.~$\bm{19}$.} The evolution of Riemann invariants, with solid red lines denoting the interaction region.}
\end{figure}
\begin{figure}[htbp]
\centering
\setcounter{subfigure}{0}
\subfigure[]{\includegraphics[width=0.333\linewidth]{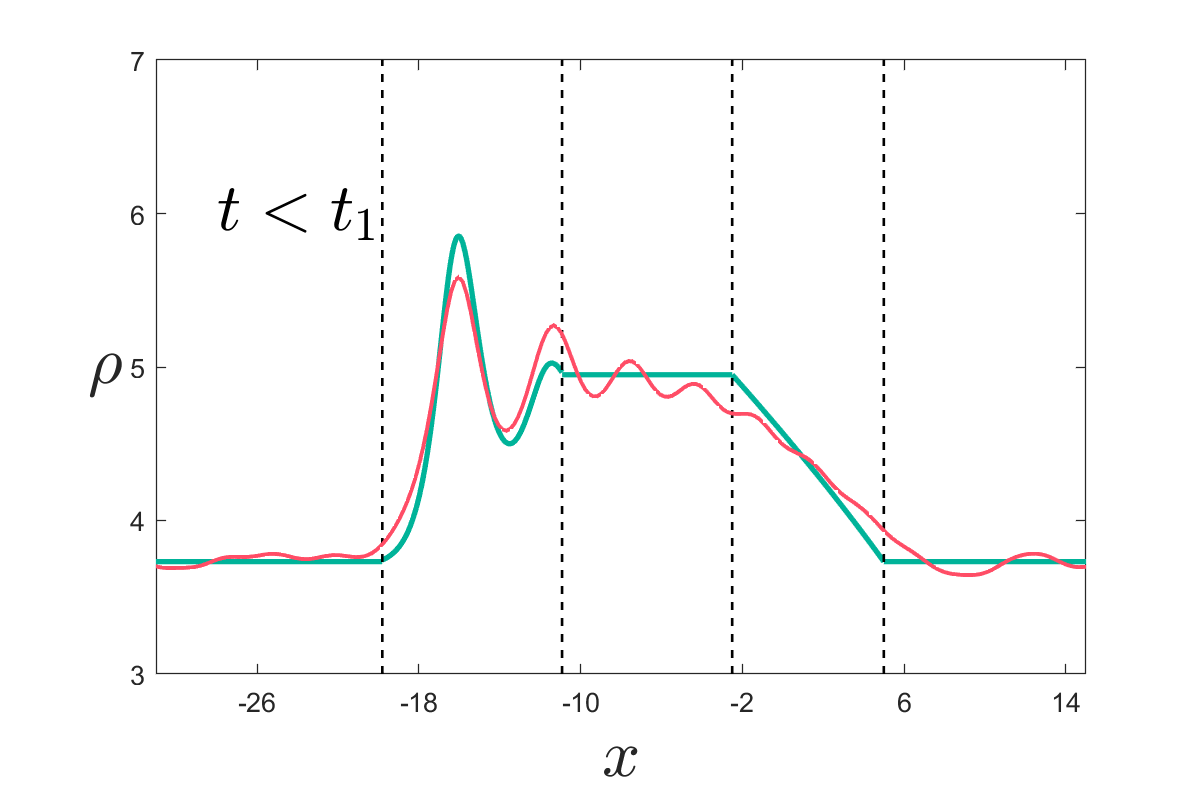}}\hfill
\subfigure[]{\includegraphics[width=0.333\linewidth]{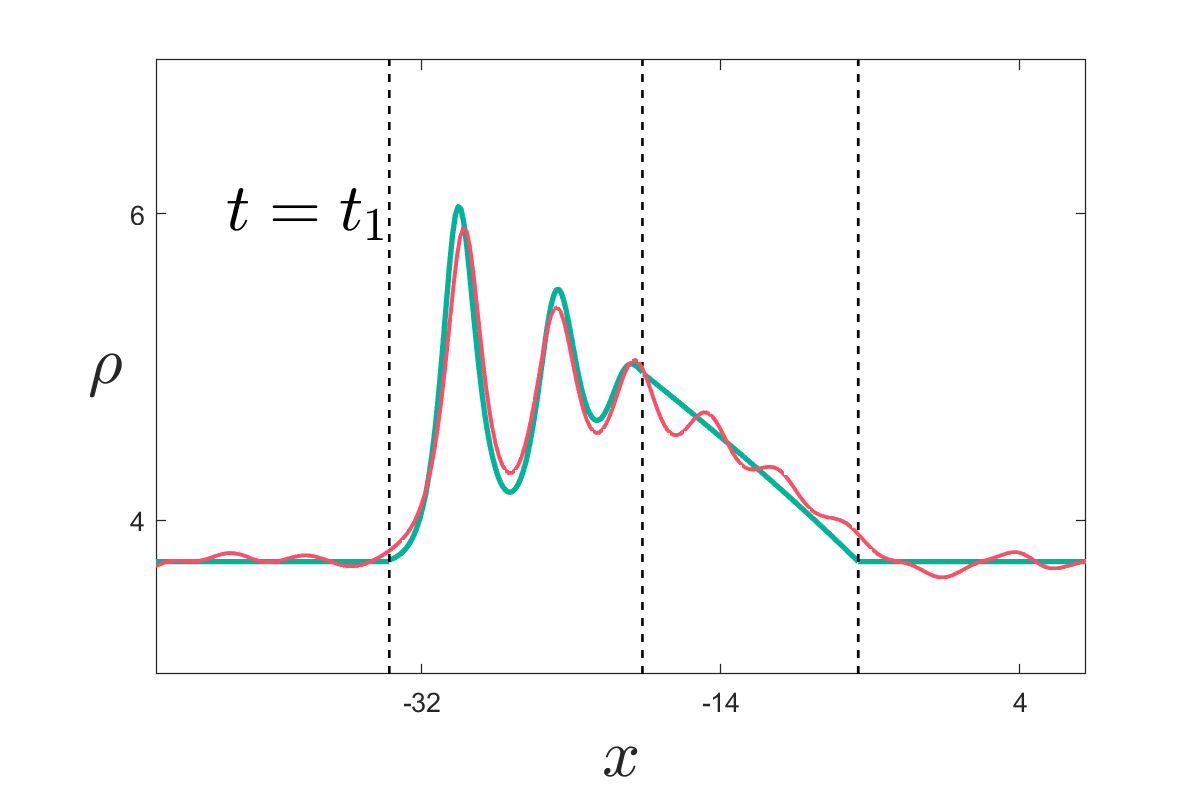}}\hfill
\subfigure[]{\includegraphics[width=0.333\linewidth]{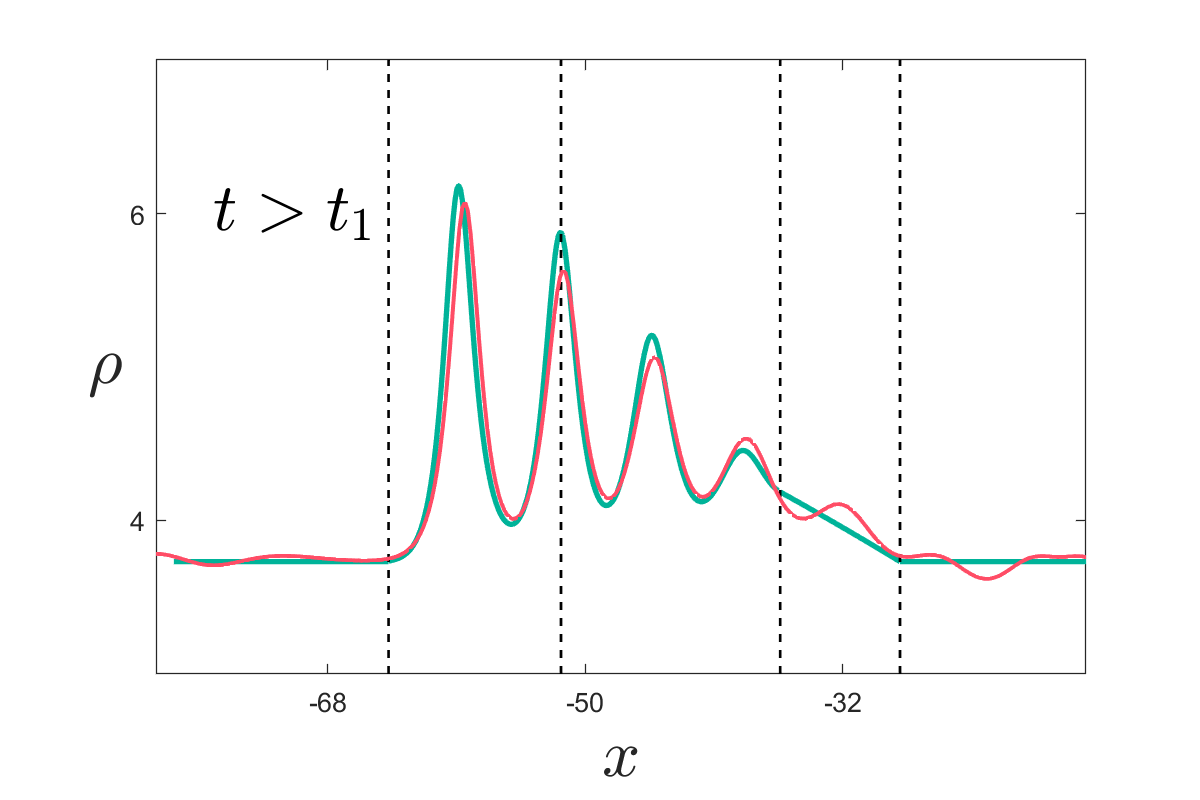}}\hfill
\subfigure[]{\includegraphics[width=0.333\linewidth]{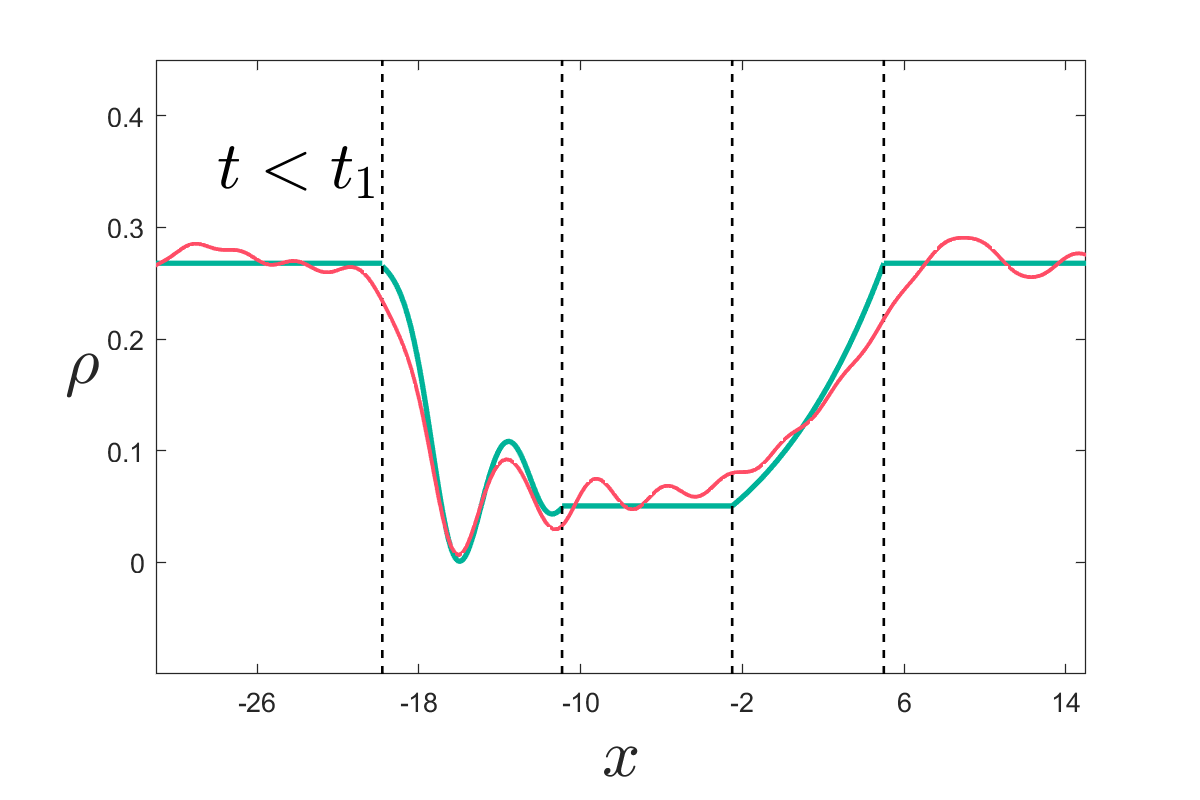}}\hfill
\subfigure[]{\includegraphics[width=0.333\linewidth]{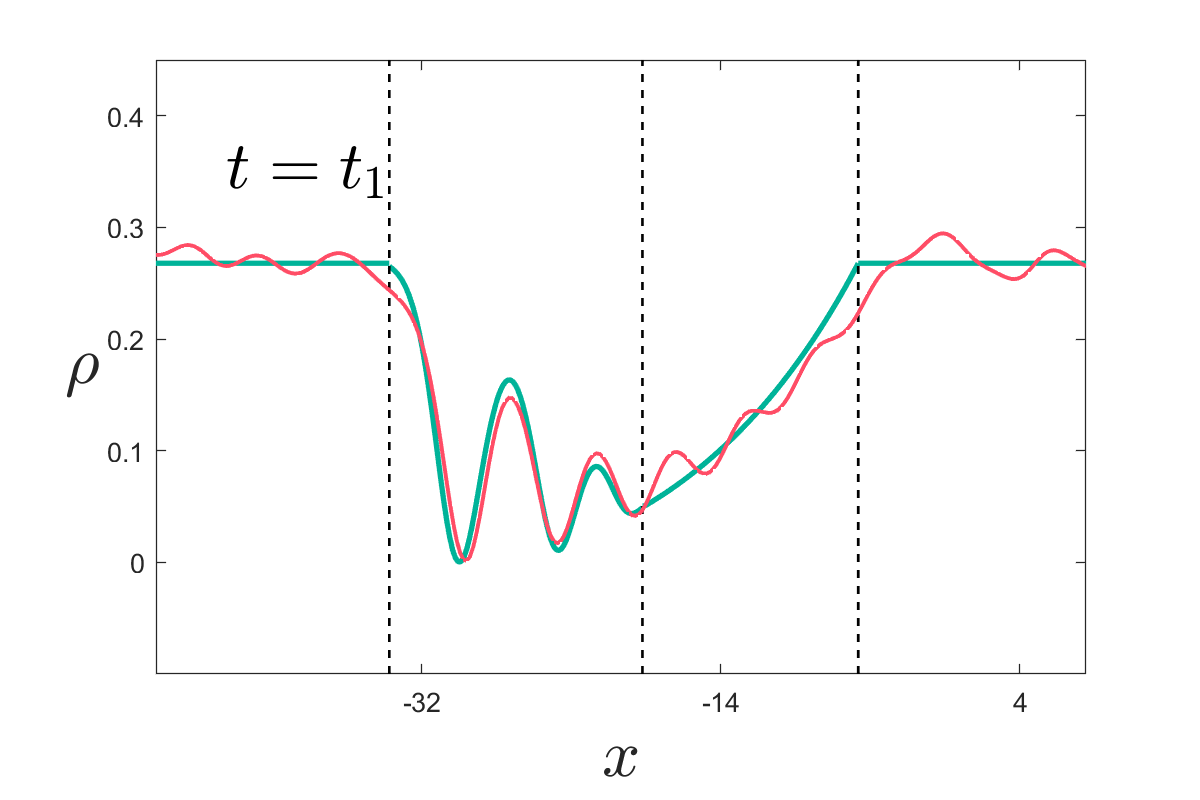}}\hfill
\subfigure[]{\includegraphics[width=0.333\linewidth]{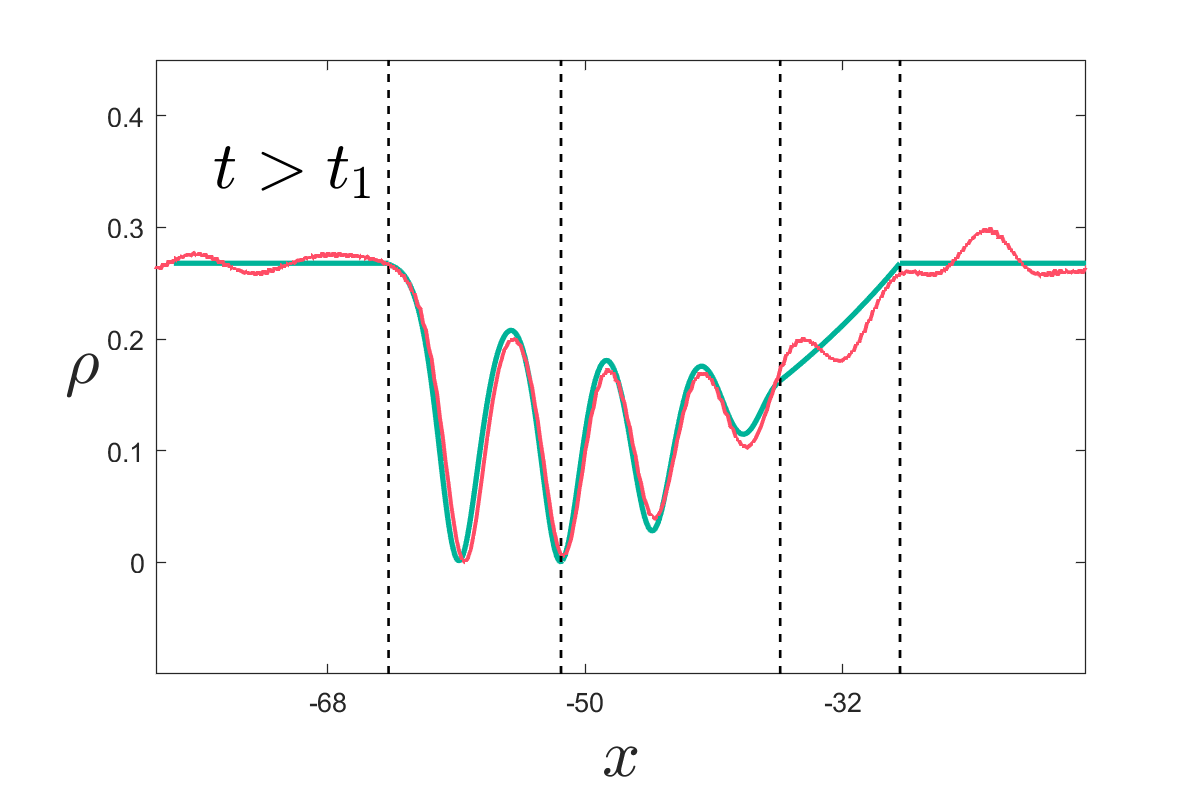}}
\flushleft{\footnotesize
\textbf{Fig.~$\bm{20}$.}  The DSW analytical solutions constructed by mapping the Riemann invariants according to the relation given in Eqs. (2.7) and (2.8) (green solid line) and the numerical simulation (red solid line) solutions on x.
}
\end{figure}

The temporal evolution profiles of the Riemann invariants and the corresponding density in this configuration are presented in Figs. 19 and 20, respectively. It can be observed that in the long time limit as $t\rightarrow +\infty $, the Riemann invariants satisfy the asymptotic behaviour 
$l_2\rightarrow l_3$, with the modulus $m\rightarrow 1$,
$$
t\approx d\frac{-2l_3+2l_{+}^{C}+\left( 2l_3-l_-^0-l_{+}^{C} \right) \ln \left( 1-m \right)}{2\pi \left( l_3-l_-^0 \right) \sqrt{-\left( \left( l_3-l_+^0 \right) \left( l_3-l_{+}^{C} \right) \right)}}\gg 1.
$$Correspondingly, the finite-gap periodic wave degenerates into a soliton train shown in Fig. 21.
\begin{figure}[htbp]
\centering
\setcounter{subfigure}{0}
\subfigure[]{\includegraphics[width=0.5\textwidth, height=0.25\textwidth]{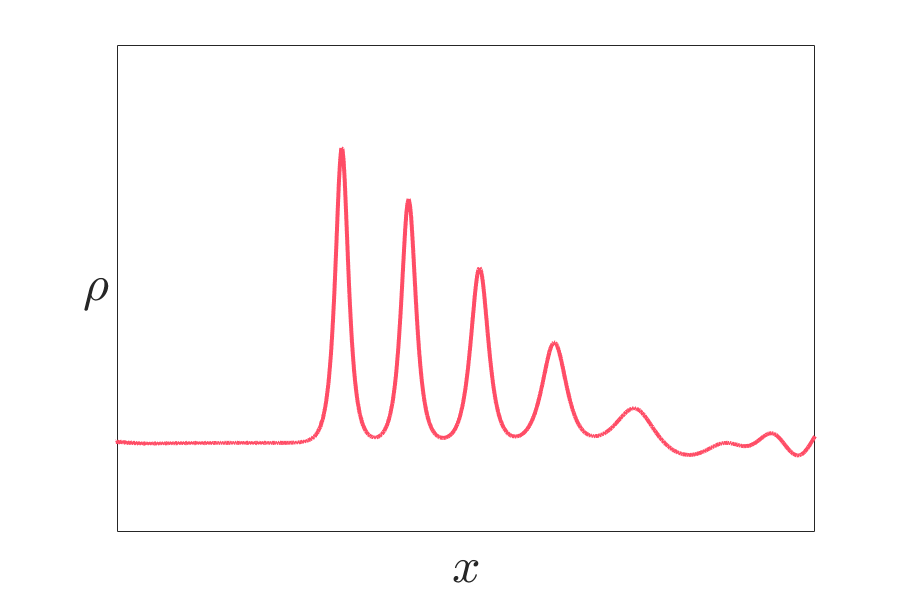}}\hfill
\subfigure[]{\includegraphics[width=0.5\textwidth, height=0.25\textwidth]{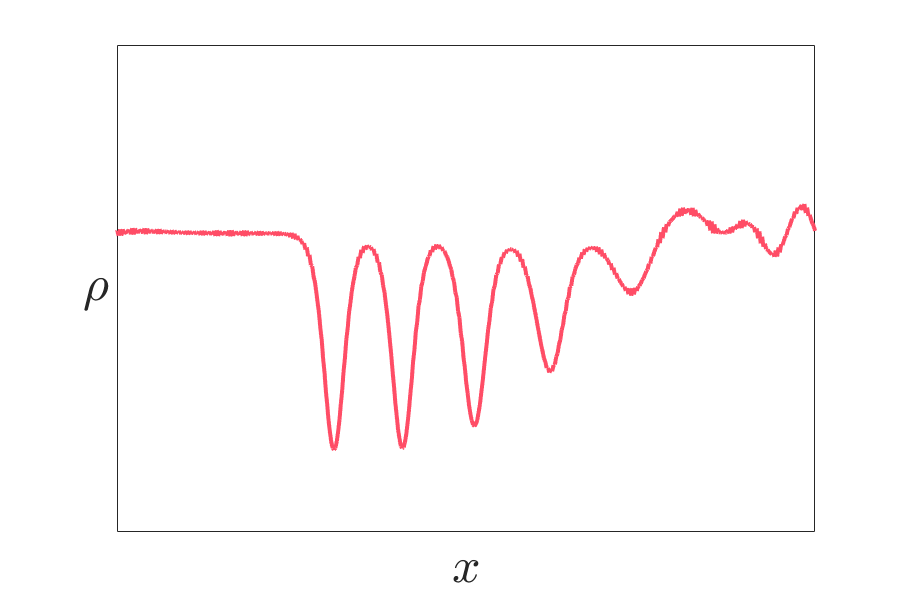}}
\flushleft{\footnotesize
\textbf{Fig.~$\bm{21}$.} The evolution of the soliton train in the long time limit as $t\rightarrow +\infty $.}
\end{figure}

\vspace{5mm}\noindent\textbf{4.3 RW overtakes CDSW}
\hspace*{\parindent}\\

Configuration: Suppose the CDSW and RW are generated at 
$(0,0)$ and $(d,0)$ on the $(x,t)$ plane at $t=0$, respectively. Their respective expanding regions given by $x_1^-(t)<x(t)<x_1^+(t)$ and $x _2^-(t)<x(t)<x_2^+(t)$. At the origin, the Riemann invariants exhibit no jump, which gives rise to the formation of a CDSW. We construct the following initial data:
\begin{equation}
\begin{aligned}
l_-(x,0)=l_-^0,
\quad \text{and} \quad
l_+(x,0)=
\begin{cases} 
l_+^L, & x<d, \\
l_+^R, & x>d,
\end{cases}
\end{aligned}
\end{equation}
where $l_+^R<l_+^L$.

$(i)$ Before interaction $(t<t_1)$:

We can obtain the solutions of the Whitham equations of the CDSW structure
\begin{equation}
\begin{aligned}
&l_1= l_-^0 ,~l_2 = l_+^L,~l_3=l_4,\\
V _ { 4 } =&~V_3= \frac{l_-^{02}-2l_-^0l_+^L+4l_-^0l_4-8l_4^2}{2(l_-^0+l_+^L-2l_4)},
\end{aligned}
\end{equation}
The boundaries of the CDSW are determined as
\begin{equation}
x _ { 1 } ^{-} = \frac {l_-^{02} + 2 l_-^0l_+^L - 3 l_+^{L2}} {2(l_-^0-l_+^L)} t , ~x _ { 1 } ^{+} =  \frac { l_-^{02}-2l_-^0l_+^L+l_+^{L2}} {2(l_-^0 -l_+^L} t .
\end{equation}
\begin{figure}[htbp]
\centering
\setcounter{subfigure}{0}
\subfigure[]{\includegraphics[width=0.79\linewidth]{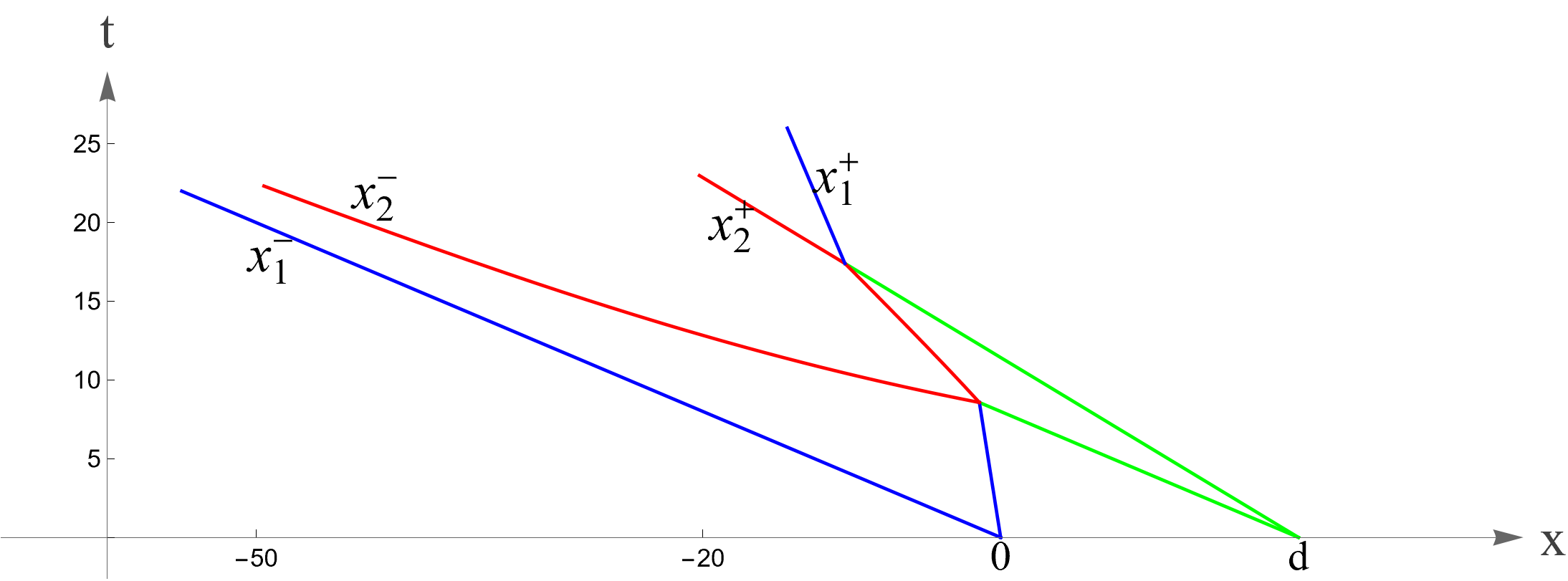}}\hfill
\subfigure[]{\includegraphics[width=0.25\linewidth]{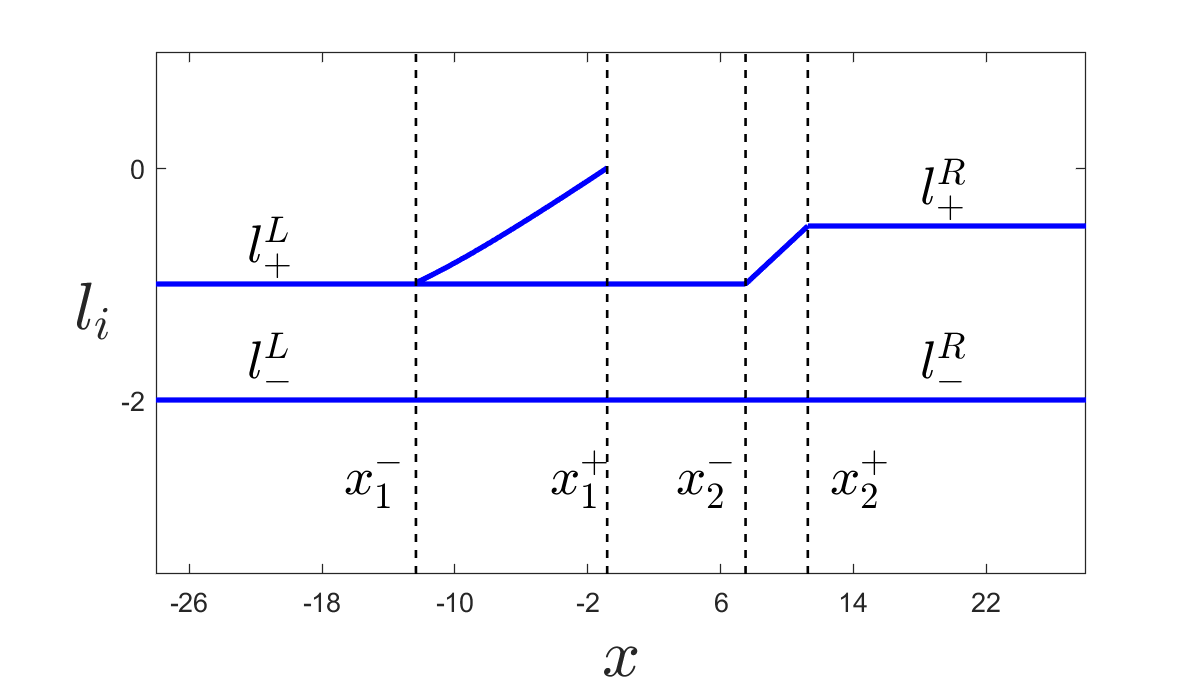}}\hfill
\subfigure[]{\includegraphics[width=0.25\linewidth]{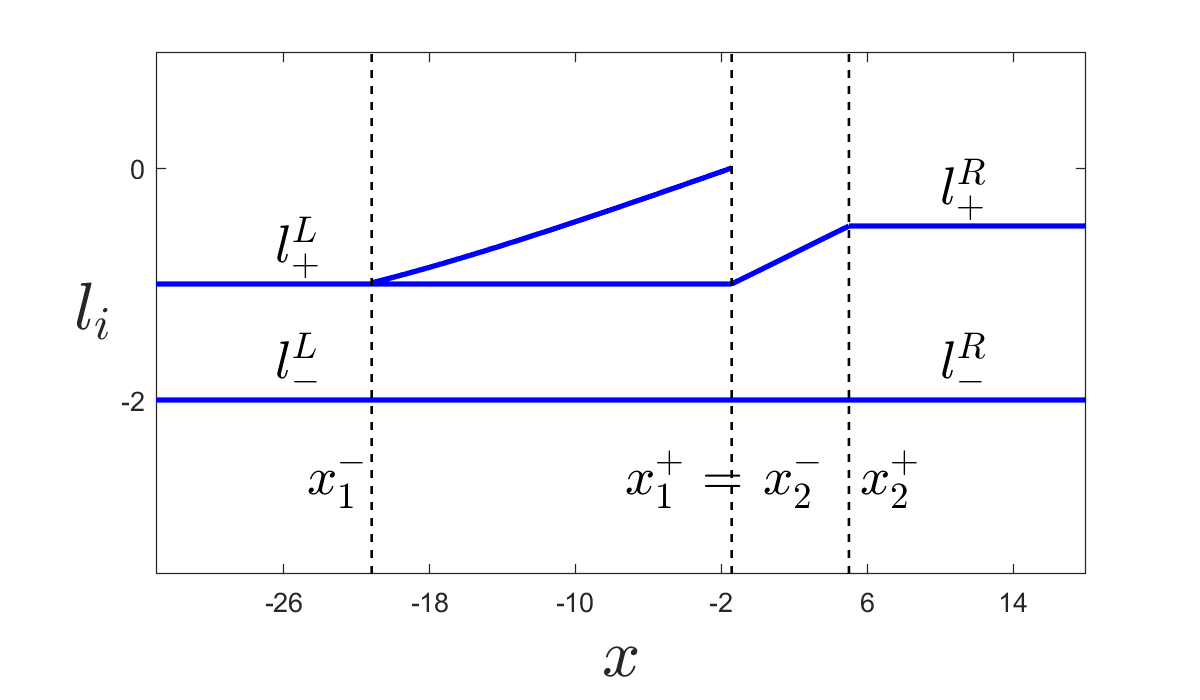}}\hfill
\subfigure[]{\includegraphics[width=0.25\linewidth]{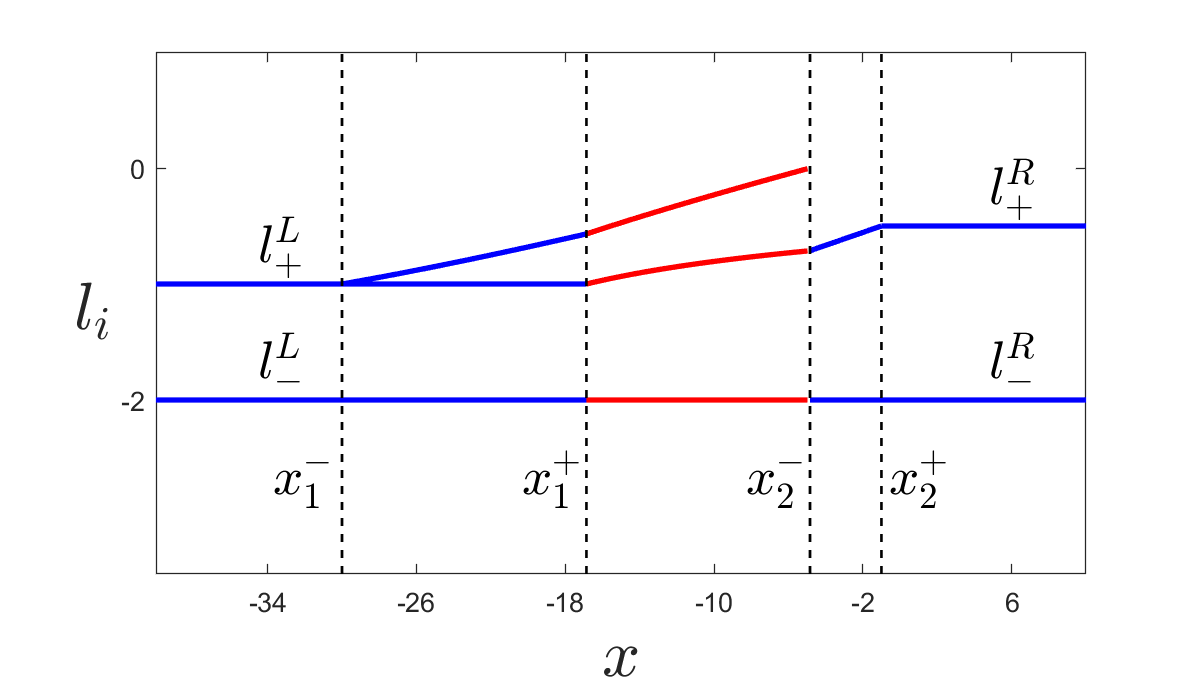}}\hfill
\subfigure[]{\includegraphics[width=0.25\linewidth]{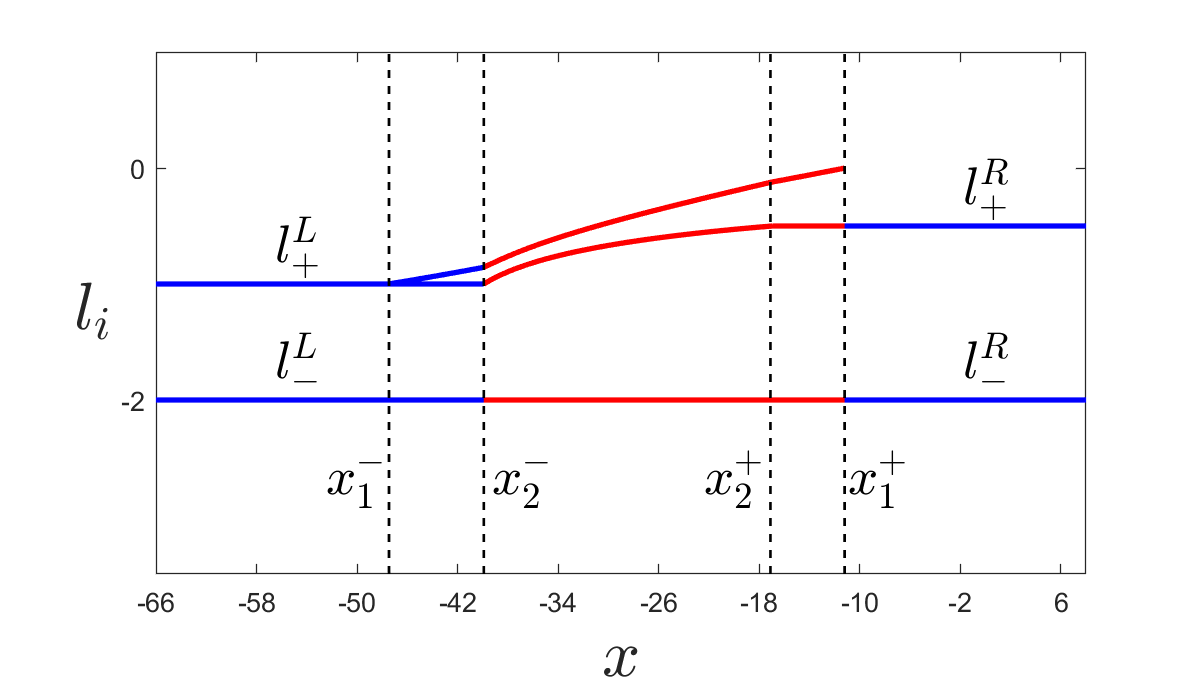}}
\flushleft{\footnotesize
\textbf{Fig.~$\bm{22}$.} (a) presents the boundaries of distinct wave regions throughout the evolution: blue lines indicate the boundaries of the CDSW, green lines indicate the boundaries of the RW, and red lines indicate the
boundaries of the interaction region. (b-f)The evolution of Riemann invariants, with solid red lines denoting the interaction region.}
\end{figure}
\begin{figure}[htbp]
\centering
\setcounter{subfigure}{0}
\subfigure[]{\includegraphics[width=0.25\linewidth]{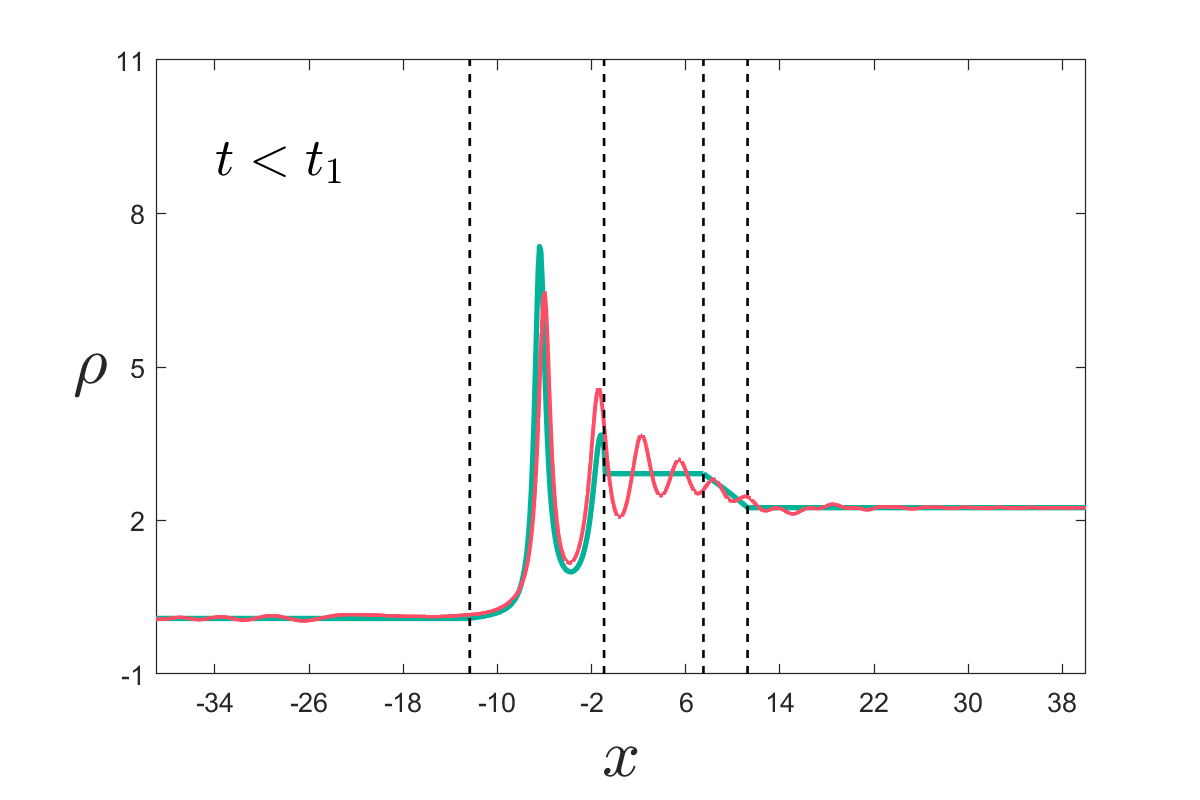}}\hfill
\subfigure[]{\includegraphics[width=0.25\linewidth]{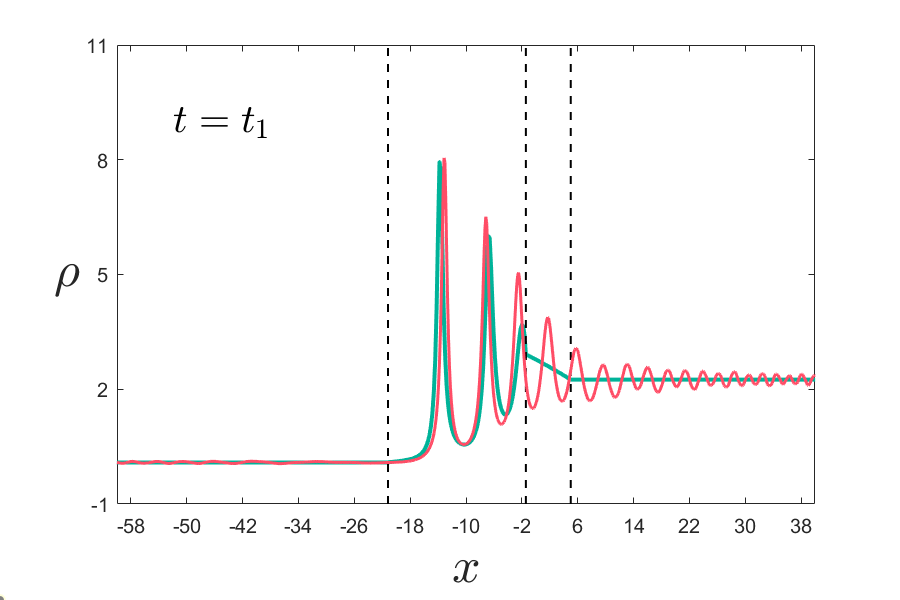}}\hfill
\subfigure[]{\includegraphics[width=0.25\linewidth]{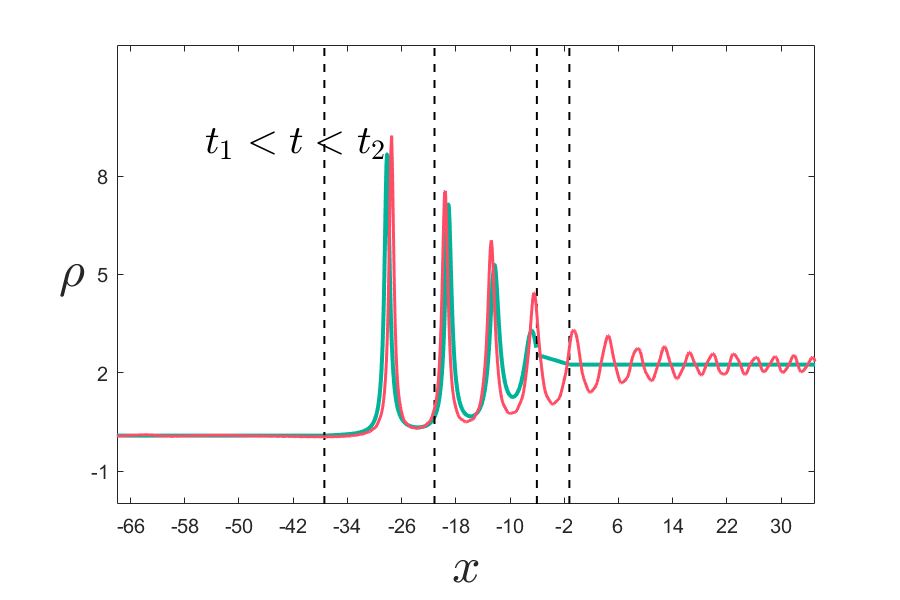}}\hfill
\subfigure[]{\includegraphics[width=0.25\linewidth]{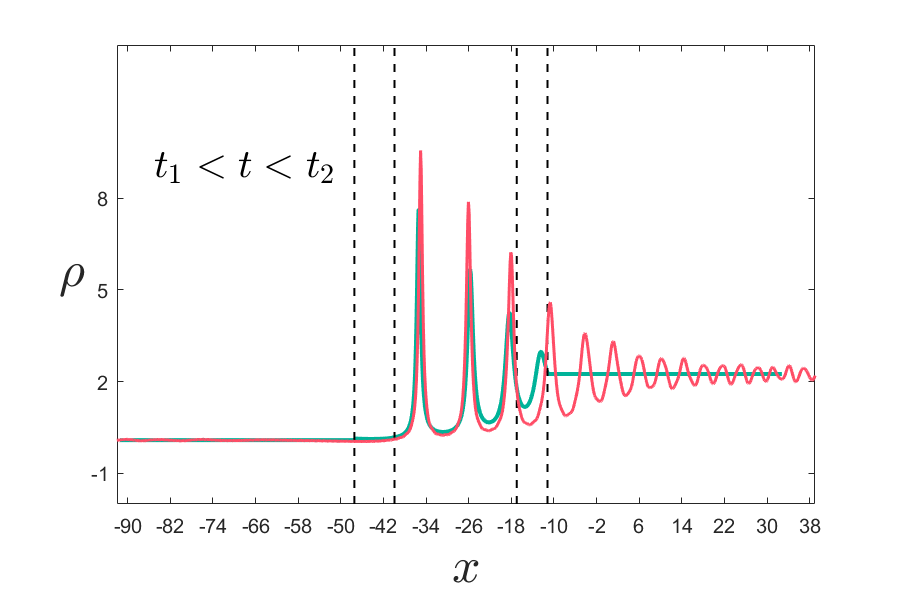}}\hfill
\subfigure[]{\includegraphics[width=0.25\linewidth]{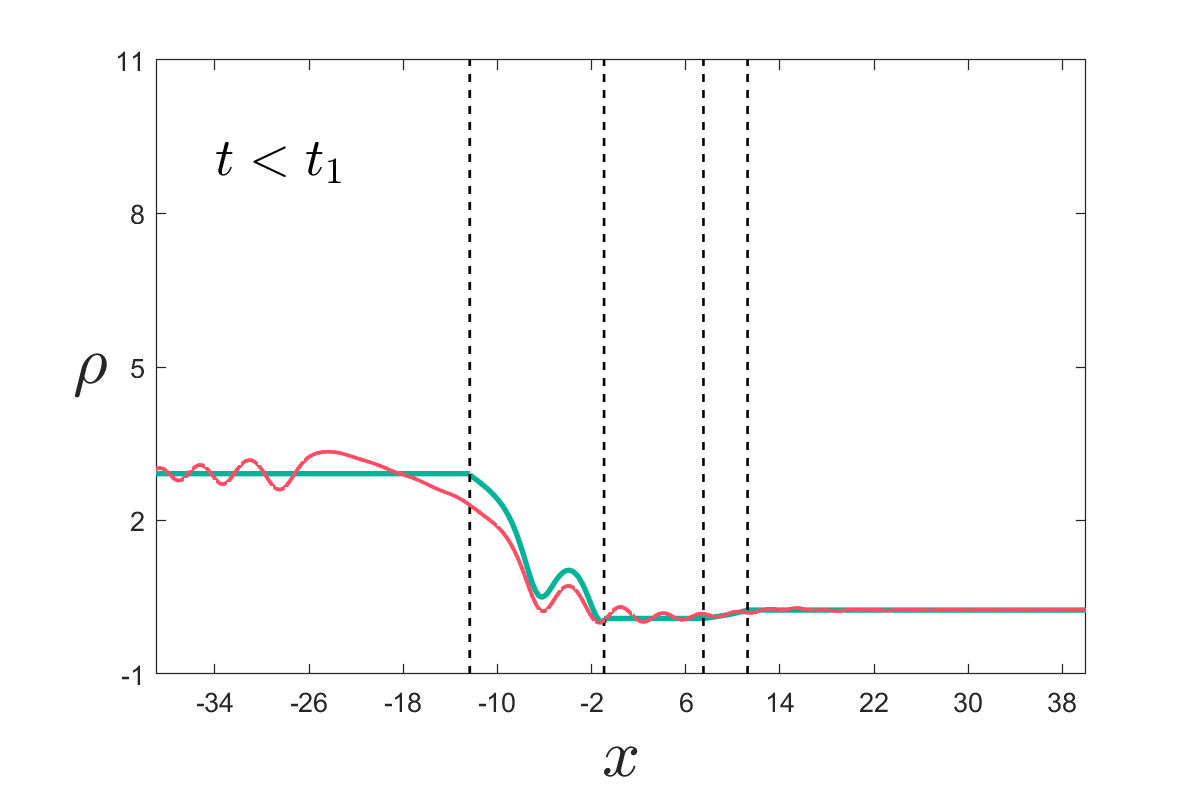}}\hfill
\subfigure[]{\includegraphics[width=0.25\linewidth]{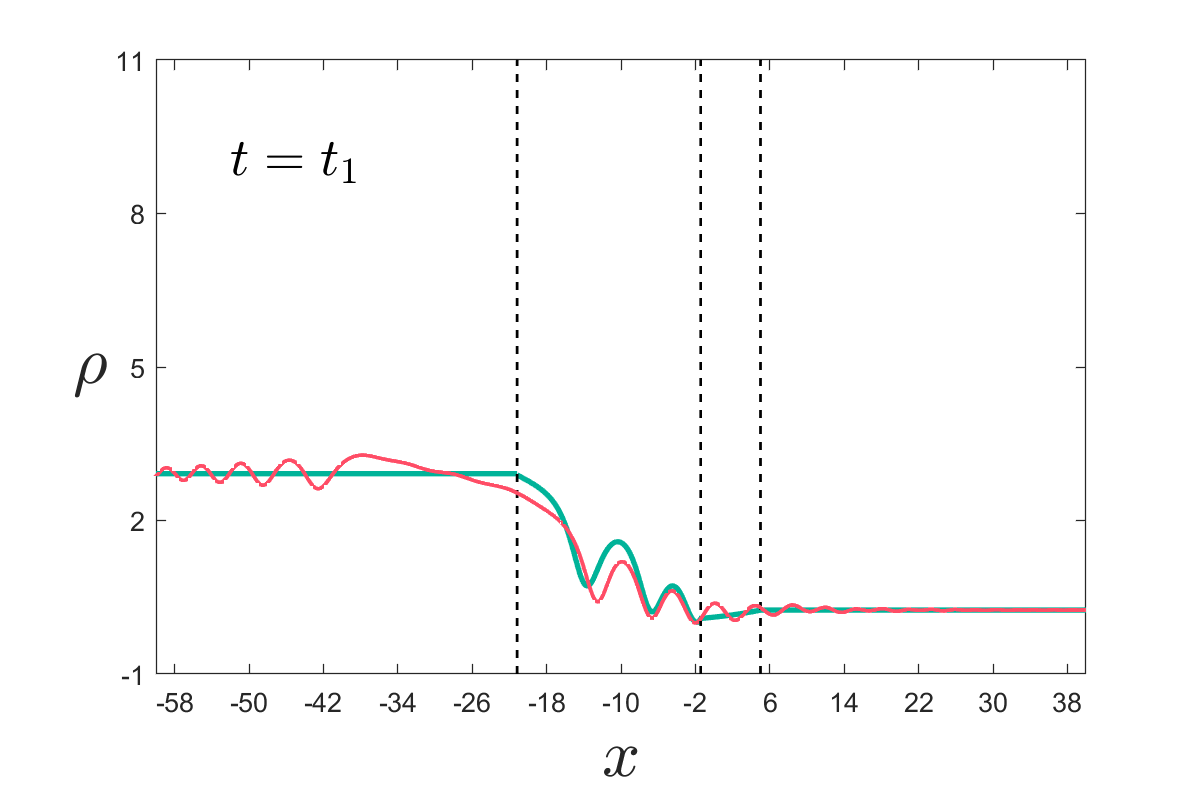}}\hfill
\subfigure[]{\includegraphics[width=0.25\linewidth]{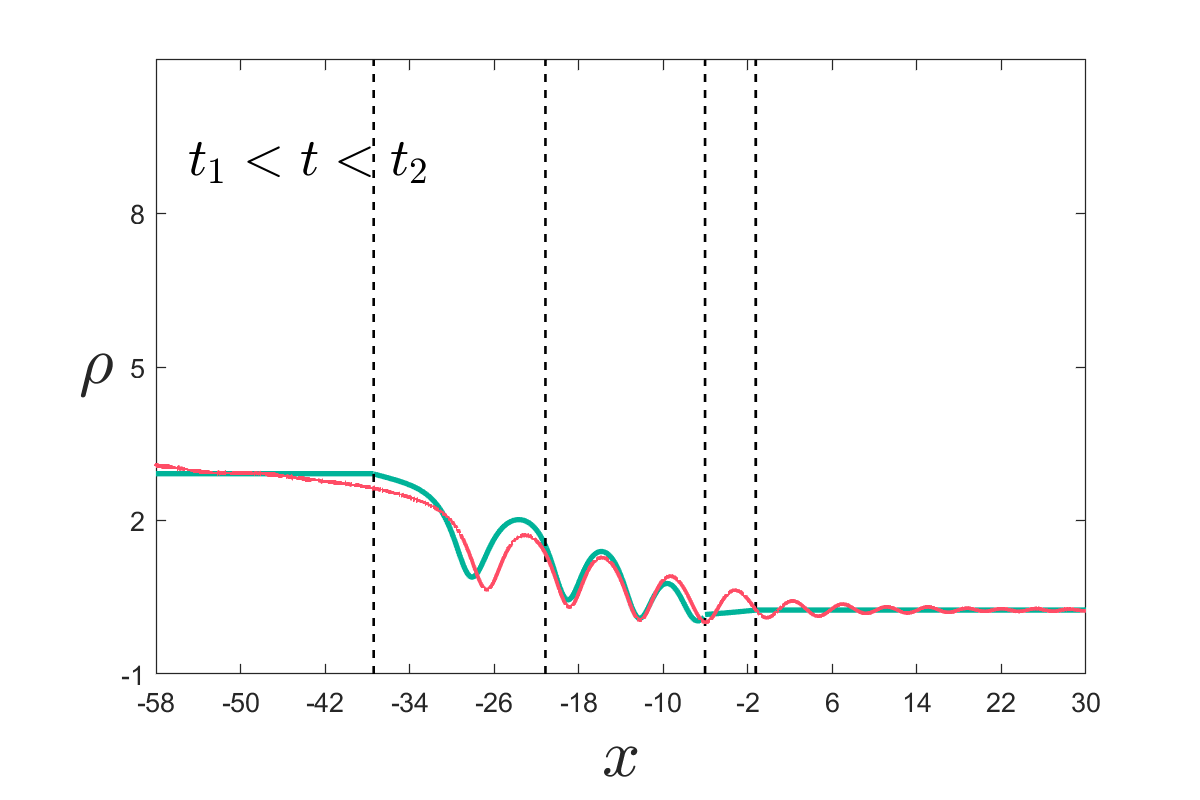}}\hfill
\subfigure[]{\includegraphics[width=0.25\linewidth]{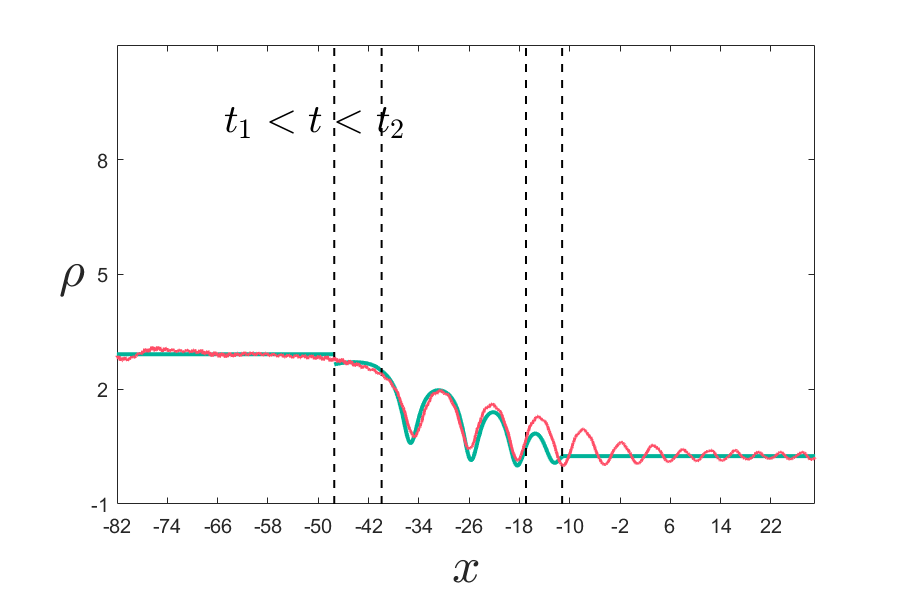}}
\flushleft{\footnotesize
\textbf{Fig.~$\bm{23}$.}  The CDSW analytical solutions constructed by mapping the Riemann invariants according to the relation given in Eqs. (2.7) and (2.8) (green solid line) and the numerical simulation (red solid line) solutions on x.
}
\end{figure}
\begin{figure}[htbp]
\centering
\setcounter{subfigure}{0}
\subfigure[]{\includegraphics[width=0.5\textwidth, height=0.25\textwidth]{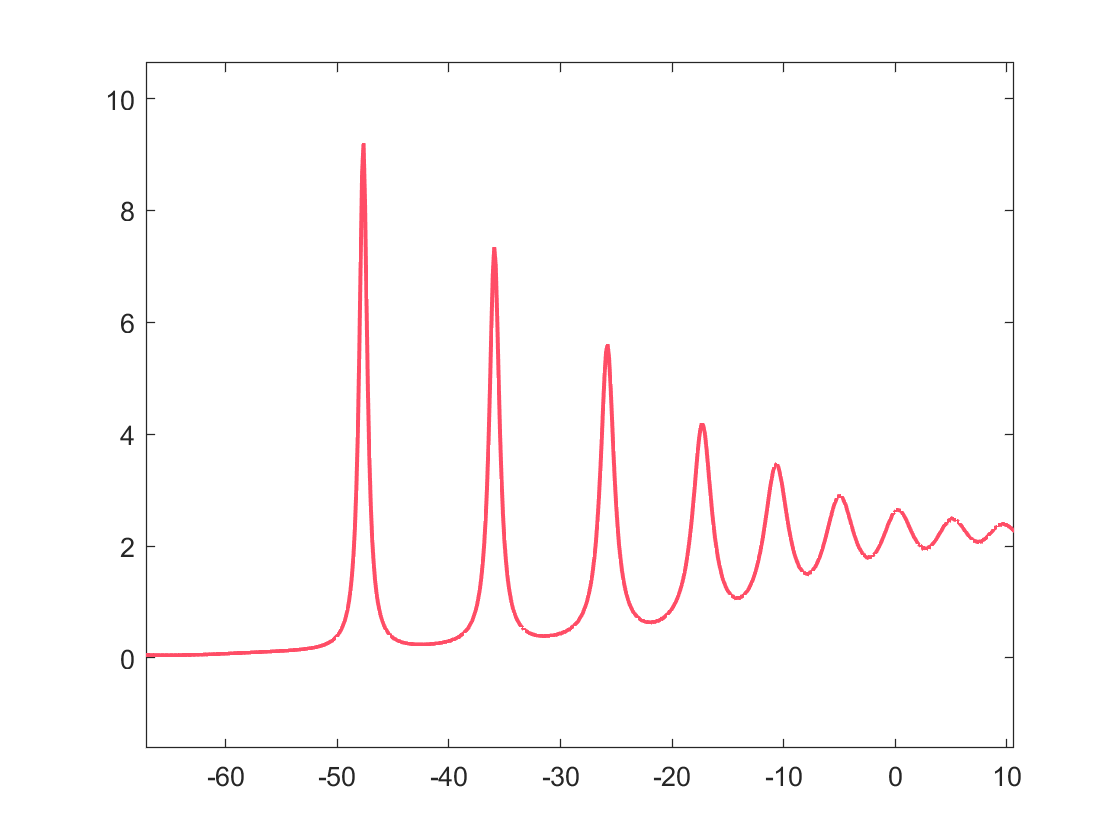}}\hfill
\subfigure[]{\includegraphics[width=0.5\textwidth, height=0.25\textwidth]{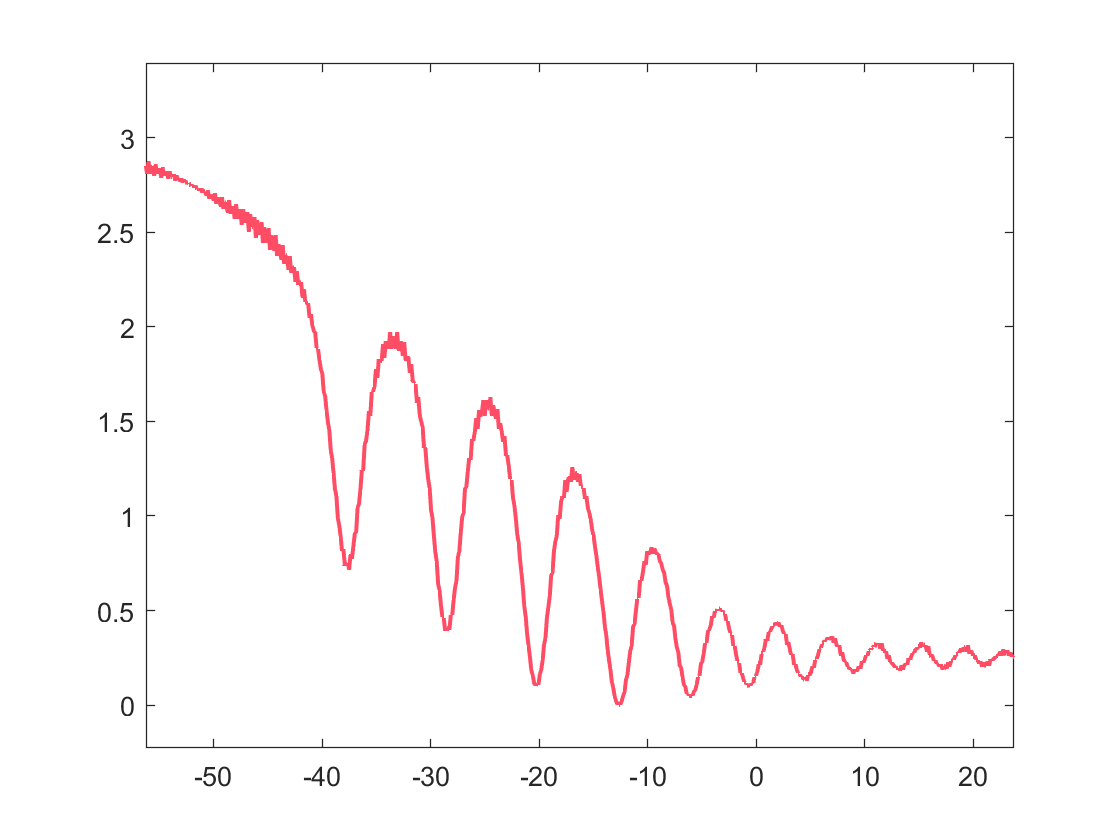}}
\flushleft{\footnotesize
\textbf{Fig.~$\bm{24}$.} The evolution of the algebraic soliton train in the long time limit as $t\rightarrow +\infty $.}
\end{figure}
\begin{figure}[htbp]
\centering
\setcounter{subfigure}{0}
\subfigure[]{\includegraphics[width=0.333\linewidth]{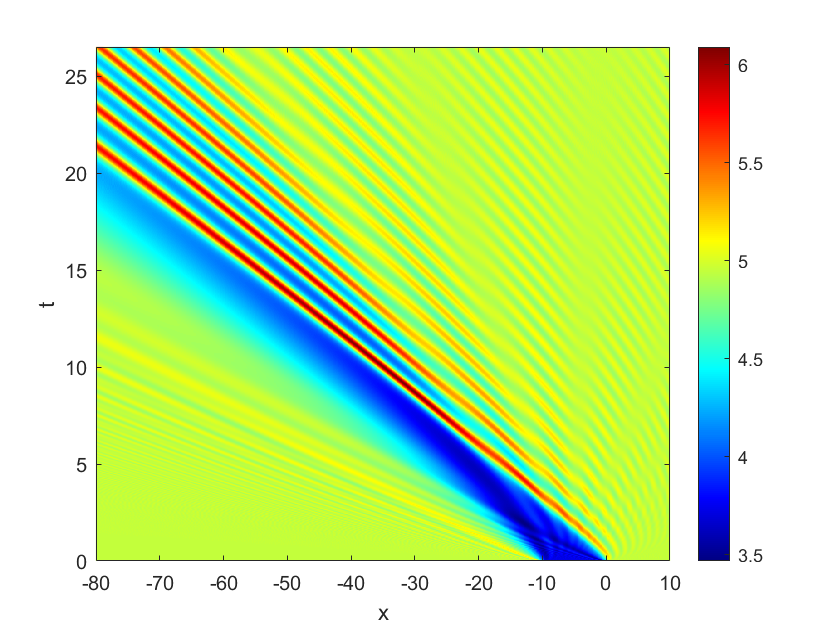}}\hfill
\subfigure[]{\includegraphics[width=0.333\linewidth]{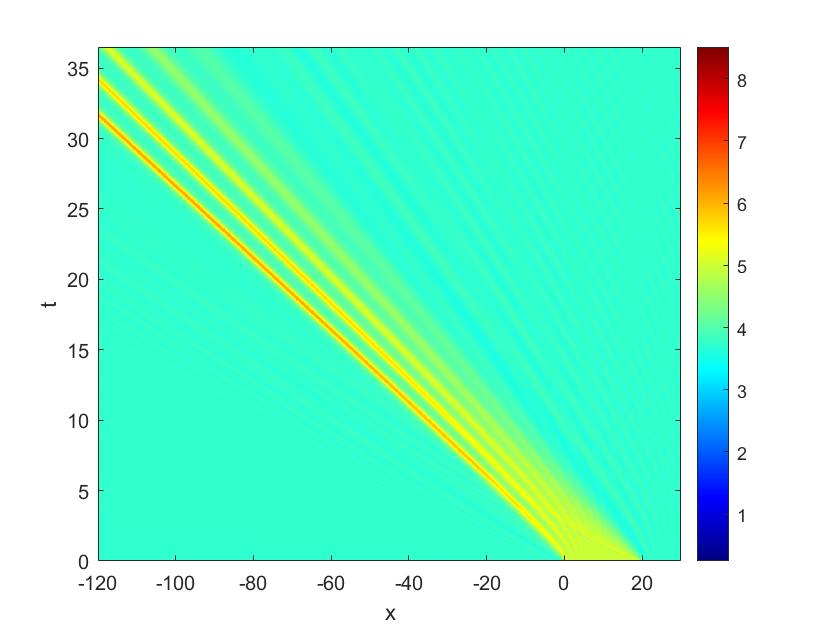}}\hfill
\subfigure[]{\includegraphics[width=0.333\linewidth]{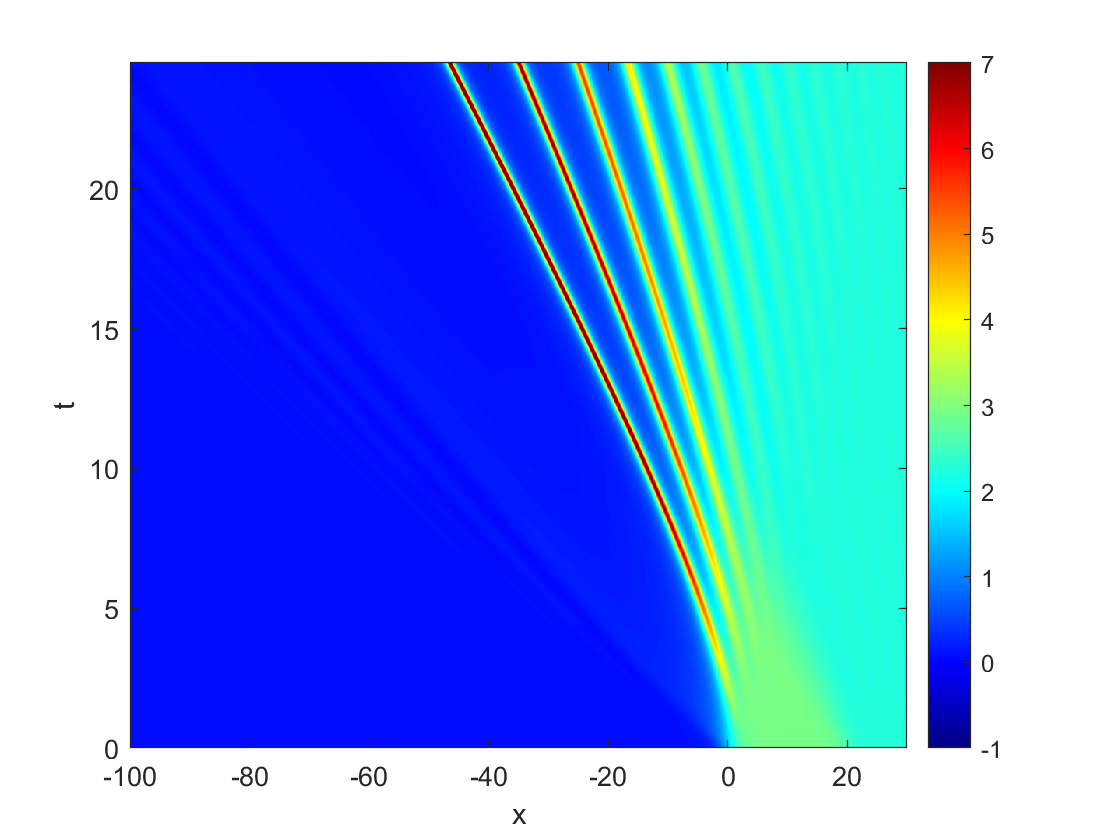}}\hfill
\subfigure[]{\includegraphics[width=0.333\linewidth]{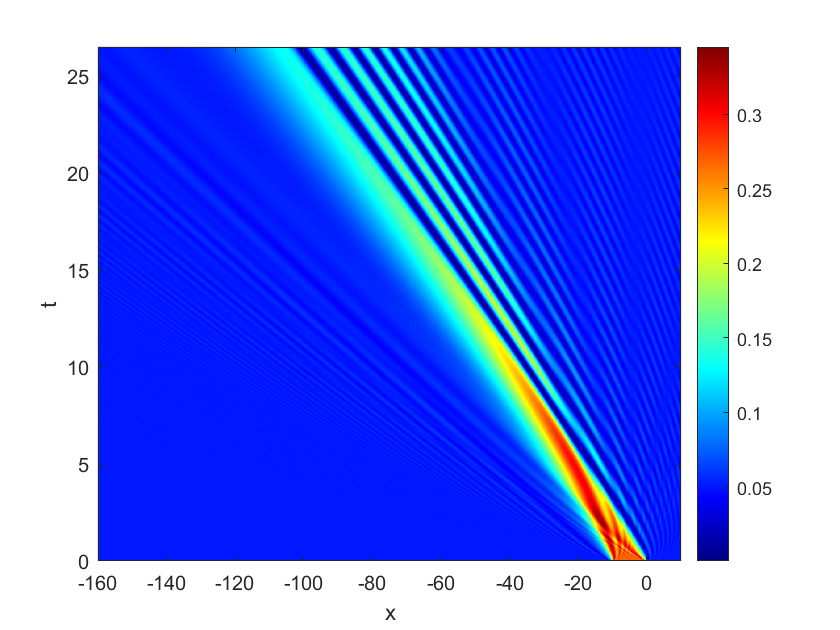}}\hfill
\subfigure[]{\includegraphics[width=0.333\linewidth]{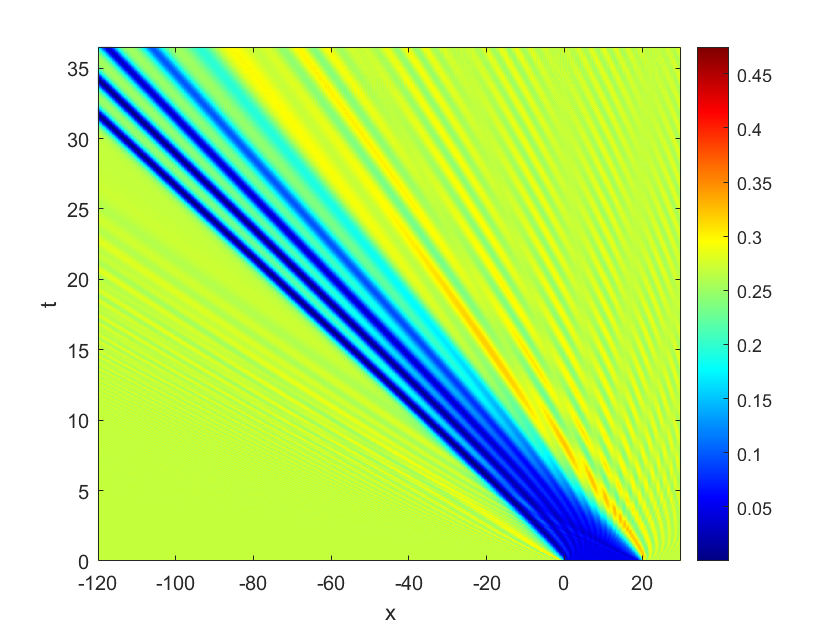}}\hfill
\subfigure[]{\includegraphics[width=0.333\linewidth]{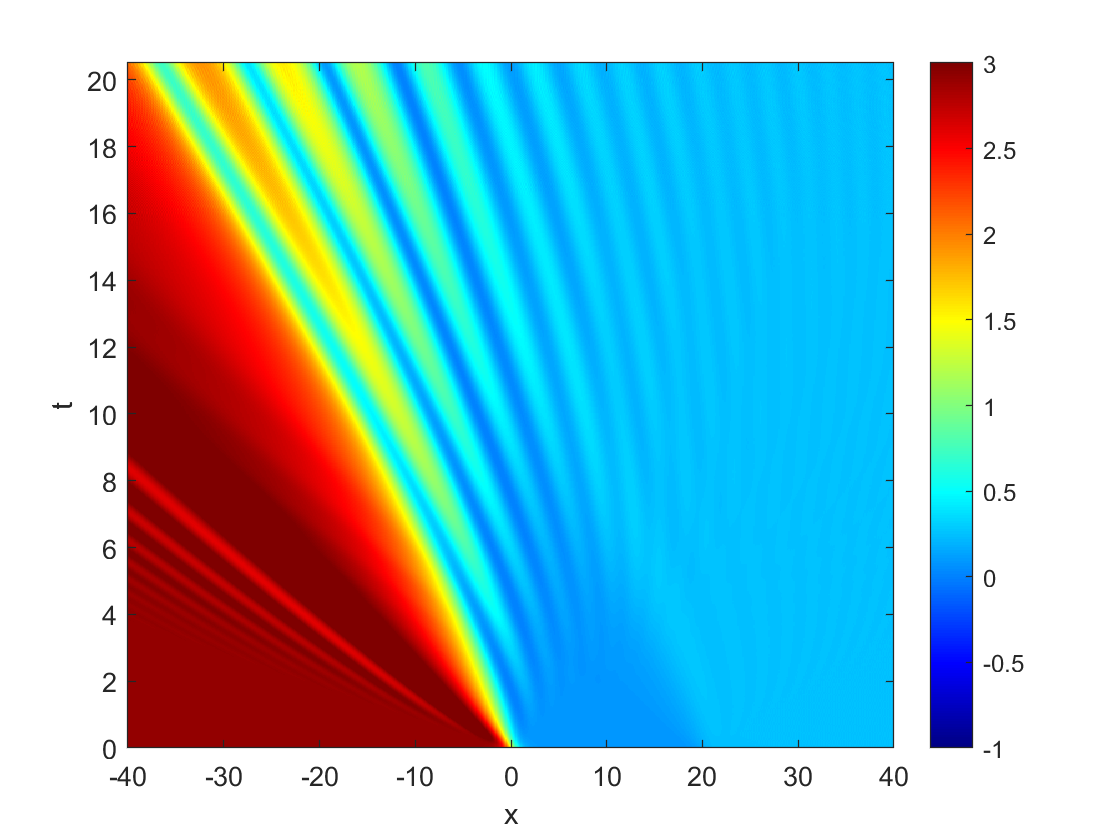}}
\flushleft{\footnotesize
\textbf{Fig.~$\bm{25}$.} The density figure of the two-wave merging regime after interaction.}
\end{figure}

The RW solution takes the following form
\begin{equation}
\begin{aligned}
l_-= l_-^0 , ~\frac{x-d}{t}=&V _ { + } (l_+ ,l_- ) = \frac{3l_++l_-^0}{2}.
\end{aligned}
\end{equation}
Its two boundaries are given by
\begin{equation}
x _ { 2 } ^{-} = d+ \frac { 3 l_+^L + l_-^0 } { 2 }t ,\   x _ { 2 } ^{+} = d + \frac { 3 l_+^R + l_-^0 } { 2 } t .
\end{equation}

The corresponding time $t_1$ and position $x_1$ are expressed as
\begin{equation}
t _ { 1 } =- \frac {d(l_-^0+l_+^L)} { l_-^0(3l_+^L+l_-^0) } , ~x _ { 1 } = -d \frac {(l_-^{0}-l_+^{L})^2} {2l_-^0(3l_+^L+l_-^0)}.
\end{equation}

$(ii)$ Being interaction $(t>t_1)$:

The modulated solution within the interaction zone is expressed as
\begin{equation}
\begin{aligned} 
&l_ 1= l_-^0 ,~l_ { 3 } = l_4,  \\ 
x - V _ {2, 4}t&= \left( 1 - \frac { \mathfrak { L } } { \partial _ { 2 , 4 } \mathfrak { L } } \partial _ { 2 , 4 } \right) f ( l _ { 2 } , l _ { 4 } ) , 
\end{aligned}
\end{equation}
where
\begin{equation}
f ( l_ { 2 } , l_ { 4 } ) = \frac {- 2 d~l_+^L} { \pi \sqrt { - l_ { 2 } ( l _ {4 }-l_+^L) } } ( \Pi _ { 1 } ( s , z ) - K( z ) ) ,
\end{equation}
and
$$
z = \frac { l_ { 4 } ( l_ { 2 } -l_+^L ) } {l_ { 2 } ( l_ { 4 } -l_+^L ) } ,~ s = \frac { l_ { 2 } -l_+^L } {l_ { 2 } } .
$$

The detailed plots are presented in Figs. 22 and 23. As $t\to\infty$, one can see $l_2\to l_3=l_4$ from the solution (4.29)
$$
t\approx d\frac{2l_{+}^{L}\ln \left( 1-m \right)}{3\pi l_4\sqrt{-l_4\left( l_4-l_{+}^{L} \right)}}\gg 1,
$$
and the modulated interaction region degenerates into an algebraic soliton train (see Fig. 24). We present the density profiles for three distinct cases in the merging regime (see Fig. 25).

Up to this point, we have completed our discussion of all two-step configurations. It can be seen that when the two steps of the wave structures correspond to two distinct Riemann invariants, one wave always overtakes the other, the two waves interact, and they eventually separate. By contrast, when both steps are associated with a single Riemann invariant, one wave catches up with the other and remains in sustained interaction without subsequent separation. All these distinct behavioural regimes stem from the differences in the characteristic velocities of the waves.

\vspace{7mm}\noindent\textbf{5 Conclusion}
\hspace*{\parindent}
\renewcommand{\theequation}{5.\arabic{equation}}\setcounter{equation}{0}\\

Based on Whitham modulation theory, we have conducted analytical and numerical investigations into the interaction between a RW and a classical DSW or a CDSW within the framework of the GI equation. The main conclusions are summarized below.

First, through constructing two-step piecewise constant initial conditions, we have fully covered all admissible configurations of the interaction between a single RW and a single DSW or CDSW. It is found that the interaction gives rise to two distinct evolutionary scenarios. In the separation scenario, the two waves exchange their spectral parameters and then propagate independently with modified phases. In particular, the mapping corresponding to the Riemann invariants of the RW is altered after the interaction between the RW and the CDSW. In the merging scenario, the wave structures undergo sustained interaction without separation. They evolve into a soliton train or a small-amplitude harmonic wave train in the long-time limit for DSW. And for CDSW, the resulting modulation region asymptotically into an an algebraic soliton train. The evolutionary outcome of the interaction is determined by the initial states, and analytical expressions for the wave boundaries at each evolutionary stage are also derived in this work.

Second, we have applied Whitham modulation theory to wave interaction problems in non-convex hydrodynamics, thereby establishing a systematic theoretical framework for characterizing the complete interaction dynamics. The full-stage evolution of the modulation region can be obtained by applying generalized hodograph transformation and establishing the corresponding boundary matching conditions. The resulting analytical predictions are in good agreement with numerical simulations, thereby confirming the validity of the modulation description in both the separation and merging regimes. It should be noted that the present study on two-wave interaction only involves one-phase modulation solutions. Extending the analysis to interactions involving two-phase or multi-phase solutions remains an important problem for future work.

This work advances the theory of nonlinear wave dynamics in non-convex dispersive systems, and provides theoretical support for experimental observation and manipulation in fields such as plasma physics and nonlinear optics.
\\

\noindent\textbf{Acknowledgments}
\hspace*{\parindent}

We express our sincere thanks to each member of our discussion group for their suggestions. This work has been supported by the Fund Program for the National Natural Science Foundation of China under Grant No. 12575005, the Shanxi Province Science Foundation under Grant No. 202303021221031, and the Research Project Supported by Shanxi Scholarship Council of China under Grant No. 2024-033.

\end{document}